\documentclass[twocolumn,
amsmath,amssymb,
 aps,
superscriptaddress,
altaffilletter
]{revtex4-2}

\usepackage{graphicx}\usepackage{dcolumn}\usepackage{bm}\usepackage{siunitx}
\usepackage{ifthen}
\usepackage[colorlinks = true, linkcolor = blue,]{hyperref}
\usepackage{xcolor}
\usepackage{isotope}
\usepackage{multirow}
\usepackage{datetime}

\usepackage{lineno}

\usepackage[super]{nth}
\usepackage{float}
\usepackage{booktabs}
\usepackage{xspace}
\usepackage[symbol]{footmisc}

\newcommand{\amu}{\ensuremath{a^{}_{\mu}}\xspace}
\newcommand{\gm}{\ensuremath{g\!-\!2}\xspace}
\newcommand{\gmtwo}{\gm}
\newcommand{\oa}{\ensuremath{\omega^{}_a}\xspace}

\newcommand{\opprimetilde}{\ensuremath{\tilde{\omega}'^{}_p}\xspace}

\newcommand{\RunOne}{Run-1\xspace}
\newcommand{\RunTwo}{Run-2\xspace}
\newcommand{\RunThree}{Run-3\xspace}

\newcommand{\RunOneA}{Run-1a\xspace}
\newcommand{\RunOneB}{Run-1b\xspace}
\newcommand{\RunOneC}{Run-1c\xspace}
\newcommand{\RunOneD}{Run-1d\xspace}
\newcommand{\RunThreeA}{Run-3a\xspace}
\newcommand{\RunThreeB}{Run-3b\xspace}
\newcommand{\RunTwoThree}{Run-2/3\xspace}

\newcommand{\authornote}[1]{{\let\thempfn\relax \footnotetext[0]{$\diamond${ }#1}}}

\begin{document}

\preprint{APS/123-QED}

\title{Detailed Report on the Measurement of the Positive Muon Anomalous Magnetic Moment to 0.20\,ppm}
\clearpage{}
\renewcommand{\thefootnote}{\fnsymbol{footnote}}
\affiliation{Argonne National Laboratory, Lemont, Illinois, USA}
\affiliation{Boston University, Boston, Massachusetts, USA}
\affiliation{Brookhaven National Laboratory, Upton, New York, USA}
\affiliation{Budker Institute of Nuclear Physics, Novosibirsk, Russia}
\affiliation{Center for Axion and Precision Physics (CAPP) / Institute for Basic Science (IBS), Daejeon, Republic of Korea}
\affiliation{Cornell University, Ithaca, New York, USA}
\affiliation{Fermi National Accelerator Laboratory, Batavia, Illinois, USA}
\affiliation{INFN, Laboratori Nazionali di Frascati, Frascati, Italy}
\affiliation{INFN, Sezione di Napoli, Naples, Italy}
\affiliation{INFN, Sezione di Pisa, Pisa, Italy}
\affiliation{INFN, Sezione di Roma Tor Vergata, Rome, Italy}
\affiliation{INFN, Sezione di Trieste, Trieste, Italy}
\affiliation{Department of Physics and Astronomy, James Madison University, Harrisonburg, Virginia, USA}
\affiliation{Institute of Physics and Cluster of Excellence PRISMA+, Johannes Gutenberg University Mainz, Mainz, Germany}
\affiliation{Joint Institute for Nuclear Research, Dubna, Russia}
\affiliation{Department of Physics, Korea Advanced Institute of Science and Technology (KAIST), Daejeon, Republic of Korea}
\affiliation{Lancaster University, Lancaster, United Kingdom}
\affiliation{Michigan State University, East Lansing, Michigan, USA}
\affiliation{North Central College, Naperville, Illinois, USA}
\affiliation{Northern Illinois University, DeKalb, Illinois, USA}
\affiliation{Regis University, Denver, Colorado, USA}
\affiliation{School of Physics and Astronomy, Shanghai Jiao Tong University, Shanghai, China}
\affiliation{Tsung-Dao Lee Institute, Shanghai Jiao Tong University, Shanghai, China}
\affiliation{Institut f\"ur Kern- und Teilchenphysik, Technische Universit\"at Dresden, Dresden, Germany}
\affiliation{Universit\`a del Molise, Campobasso, Italy}
\affiliation{Universit\`a di Udine, Udine, Italy}
\affiliation{Department of Physics and Astronomy, University College London, London, United Kingdom}
\affiliation{University of Illinois at Urbana-Champaign, Urbana, Illinois, USA}
\affiliation{University of Kentucky, Lexington, Kentucky, USA}
\affiliation{University of Liverpool, Liverpool, United Kingdom}
\affiliation{Department of Physics and Astronomy, University of Manchester, Manchester, United Kingdom}
\affiliation{Department of Physics, University of Massachusetts, Amherst, Massachusetts, USA}
\affiliation{University of Michigan, Ann Arbor, Michigan, USA}
\affiliation{University of Mississippi, University, Mississippi, USA}
\affiliation{University of Virginia, Charlottesville, Virginia, USA}
\affiliation{University of Washington, Seattle, Washington, USA}
\affiliation{City University of New York at York College, Jamaica, New York, USA}
\author{D.~P.~Aguillard} \affiliation{University of Michigan, Ann Arbor, Michigan, USA}
\author{T.~Albahri} \affiliation{University of Liverpool, Liverpool, United Kingdom}
\author{D.~Allspach} \affiliation{Fermi National Accelerator Laboratory, Batavia, Illinois, USA}
\author{A.~Anisenkov} \altaffiliation[Also at ]{Novosibirsk State University.} \affiliation{Budker Institute of Nuclear Physics, Novosibirsk, Russia}
\author{K.~Badgley} \affiliation{Fermi National Accelerator Laboratory, Batavia, Illinois, USA}
\author{S.~Bae{\ss}ler} \altaffiliation[Also at ]{Oak Ridge National Laboratory.} \affiliation{University of Virginia, Charlottesville, Virginia, USA}
\author{I.~Bailey} \altaffiliation[Also at ]{The Cockcroft Institute of Accelerator Science and Technology, Daresbury, United Kingdom.} \affiliation{Lancaster University, Lancaster, United Kingdom}
\author{L.~Bailey} \affiliation{Department of Physics and Astronomy, University College London, London, United Kingdom}
\author{V.~A.~Baranov\textsuperscript{\dag}} \affiliation{Joint Institute for Nuclear Research, Dubna, Russia}
\author{E.~Barlas-Yucel} \affiliation{University of Illinois at Urbana-Champaign, Urbana, Illinois, USA}
% ...
\footnotetext[2]{Deceased.}
\renewcommand{\thefootnote}{\arabic{footnote}}
\collaboration{The Muon \gmtwo Collaboration} \noaffiliation
\vskip 0.25cm

\clearpage{}

\date{\today}

\begin{abstract}
We present details on a new 
measurement of the muon magnetic anomaly, $a_\mu = (g_\mu -2)/2$. The result is based on positive muon data taken at Fermilab's Muon Campus during the 2019 and 2020 accelerator runs.  The measurement uses $3.1$ GeV$/c$ polarized muons stored in a $7.1$-m-radius storage ring with a \SI{1.45}{T} uniform magnetic field. The value of $ a_{\mu}$ is determined from the measured difference between the muon spin precession frequency and its cyclotron frequency.
This difference is normalized to the strength of the magnetic field, measured using Nuclear Magnetic Resonance (NMR).  The ratio is then corrected 
 for small contributions 
 from beam motion, beam dispersion, and transient magnetic fields.  We measure $a_\mu = 116 592 057 (25) \times 10^{-11}$ (\SI{0.21}{ppm}).  This is the world's most precise measurement of this quantity and represents a factor of $2.2$ improvement over our previous result based on the 2018 dataset. In combination, the two datasets yield $a_\mu(\text{FNAL}) = 116 592 055 (24) \times 10^{-11}$ (\SI{0.20}{ppm}).
Combining this with the measurements from Brookhaven National Laboratory for both positive and negative muons, the new world average is $a_\mu$(exp)$ = 116 592 059 (22) \times 10^{-11}$ (\SI{0.19}{ppm}).
\end{abstract}

\maketitle

\tableofcontents

\section{Introduction} \label{sec:intro}
The anomalous magnetic moment of a charged lepton arises from radiative corrections and interactions with virtual particles. It can be calculated for Standard Model (SM) interactions with high precision. Measurements of the muon magnetic anomaly, expressed as $a_{\mu} = (g_{\mu}-2)/2$, with similar or greater precision thus challenge the SM calculations and probe possible Beyond the Standard Model (BSM) physics. Measurement of the electron $a_e$ provides a 0.13-ppt determination of $g_e$, which is mostly sensitive to electromagnetic interactions~\cite{PhysRevLett.130.071801}. The muon, due to its greater mass, is approximately $43000$ times more sensitive to BSM interactions of new heavy particles. 

In a series of measurements with both positive and negative muons, the E821 collaboration at Brookhaven National Laboratory (BNL) determined $a_\mu$ with a relative precision of 0.54 ppm~\cite{PhysRevD.73.072003} and found a discrepancy with the SM calculation of about three standard deviations at the time. 
Improved precision of the SM prediction in subsequent years led to increased significance, and $a_\mu$ became one of the largest measured discrepancies with the SM and a possible signal of BSM physics
~\cite{Stockinger:2006zn,Czarnecki:2001pv}. 
On April 7,  2021, the Muon \gm Collaboration 
released the first result for $a_\mu$ based on the \RunOne 2018 data campaign at Fermilab~\cite{Run1PRL,Run1PRDomegaa, Run1PRAField,Run1PRAB}, which was consistent with the BNL results.
Meanwhile, newer SM calculations~\cite{Borsanyi:2020mff} challenge the 2020 \gm Theory Initiative White Paper~\cite{Aoyama:2020ynm} recommendation.
In 2023, the collaboration published the Run-2/3 result~\cite{PhysRevLett.131.161802}. This paper provides the analysis details of that result.

The magnetic anomaly of \SI{3.1}{GeV} muons is measured in 
a magnetic storage ring with a uniform vertical magnetic field $\vec{B}$ and weakly focusing quadrupole electric fields $\vec{E}$. For $g_\mu > 2$, the muon spin precession frequency $\vec\omega_S$ is greater than the cyclotron frequency $\vec\omega_C$, resulting in
the anomalous-precession frequency
$\vec\omega_a=\vec\omega_s-\vec \omega_c$. 
For relativistic muons on the ideal orbit with a perfectly uniform magnetic field,
\begin{eqnarray}
\vec\omega_a &=&
- a_{\mu}\frac{q}{m}\vec B\nonumber\\
&+&\frac{q}{m}\Big [\ \Big (a_{\mu} - \frac{1}{\gamma^2-1}\Big )\frac{\vec{\beta}\times \vec{E}}{c}
+ 
   a_{\mu}\Big (\frac{\gamma}{\gamma+1}\Big ) (\vec{\beta} \cdot \vec{B})\vec{\beta} \Big ],\nonumber\\
   \label{eq1}
\end{eqnarray}
where $q$ is the charge, $m$ is the mass, $\beta$ is the velocity ratio with respect to the speed of light, and $\gamma$ is the Lorentz factor of the muon. 
The second term on the right-hand side, proportional to {$E$}, 
vanishes for $\gamma = \sqrt{(1+1/a_{\mu})}\approx29.3$. This corresponds to momentum $p_{0}\approx\SI{3.094}{GeV/c}$, called the ``magic momentum''.
In the absence of vertical betatron motion, the muon velocity is perpendicular to $\vec B$, leading to cancellation of the third term.

The magnitude of the measured { anomalous-precession frequency}, 
corrected for the momentum spread, betatron motion, and beam-dynamics effects is proportional to $\tilde B$, the  magnetic field magnitude averaged over the muon distribution in time and space.  
We express $\tilde B$ in terms of the measured NMR frequency of protons in a spherical water sample at a reference temperature $T_r$
\begin{equation}
\tilde\omega_p^\prime=\gamma_p^\prime(T_r)\tilde B,
    \label{eq:omega_pB}
\end{equation}
where $\gamma_p^\prime$ is the gyromagnetic ratio of protons in H$_2$O known with high precision at $T_r$.
Combining the first term on the right-hand side of Eq.~\eqref{eq1} and Eq.~\eqref{eq:omega_pB} 
 allows $a_{\mu}$ to be expressed as a ratio of frequencies,
\begin{equation} \label{eq:amutoR}
a_{\mu}\propto \frac{{\omega}_a}{\tilde{\omega}'_p(T_r)}\equiv \mathcal{R}_\mu'(T_r)
.
\end{equation}

Parity violation in the weak decay of the muon allows measurement of the anomalous-precession frequency $\omega_a$. 
In the muon rest frame, the positron emission direction correlates with the muon spin direction, most strongly for high-energy positrons. In the laboratory frame, this results in a $\omega_a$-dependent modulation of the positron energy spectrum.
Fits to the positron time distribution extract the measured frequency $\omega_{a}^m$.
Details are provided in Sec.~\ref{sec:oa}.

Five beam-dynamics-driven corrections are applied to the measured spin precession frequency $\omega_a^m$.
The electric-field correction $C_e$ accounts for the electric field contribution due to the muon momentum spread. The pitch correction $C_p$ accounts for the vertical betatron motion of the muons. $C_{ml}$ accounts for the muon losses due to the finite aperture of the storage ring. The phase acceptance correction $C_{pa}$ accounts for the injected muons' phases with respect to the detector acceptance, and finally, the differential decay corrections $C_{dd}$ account for the correlation between spin phase and momentum of the muons. 
Details are provided in Sec.~\ref{sec:bd}.

The muon-averaged magnetic field expressed in the precession frequency of shielded protons $\tilde{\omega}'_{p}$ is reconstructed from a combination of mapping and tracking the magnetic field in the muon storage region 
and weighting by the reconstructed muon distribution $M(x,y,\phi,t)$, with $x$ and $y$ the horizontal and vertical transverse coordinates, $\phi$ the azimuth in the storage ring, and $t$ the time. 
The magnetic field maps have to be corrected for transient
perturbations 
that are synchronous with the muon injection due to the eddy currents from the magnetic kick required to move the muons to stored orbit radius ($B_K$) and due to vibrations induced in the field plates of the pulsed electrostatic quadrupoles ($B_Q$). 
Details are provided in Sec.~\ref{sec:field}.

Including the corrections, we can schematically express the ratio of the measured frequencies as
\begin{eqnarray} \label{eq:R}
   \mathcal{R}_\mu'(T_r)=\frac{\omega_{a}^m\left(1+C_{e}+C_{p}+C_{m l}+C_{dd}+C_{pa}\right)}{\langle \omega_p^\prime\times M\rangle\Big (1+B_K+B_Q \Big )},
\end{eqnarray}
where $\langle \omega_p^\prime\times M\rangle$ represents the muon weighting of the magnetic field (Sec.~\ref{sec:field}).

Following an overview of the experimental setup in Sec.~\ref{sec:detector}, we describe the datasets, run conditions, and main differences compared to \RunOne in Sec.~\ref{sec:datasets}. The analysis and extraction of $\omega_a$ and beam-dynamics corrections are discussed in Sec.~\ref{sec:oa} and \ref{sec:bd}. The determination of $\tilde\omega_p^\prime$ is detailed in Sec.~\ref{sec:field}. Consistency checks over the dataset and the calculation of $a_\mu$ are presented in Sec.~\ref{sec:sliceanddice} and \ref{sec:result}, and our result is put into the context of the current SM calculation in Sec.~\ref{sec:theory}. Appendices cover details of the analyses and the combination of results.

 Throughout this paper, frequencies are expressed as angular frequencies ($\omega$ in rad/s) and rotation frequencies ($\omega/2\pi$ or $f$) as appropriate in the context.

\section{The Muon \ensuremath{\mathbf{g\!-\!2}} experimental setup and simulation packages} 
\label{sec:detector}
\subsection{Experimental setup}
The Fermilab Muon \gm (E989) Experiment uses the same magic-momentum measurement principle developed initially for the CERN III experiment~\cite{BAILEY19791}. Furthermore, the Fermilab experiment employs the same storage ring and muon injection principle of E821 at BNL \cite{PhysRevD.73.072003} but has improved instrumentation for the magnetic field and muon spin precession frequency measurements.

The superconducting storage ring magnet is made of 12 segments each consisting of a continuous iron yoke~\cite{SR-NIM}. The C-shape of the magnet cross-section faces the interior of the ring so that positrons from muon decay, which spiral inward, can travel unobstructed by the magnet yoke to detectors placed around the interior of the storage ring. 
The strong vertical magnetic field is generated by four liquid helium-cooled superconducting coils and shaped by 36 high-purity iron pole pieces on top and the bottom of the opening.
To improve the field uniformity, edge shims and iron foils are used to control the transverse gradients and fine tune the magnetic field over the entire azimuthal and transverse storage volume. A set of magnetic coils with individually controlled currents run parallel to the muon beam above and below the vacuum chambers and are trimmed to achieve field uniformity in the storage region to better than one part per million~\cite{Run1PRAField} averaged around the ring.
The magnet power supply is adjusted continuously by a feedback system that stabilizes the field measured by NMR probes. This compensates for effects such as the thermal expansion of the ring.

Every \SI{1.4}{s}, a burst of eight bunches or fills every \SI{10}{ms}, followed by the same pattern approximately \SI{267}{ms} later, of ${\cal O}(10^5)$ {$\sim$}\SI{96}{\percent} polarized positive muons are delivered to the storage ring~\cite{PhysRevAccelBeams.20.111003}.
The initial momentum distribution of a fill has a width of \SI{1.6}{\percent} centered on the magic momentum of $p_0=\SI{3.094}{GeV/c}$.
Five collimators are positioned inside the storage ring to confine stable muon orbits within a torus of major radius $R \approx R_0$ and minor radius $r \approx \SI{4.5}{cm}$. Per fill, approximately 5000 muons with a momentum spread around \SI{0.15}{\percent} RMS are stored for up to \SI{700}{\micro\second}.
The central orbit radius is $R_{0}=\SI{7.112}{\m}$, with a cyclotron period of $T_c$=\SI{149.1}{ns} at $B=\SI{1.451}{T}$.

Before entering the storage ring, the muon beam passes through a scintillator detector and three scintillating fiber detectors. The scintillator detector is a 1-mm-thick plastic scintillator coupled via light guides to two photomultiplier tubes (PMTs). This detector provides the time reference (called $T_0$) for each fill, the time profile of the beam,  and the integrated beam intensity used for determining the beam storage efficiency and performing quality monitoring. 
After the $T_0$ detector, the muons pass through three scintillating fiber detectors that measure the horizontal and vertical beam profile before and after the injection. They comprise the Inflector Beam Monitoring System (IBMS). The first two are made of a $16 \times 16$ grid of 0.5mm-diameter scintillating fibers read out by \SI{1}{\milli\meter\squared} silicon photo-multipliers (SiPMs). The third IBMS detector (IBMS3) only has the vertical fibers to measure the horizontal plane profile. It can be deployed to either measure the profile at injection or multiple turns into beam storage. During normal data taking it is in a retracted position to avoid degrading the beam. 

Muons tangentially enter the storage ring from a low-field region through a superconducting inflector magnet.
This inflector magnet cancels the storage ring magnetic field locally and provides a virtually field-free injection channel.
The particles are displaced \SI{77}{mm} radially outward from the radial center of the storage region and are not on trajectories suitable for storage in the ring. A set of three fast non-ferric pulsed magnetic kickers is placed a quarter turn downstream from the injection point. 
The kickers are composed of three 1.27-m-long aluminum plates. Pulsing the kickers at $\sim$\SI{4.3}{kA} during the first turn after injection reduces the total magnetic field in the kicker region. This brief reduction deflects the muons onto the radially centered trajectory. Ideally, this pulse would last \SI{120}{ns}, which is a typical length of injected muon bunches. However, significant upgrades to the system were required to reach a FWHM around the cyclotron period to minimize the kick on the second turn. In addition, reflections and eddy currents are induced that have been the subjects of extensive dedicated studies. Detailed characterization of the kicker system and the upgrade effort are described in Ref.~\cite{schreckenberger_fast_2021}.

Four electrostatic quadrupoles (ESQs) distributed around the storage ring provide vertical focusing. Each ESQ has a long (spanning \SI{26}{\degree}) and a short (spanning \SI{13}{\degree}) section. 
The ESQ plates are charged before each beam injection, remain powered for about \SI{700}{\micro\second} after beam injection, and get discharged after the fill. Pulsing is required to ensure a stable operation voltage.  
Muons can be stored for up to ten times the muon lab-frame lifetime. 
The pulsing of the ESQ plates results in resonant mechanical vibrations that cause magnetic field perturbations synchronous to the muon injection that have been measured to determine a correction to the muon-averaged magnetic field.

A set of four fiber-detector arrays (harps) positioned 
around the ring monitors the beam profile and motion directly in the storage region. The fiber harps comprise horizontal and vertical planes of scintillating fibers that destructively measure the stored muons and can be inserted for dedicated systematic runs.  
Fiber-harp data are used to measure the beam momentum distribution, the cyclotron frequency, and the debunching of the muon beam during a fill.

The magnetic field is determined by mapping within the storage volume and tracking during muon storage and data taking. Mapping is accomplished with a trolley consisting of $17$ NMR probes housed in a movable aluminum shell that is pulled through the storage ring on rails. It measures with centimeter-scale spacing in both azimuthal and transverse directions. A high-purity calibrated water NMR probe, mounted on a 3D movable arm~\cite{FlayPP}, calibrated the trolley probes in the storage ring vacuum before \RunTwo and after \RunThree. The trolley is removed from the storage volume during data taking, and an array of 378 NMR probes, called fixed probes, help track the field. The fixed probes are located in grooves on the outer surfaces of the vacuum chambers above and below the storage volume. While the trolley is mapping the field, fixed probe measurements and trolley measurements are synchronized. The entire chain of NMR measurements is calibrated to provide the precession frequency of shielded protons in a spherical water sample at \SI{34.7}{\degree C}.

The positrons from stored positive muon decays are detected in $24$ calorimeter stations 
located equidistantly around the interior arc of the storage ring vacuum chamber. These calorimeters use lead fluoride (PbF$_2$) crystals as Cherenkov radiators from which signals are read out via SiPMs \cite{Kaspar_2017,KHAW2019162558,ANASTASI201786}.
Each calorimeter consists of a $6\times 9$ (H$\times$W) array of PbF$_2$ crystals. Each crystal block is  \SI{14}{\centi\meter} (15 radiation lengths) long with a \SI{2.5}{cm} square cross-section. 
In addition to the excellent spatial resolution produced by crystal segmentation, the calorimeters provide sub-ns timing resolution to distinguish individual positron events.
A laser-based gain monitoring system~\cite{Anastasi_2019} is employed to continuously measure the calorimeter response to obtain energy measurements that are stable with respect to the hit rate and the environmental conditions.

An in-vacuum tracking system based on straw trackers~\cite{King_2022} is installed at two locations around the storage ring just upstream of a calorimeter to track muon decay electrons headed for the calorimeters. The trackers are used to monitor the beam distribution ($M^{T}(x,y,t)$) in the storage ring in the proximity of the two tracking stations. These stations are composed of 32 planes of straw-tube detectors assembled into eight modules. The straw tubes are filled with Argon-Ethane gas, and a thin tungsten wire positioned along the central axis of each straw collects the drift electrons arising from the ionization induced by a passing positron. Tracks are reconstructed by registering hits across multiple planes, and the track reconstruction facilitates both a measurement of the positron momentum and extrapolation to the muon decay vertex.
\subsection{Simulation packages}
A suite of different simulation packages was developed to validate analysis tools. 
Simulation results from the three compact packages are cross-checked against each other. Each package's toolkit provides unique properties, which lead to specific advantages or shortcomings depending on the analysis. For example, \texttt{gm2ringsim} models with high fidelity the material interactions that determine the properties of the stored beam, whereas symplectic tracking for long-term beam effects is verified with the \texttt{COSY-INFINITY} and \texttt{BMAD} models. Below, we describe the main characteristics of each simulation package. For comparisons of the simulation packages, please refer to Ref.~\cite{Run1PRAB}.

\texttt{gm2ringsim} is a model of the \gm injection line and storage ring that has been implemented in the \texttt{GEANT4} simulation framework~\cite{ALLISON2016186,1610988,AGOSTINELLI2003250}. The model consists of a full description of the material structures, as well as the particle detectors that reconstruct the kinematics of the muons and decay positrons~\cite{Run1PRAB}. The \texttt{gm2ringsim} package includes several particle guns, one that allows for high-fidelity production of decay positrons within the ring and one that allows for muon production, propagation, and decay through the full injection channel. Runge-Kutta integration methods are used to numerically integrate a particle's equation of motion and propagate it through electromagnetic fields and across detector boundaries. The parallel world functionality is used to insert ``virtual'' tracking planes into the ring, without adding any material. These planes allow for the reconstruction of the motion of the injected particles as they circulate within the ring. 
The non-symplectic nature of \texttt{GEANT4} did not cause any issues for the systematic errors presented.

The \texttt{COSY}-based model~\cite{tarazona2019muon} is a data-driven computational representation of the storage ring in \texttt{COSY INFINITY}~\cite{MAKINO2006346}. The magnetic field in the storage volume is an implementation of the azimuthally dependent set of multipole strengths from the experimental data, described as a series of magnetic multipole lattice elements. An optical element superimposed on the magnetic field recreates the ESQ stations. The high-order coefficients of the electrostatic potential's transverse Taylor expansion produce the non-linear action of the ESQ on the beam's motion. A recursive iteration of the horizontal midplane coefficients, modeled with conformal mapping methods to satisfy Laplace's equation in curvilinear optical coordinates, provides these coefficients. The boundary element method is utilized in \texttt{COULOMB}'s field solver to recreate the ESQ's effective field boundary and fringe fields in the model. The \texttt{COSY}-based model calculates lattice configurations, Twiss parameters, betatron tunes, closed orbits, and dispersion functions of the storage ring.

A third model based on \texttt{BMAD}~\cite{SAGAN2006356} models the injection line and storage ring, which are arranged as a series of guide field elements referred to as the lattice. The electromagnetic fields of the elements are represented as field maps, or multipole expansions. Particles are tracked by Runge-Kutta or symplectic integration of the equations of motion as required. Muon spin is likewise propagated by numerical integration. Multiple scattering is included at the entrance and exit windows of the inflector and the outer ESQ plate through which particles are injected into the ring. Otherwise, element boundaries are considered apertures, and particles incident on those boundaries are lost. Calorimeters and trackers are represented as simple markers that indicate particle phase space coordinates. \texttt{BMAD} library routines are used to compute beam parameters like beta-functions, chromaticity, dispersion, emittance, etc.

\section{Datasets and run conditions}

\subsection{Datasets}
\label{sec:datasets}

\RunTwo and \RunThree data were acquired from March to July 2019 and November 2019 to March 2020, respectively.  The data are divided into 9 and 13 data subsets labeled 2A-2I and 3A-3O for \RunTwo and \RunThree, respectively.
Four data subsets (2A, 2I, 3A, and 3H) were excluded from the measurement analysis because systematic studies dominated the periods. 
The improved stability of the hardware conditions with respect to \RunOne allowed multiple datasets to be combined in the $\omega_a^{m}$ analysis to leverage the higher statistics and minimize the statistical uncertainties of some systematic effects. 
 The smaller data partitions are combined into the following datasets: \RunTwo = [2B-2H], \RunThreeA = [3B-3G, 3I-3M], and \RunThreeB = [3N-3O]. The three datasets have different beam storage characteristics, ESQ voltage, and kicker strength. The data were hardware-blinded by hiding the true value of the calorimeter digitization clock frequency. This blinding factor was different for \RunTwo and \RunThree.
 In \RunTwo, we performed 25 trolley runs and tracked 17 field periods, and in \RunThree, we performed 44 trolley runs and tracked 34 field periods. In each case, only two field periods did not receive a terminal trolley run.

\begin{figure}[hbt!]
    \centering
    \includegraphics[width=\linewidth]{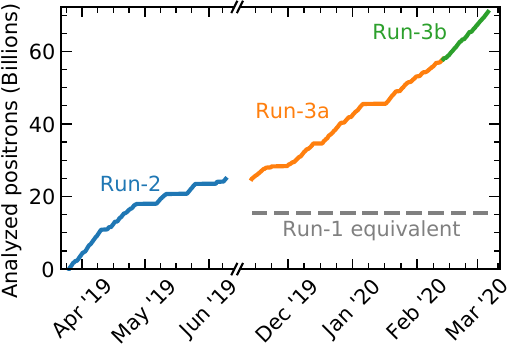}
    \caption{Muon-decay positrons accumulated in \RunTwo and \RunThree after DQC. Positrons with \SI{1}{GeV}$<E<$\SI{3}{GeV} hitting the calorimeters $t>$\SI{30}{\micro\second} after injection are shown. The \RunOne equivalent ($15.4\times10^9$) is shown for comparison.}
    \label{fig:ctags}
\end{figure}

Muon-decay positrons included in the final datasets are selected according to Data Quality Cuts (DQC) based on the quality of fills 
and magnetic field stability. 
Selection criteria for good fills include 
 the kick amplitude and timing, beam profiles, and presence of laser synchronization pulses. DQC are based on the average rate of lost muons, the number of positrons detected, and the quality of the magnetic field and monitor data.
 DQC selection criteria are chosen so that the muon storage conditions are uniform across each of the combined datasets. Overall, roughly \SI{20}{\percent} of the time periods have been discarded, most of them containing zero or few positron events, which corresponds to $\sim$\SI{2}{\percent} of the total data.
The detector and magnetic field DAQ systems are separate and not synchronized, resulting in short periods between field DAQ runs where the precession data would not have corresponding field data. Elimination of those time periods reduces the precession data by $\sim$\SI{0.3}{\percent}.
Magnetic field quality criteria excluded muon data collected from occasional sudden changes of the magnetic field, probably due to magnet component movement, large field oscillations with a period around two minutes related to variations of the superconducting coils' cryogenics, and rare spikes related to the NMR probes used in the magnetic-field stabilization system.
Figure~\ref{fig:ctags} shows the accumulated positrons for \RunTwo and \RunThree after DQC. In total, $71 \times 10^{9}$ positrons with an energy above \SI{1}{GeV} were accumulated.

\subsection{Run conditions: \RunTwoThree vs \RunOne} \label{sec:conditions}

Table~\ref{tab:datasets} presents the number of fills and reconstructed positrons with energies between 1 and \SI{3}{GeV} along with the field indices and kicker strengths for the \RunOne and \RunTwoThree datasets.

\begin{table}[!ht]
	\centering
     \caption{Dataset statistics and hardware conditions for \RunTwoThree compared to \RunOne. The number of analyzed positrons (e$^+$) represents the statistics used in the final $\omega_a^{m}$ fits. 
     }
	\label{tab:datasets}
	\begin{ruledtabular}
        \begin{tabular}{ccccc}
		Dataset&Fills ($\times 10^6$)&e$^+$ ($\times 10^9$)&Field index&Kicker (kV)\\\hline
		\RunOneA&1.51&2.0&0.108&130\\
		\RunOneB&1.96&2.8&0.120&137\\
		\RunOneC&3.33&4.3&0.120&130\\
		\RunOneD&7.33&6.3&0.107&125\\\hline
		\RunTwo&18.60&24.7&0.108&142\\
		\RunThreeA&33.53&33.1&0.107&142\\
		\RunThreeB&11.55&11.9&0.108&161\\
	\end{tabular}
        \end{ruledtabular}
\end{table}

Significant improvements and changes for \RunTwoThree with respect to \RunOne~\cite{Run1PRAB}, include the following:

\begin{itemize}

\item During \RunOne, two resistors electrically connected to the upper and lower plates of the long section of the first ESQ after injection (Q1L) were damaged. Replacing the resistors after \RunOne improved the stability of radial and vertical beam positions.
This significantly reduces the phase acceptance correction in \RunTwoThree.

\item For \RunTwo\ and \RunThreeB\ the operational high-voltage set points for the ESQ system were lowered by \SI{0.1}{kV} to avoid
betatron resonances for beam stability. 
This shift reduced the muon losses by roughly \SI{20}{\percent}.

\item While in \RunOne only two collimators were used, all five collimators were used in \RunTwoThree, which led to better beam scraping and further reduced the effect of muon losses during storage.

\item The kicker strengths for \RunOne and \RunTwo were limited to \SI{142}{kV} by the use of A5596 cables~\footnote{\url{https://timesmicrowave.com/}}. As a result, the beam was not perfectly centered in the storage region. At the end of \RunThreeA, the cables were upgraded~\footnote{with Dielectric Sciences DS2264; \url{https://www.dielectricsciences.com/}} and the kicker voltage was increased to \SI{161}{kV} in \RunThreeB to achieve a more optimal kick. This results in a better-centered muon beam, reducing the E-Field correction~\cite{schreckenberger_fast_2021}.

\item Between \RunOne and \RunTwo, the magnet yokes were covered with a thermal insulating blanket to mitigate day-night field oscillations due to temperature drifts. In addition, the experimental hall's air conditioning system was upgraded after \RunTwo to further stabilize the temperature of both the magnet yokes and the detector electronics to better than $\pm\SI{0.5}{\celsius}$. Figure~\ref{fig:temperature} shows the stability improvement for both the magnet and the calorimeter SiPMs since \RunOne. 

\begin{figure}[hbt!]
    \centering
    \includegraphics[width=\linewidth]{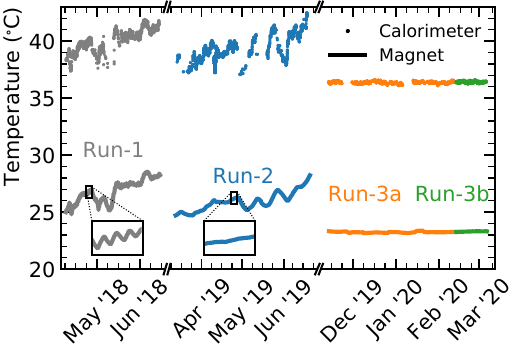}
    \caption{Temperature of the calorimeter SiPMs (small dots) and the magnet yokes (thicker lines) across \RunOne, \RunTwo, and \RunThree. The two inserts show a box of four days with a temperature range of \SI{1}{\celsius}. The magnet thermal insulating blanket installed after \RunOne reduced the day-night oscillations of the magnet temperature. The upgraded air conditioning system greatly improved the long-term stability of both the calorimeters and magnet temperature after \RunTwo.}
    \label{fig:temperature}
\end{figure}

\item In \RunTwoThree, the magnetic field hardware operation procedures improved compared to \RunOne. The more standardized and automated procedures, especially for trolley runs, made measurements and monitoring of the magnetic field faster and more reliable. 
In addition, the magnet power supply feedback loop was optimized during \RunTwo to suppress oscillations in the magnetic field more efficiently and better decouple from higher-order moment changes.

\item 
For \RunTwoThree, modifications were made to the real-time processing
of the digitized waveforms from the calorimeter crystals that are utilized in the positron-based analyses.
In \RunOne, when an individual crystal exceeded a preset threshold, the digitized waveforms of all $54$ crystals of the associated calorimeter were recorded (see Ref.\ \cite{Run1PRDomegaa} for details).
In \RunTwoThree, when an individual crystal exceeded a preset threshold, 
only the above-threshold crystals and their neighboring crystals were recorded. This change permitted data collection of positron-based data at higher rates.

\item 
For \RunTwoThree, modifications were also made to the real-time processing
of the digitized waveforms from the calorimeter crystals that are utilized in the energy-based analyses.
In \RunOne, the raw ADC samples from each calorimeter crystal were summed into 75~ns-binned histograms. These per-crystal histograms were then stored for each fill (see Ref.\ \cite{Run1PRDomegaa} for details).
In \RunTwoThree, the raw ADC samples from each calorimeter crystal were summed into 18.5~ns binned-histograms. These per-crystal histograms was then accumulated for 4 fills and stored for every fourth fill. 
These changes permitted the acquisition of energy-based data with a finer time binning and a greater time range. 

\item During \RunTwo (i.e., after dataset 2E), a wedge absorber for muon momentum-spread reduction was installed in the incident muon beamline~\cite{prab_wedges}.

\end{itemize}

\subsection{Beam storage conditions}
\label{sec:beamdynamicsintro}

Many of the changes listed in the last chapter define the beam dynamics conditions in the storage ring. The main characteristics, such as typical beam oscillation frequencies, muon losses, and beam distributions, are described in the following subsections. 

\subsubsection{Beam oscillation frequencies}
\label{sec:beamdynamicintro:freqs}
The 120-ns duration of muon injection causes a modulation of positron hits in individual detectors with a cyclotron period $T_c$.
Due to the momentum spread of the stored muons with $p=m_\mu c/\sqrt{a_\mu}\pm 0.15\%$, this initial bunching is gradually debunched~\cite{Run1PRDomegaa}.

The muons stored in the ring follow both radial and vertical betatron oscillations with frequencies ($f_x, f_y$) determined by the configuration of the guide fields, characterizing the transverse motion along the azimuth of the ring. 
In addition, the beam widths (frequencies $2f_x, 2f_y$) and centroids of the stored muons follow the optical lattice (with azimuthal variations smaller than 3\%) and closed orbits.

The observed time distribution in a detector is perturbed by these beam oscillations through their coupling to the detector acceptance.
In practice, the radial centroid oscillation ($f_x$) dominates the radial perturbations, and the vertical width oscillation (2$f_y$) dominates the vertical perturbation.

Since muons pass each detector once every cyclotron period, the radial centroid oscillation is observed at an aliased frequency, dubbed coherent betatron oscillation (CBO), $f_{CBO} = f_c - f_x$.
A substantial cancellation of cyclotron period modulation, called fast rotation, is achieved by histogramming data with bin widths as close as achievable to the cyclotron period.
Such a binning causes any frequency that exceeds the Nyquist limit $f_c / 2$ to also be aliased.
The vertical width oscillation appears in the histogram aliased to $f_{VW} = f_c - 2 f_y$.
Table~\ref{t:beamfrequencies} is a summary of these frequencies for the field index (see Ref.~\cite{berz2015introduction}) $n = 0.108$.

\begin{table*}
\centering
\caption{Compilation of frequencies and periods of important beam oscillations for the field index $n = 0.108$ (the anomalous precession frequency $fa$ and cyclotron frequency $f_c$ are given for comparison). Columns 1 and 2 denote the frequency and its symbol. 
Column 3 gives the relation of the beam frequency to the field index $n$, cyclotron frequency $f_c$, and betatron frequencies $f_x$, $f_y$, in the continuous ESQ approximation. Columns 4 and 5 list the numerical values of the frequencies and periods for a field index $n = 0.108$ in the continuous ESQ approximation. Note that the measured frequencies differ slightly from the continuous ESQ approximation frequencies. }
\label{t:beamfrequencies}
\begin{ruledtabular}
\begin{tabular}{llccc}
Term & Symbol & Field index & Freq.\ (MHz) & Period (\SI{}{\micro\second})  \\
&  & relation & $n = 0.108$ & $n = 0.108$  \\
\hline
g$-$2 & $f_a$ & & 0.229 & 4.37  \\
Cyclotron & $f_c$ & & 6.70 & 0.149 \\
Horizontal betatron & $f_x$ & $\sqrt{1 - n} f_c$ & 6.33 & 0.158 \\
Vertical betatron & $f_y$ & $\sqrt{n} f_c$  & 2.20 & 0.454 \\
Coherent betatron & $f_{CBO}$ & $f_c - f_x$ & 0.372 & 2.69 \\
Vertical waist & $f_{VW}$ & $f_c  - 2 f_y$  & 2.30 & 0.435 \\
\end{tabular}
\end{ruledtabular}
\end{table*}

\subsubsection{Muon losses}
\label{sec:beamdynamicintro:lm}
\begin{figure}
    \centering
    \includegraphics[width=\linewidth]{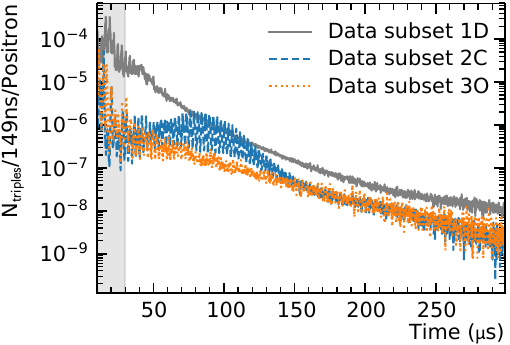}
    \caption{Muon loss time distribution L(t)  for selected \RunOne (gray), \RunTwo (blue), and \RunThree (orange) data subsets showing the reduction in losses. The values here are normalized to the number of $e^+>1.7$ GeV in each dataset. The large modulation of the muon losses with the frequency $f_{CBO}$ is a reflection of the mechanism of the losses.}
    \label{fig:muon_loss_changes_run_123}
\end{figure}

Not all stored muons decay into positrons. Some muons impact material in the storage region, such as aperture-defining collimators, and lose energy to the point where they can no longer be stored. These muons spiral inward, and a subset of them are observed as triple-coincidences of minimum ionizing particles in adjacent calorimeters.
The muon loss spectra differ greatly between runs as seen in Figure \ref{fig:muon_loss_changes_run_123}.
 The muon loss rate was reduced by an order of magnitude between \RunOne and \RunTwo due to the repair of the damaged ESQ resistors.
The bump structure (see Sec.~\ref{sss:muonlossfit}) observed in \RunTwo between \SI{50}{\micro\second} and \SI{150}{\micro\second} was suppressed in \RunThree by better centering the vertical beam.

The presence of lost muons can bias the extraction of $\omega_a^m$ in two ways. First, a time-dependent loss of stored muons causes a time-dependent distortion of measured positrons. To avoid biasing the $\omega_a$ extraction,
the fit must therefore incorporate the effects of muon losses (see Sec.~\ref{sss:muonlossfit}). 
Second, coupling between the muon's momenta and initial spin directions can alter the measured value of $\omega_a^{m}$, as described with more details in
section~\ref{sec:bd:corr:Cml}.

\subsubsection{Beam distributions}
\label{sec:beamdynamicintro:beamdistribution}

The muon beam distribution $M(x, y, \phi)$ is reconstructed by extrapolating beam profiles measured by the two tracker stations.
The extrapolation shifts the mean and scales the transverse width of the distributions relative to the tracker station using characteristic functions obtained from the optical lattice calculated with the \texttt{COSY INFINITY}-based model of the storage ring. 

\begin{figure}[ht]
    \centering
    \includegraphics[width=0.48\columnwidth]{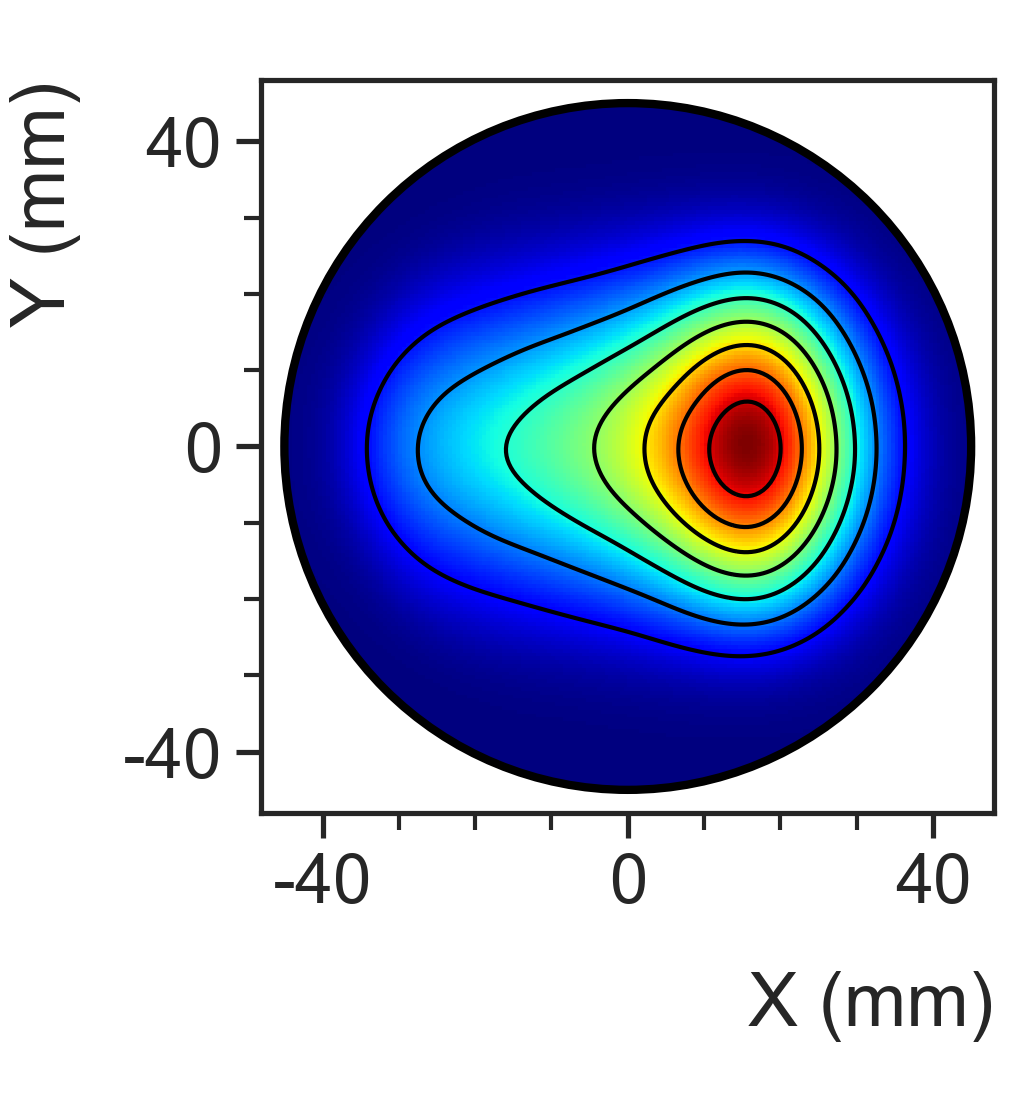}
    \includegraphics[width=0.48\columnwidth]{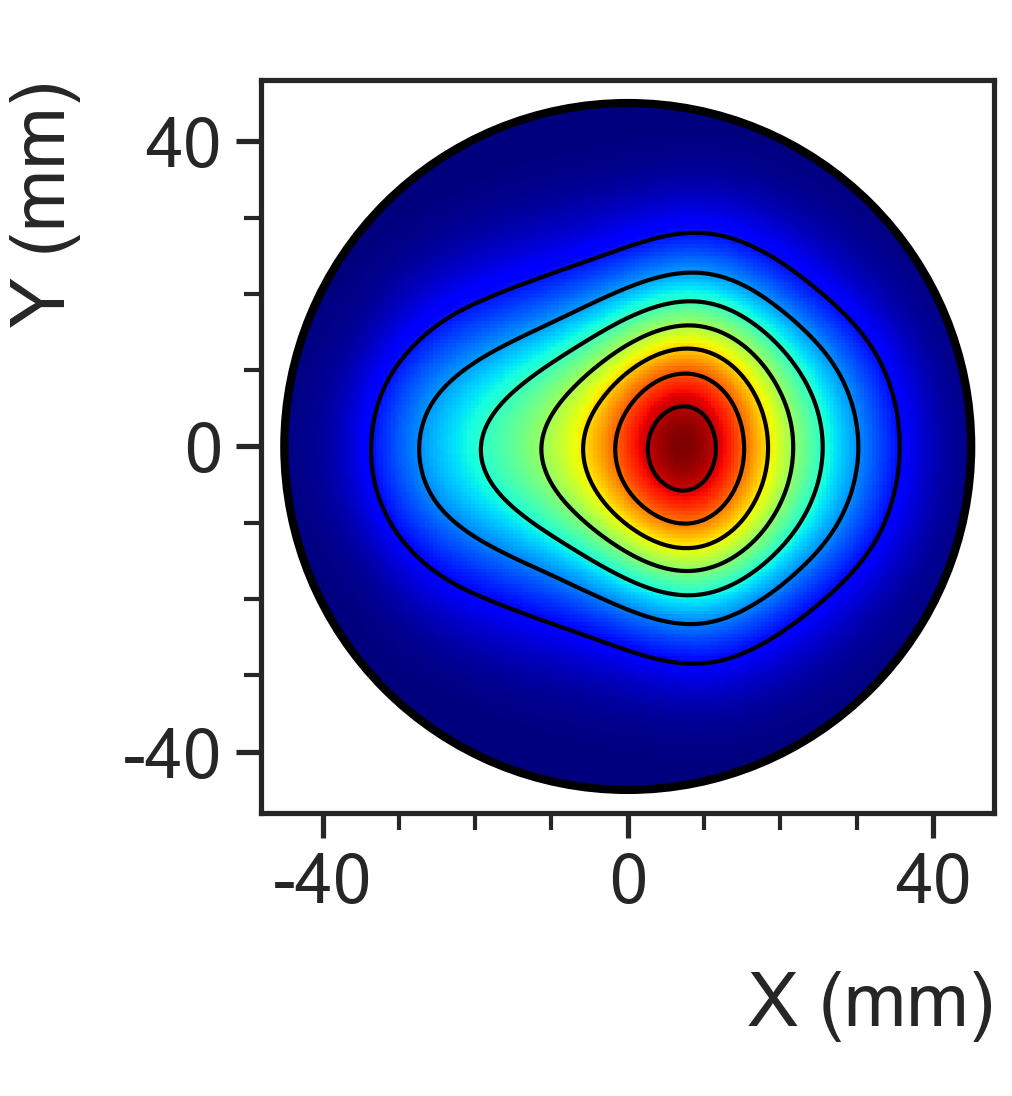}
    \caption{Azimuthally averaged muon beam distribution summed over $t>\SI{30}{\micro\second}$ ($<M(x,y)>_{\phi}$) from datasets from \RunTwo (2B) on the left and \RunThreeB (3O) on the right. The color represents the intensity, from low intensity in blue (outside) to high intensity in red (inside).}
    \label{fig:bd:beamShape}
\end{figure}

Figure \ref{fig:bd:beamShape} shows azimuthally averaged muon beam distributions based on this beam extrapolation. 
The increased kick strength in \RunThreeB moves the beam distribution closer to the center.

\section{Muon anomalous precession frequency measurement} \label{sec:oa}
\label{sec:omegaamain}
This section discusses the analysis of the
muon anomalous precession frequency, $\omega_a^m$.
It describes the time-distribution reconstructions of positron hits and integrated energy as well as the corrections and the fits that are applied to these distributions. It also discusses the $\omega_a^m$ results,
systematic uncertainties and consistency checks. We emphasize
changes since the \RunOne, $\omega_a^m$ analysis \cite{Run1PRDomegaa}.

The $\omega_a^m$ analysis was conducted by seven independent
analysis groups using a number of different strategies for the positron hit and integrated-energy reconstruction, handling of cyclotron rotation and positron pileup, and treatment of beam dynamics and muon losses.
Herein the analysis groups are denoted by Roman numerals I-VII.

\subsection{Analysis methods}
\label{ss:overviewomegaa}
The measurement benefits from multiple complementary analysis techniques that can be divided broadly into two categories. The first category is event-based
and focuses on reconstructing the energies and times of the individual decay positrons in the calorimeters.  The second category is energy-based and focuses on reconstructing the energy versus time in the calorimeters without the positron identification. For each technique, we construct a time distribution that is modulated by the anomalous precession frequency $\omega_a^m$.

\begin{figure}
\includegraphics[width=\linewidth]{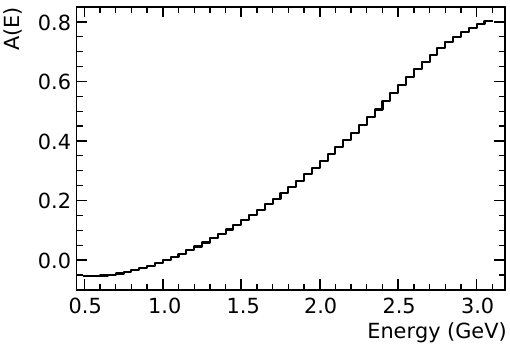}
\caption{Representative example of the measured asymmetry $A(E)$ of the anomalous precession signal versus the positron energy $E$ in the region $0.5-3.1$~GeV (for the calorimeter summed data and a selected analysis group). In the A-method, each positron is weighted by $A(E)$ to achieve the greatest possible statistical power in the anomalous precession frequency measurement. Note the measured asymmetry $A(E)$ incorporates detector acceptance effects.}
\label{f:AE}
\end{figure}

In the event-based methods, we applied two data-weighting schemes.
In the threshold analysis (denoted the T method), equal weight is given to all positrons above a fixed energy
threshold.  In the asymmetry-weighted analysis (denoted the A method), each positron is
weighted according to the decay-asymmetry corresponding to the positron's energy (see Fig.\ \ref{f:AE}).  The asymmetry-weighted analysis achieves the greatest possible statistical
power to measure the precession frequency.
The integrated-energy approach (denoted the Q method), is logically equivalent to weighting positrons with their energies even though it does not resolve individual positrons.

In a ratio method, the data are split into four subsets, two
time-shifted and two unshifted, from which a ratio histogram
is constructed. By using time shifts of one-half the anomalous
precession period, the $\omega_a^m$ modulation is preserved
while slow-time variations are mitigated. See Ref.\ \cite{Run1PRDomegaa} for the details of the construction of the ratio histogram.

\subsection{Reconstruction approaches}
\label{ss:reconstructionapproaches}

For the event-based analyses, we used two distinct
reconstruction schemes: a local-fitting approach and a global-fitting approach.
The local-fitting approach was used by the groups I through IV and the global-fitting approach was used by groups V and VI.
An important difference between these two approaches was the inclusion or exclusion of spatial separation of positron hits in the fitting procedure (see Ref.~\cite{Run1PRDomegaa} for details).

The local-fitting approach involves individually fitting the
waveform from each crystal. Each crystal waveform is first fit to an empirically-determined pulse template to determine its time and energy. The crystal hits occurring in a given time window are then clustered into positron candidates. The cluster time was defined as the time of the crystal hit with the largest energy, and the cluster energy was defined as the sum of the clustered crystal energies.

The global-fitting approach involves simultaneously fitting the waveforms from 3×3 crystal arrays that are centered on the highest-energy crystal. The
3$\times$3 waveforms are simultaneously fit to empirically-determined pulse templates to determine a single shared fitted time and individual crystal energies (see Ref.~ \cite{Run1PRDomegaa} for the details of the construction of the templates). The cluster time was defined as the single shared fitted time and the cluster energy as the sum of the contributing crystal energies.

The group VII, energy-based reconstruction involves the construction of a time distribution of the deposited energy in each calorimeter.
The approach utilizes a rolling pedestal
with a low-energy threshold
in order to extract the integrated energy
and mitigate any pedestal variations
(see Ref.~\cite{Run1PRDomegaa} for details).
It negates the need for
fitting and clustering of crystal pulses
and decision making in positron identification.
Although statistically less powerful,
its value lies in utilizing different raw data, applying different reconstruction procedures, and inheriting different systematic uncertainties.

\subsection{Data corrections}
\label{ss:wacorrections}

The analysis methods (Sec.~\ref{ss:overviewomegaa}) 
and reconstruction approaches (Sec.~\ref{ss:reconstructionapproaches})
are used to build time distributions 
of positrons hits or integrated energy.
Before fitting the time distributions to extract $\omega_a^m$ we apply several corrections.

One correction applied to the raw data,
accounts for any gain changes in the calorimeter electronics.
Another correction applied to the time histograms,
removes the distortions arising from positron pileup.
A final correction treats the imprint on the data of the cyclotron rotation of the stored beam.
These corrections are described below.

\subsubsection{Gain corrections}
\label{sss:gaincorrections}

The calorimeter SiPMs and readout electronics suffer from gain
fluctuations on multiple timescales from various physical effects.  At the
longest timescales, temperature variations in the experimental hall lead to gain
changes over days or longer (long-term gain correction).  Within a
muon fill, the initial beam flash causes an immediate gain
sag with gradual gain recovery that impacts all calorimeters but especially those
near the inflector (in-fill gain correction). 
At the shortest
timescales, the SiPM pixel deadtime causes a short-term gain sag 
if a second positron is recorded just after an earlier positron (short-term gain correction).

These effects are corrected using dedicated studies
with a laser calibration system Ref.~\cite{lasersystem}. One improvement since Run-1 is the treatment of the temperature dependence of the short-term gain corrections.

Note that the significant improvement in the temperature stability of the experimental hall from \RunTwo to \RunThree (see Fig.\ \ref{fig:temperature}),  reduced the size of long-term gain corrections and limited the need for temperature-dependent, short-term gain corrections in \RunThree.

\subsubsection{Pileup corrections}
\label{sss::PUcorrections}

For event-based analyses, it is generally not possible to resolve positron hits in the same calorimeter crystal within a 1.25-ns time interval (we note that the spatial resolution of the global-fitting approach can sometimes identify such pileup events).
Consequently,
such close-in-time positrons are summed and treated as a single positron with the
summed energy of the true positrons.  Since the likelihood of positron pileup will
decrease during the muon fill, this potentially biases the $\omega_a$ extraction.

To account for pileup, the raw time distribution is corrected through a data-driven, statistical reconstruction of a pileup time distribution.
Three methods were used in building the pileup distribution: the so-called empirical, semi-empirical, and shadow window methods.
All three methods model the effects of
pileup by computing the difference between the reconstructed energy-time distributions of unresolved positrons and resolved positrons.
This pileup time distribution is then subtracted from the raw time distribution.

The pileup modelling is achieved by superimposing data from the same calorimeter with a one cyclotron period delay from the reconstructed positron. This separation randomly samples the calorimeter data with a similar rate. The initial reconstruction provides the individual positrons before the data superposition.

In practice, this superposition
of data can be performed at the level of the digitized waveforms, crystal hits,
and reconstructed positrons. These levels correspond to the aforementioned empirical, semi-empirical, and shadow window methods, respectively
\footnote{In superimposing waveforms, the waveforms are first superimposed and then the full positron reconstruction is rerun. In superimposing the crystal hits, the hits are first superimposed and then the positron clustering stage is rerun.}. An improvement on Run-1 was the handling of triple pileup in most Run-2/3 analyses. 

\begin{figure}
\includegraphics[width=\linewidth]{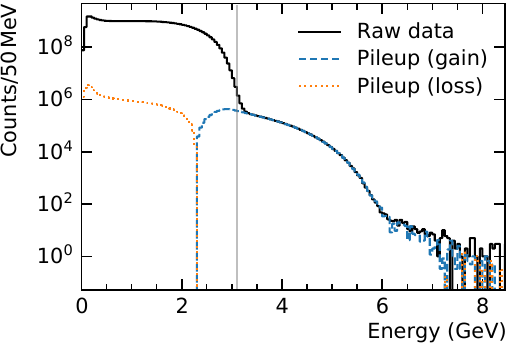}
\caption{Illustration of the reconstructed pileup correction for the empirical method. The black curve is the raw energy distribution before the pileup correction. The dashed blue (dotted orange) curves show the reconstructed gain (loss) of positron events due to positron pileup. The agreement between the black curve and the blue curve in the energy region greater than the 3.1 GeV beam energy (vertical gray line) is an indication of the quality of the pileup correction.}
\label{f:pileup}
\end{figure}

All three methods show an excellent ability to reproduce the observed
pileup energy spectrum in the energy region greater than the  3.1~GeV beam energy.
An example using the empirical method is shown in  Fig.~\ref{f:pileup}.

The energy-based analyses utilize
a non-zero energy threshold and therefore
are not completely immune to a pileup distortion.
We therefore developed a signal processing algorithm for calculating pedestals and applying thresholds that minimizes pileup effects.
The algorithm is described in~\cite{Run1PRDomegaa}.

\subsubsection{Fast-rotation handling}
\label{sss::pileuphandling}

Although the fast-rotation modulation (Sec.~\ref{sec:beamdynamicsintro}) is greatly reduced by the \SI{30}{\micro\second} start time of the $\omega_a$ fit region,
its effect is nonzero.
A substantial cancellation of fast rotation is achieved by histogramming data with bin widths as
close as possible to the cyclotron period
(\SI{149.2}{\nano\second} for the event-based analyses and \SI{150}{\nano\second} for
the energy-based analyses). A further cancellation
is achieved by summing the data from the 24 calorimeters (due to the 2$\pi$ advance of the fast-rotation
modulation around the ring circumference). These procedures were used in all the analyses.

The remaining distortion is handled by either randomizing the histogram entries
by one cyclotron period in event-based analyses or uniformly
distributing the energy entries over one cyclotron period  in energy-based analyses.

\subsection{$\mathbf{\omega_a^m}$ software blinding procedure}
\label{ss:softwareblinding}

During their analysis processes, each of the seven analysis groups were software-blinded with respect to each other ({\it  i.e.}\ in addition to the common hardware blinding).

The procedure parameterized the measured frequency $\omega_a^m$
as a fractional shift $R$ from a nominal reference frequency
$\omega_{ref} = 2 \pi \times 0.2291$ MHz, where
\begin{equation}
    \omega_a^m = \omega_{ref} \cdot ( 1 + [ R - \Delta R ] \times 10^{-6} ),
\label{e:omegaablindingequation}
\end{equation}
and $\Delta R$ is that group-dependent, software-blinding offset, which is generated within a $\pm 24$ range. The values of
$\Delta R$ were derived from group-chosen text phrases whose hash seeded a random number generator

The relative unblinding of the seven groups to a common software-blinded stage facilitated unbiased comparisons between the analyses and followed internal reviews conducted by the analysis teams.
The remaining software and hardware blindings
were not removed until the collaboration's decision to publish the result for $a_{\mu}$.

\subsection{$\mathbf{\omega_a^m}$ fitting procedure}
\label{ss:fittingprinciples}

The measured anomalous precession frequency $\omega_a^m$ was extracted
by fitting the reconstructed positron or integrated-energy time histograms
after correcting for cyclotron rotation and positron pileup.
These `$\omega_a^m$-wiggle' fits were performed using either
the \texttt{Minuit} numerical minimization package \cite{Minuit},
the \texttt{Python scipy.optimize} package \cite{Scipy},
or the \texttt{Python lmfit} package \footnote{\url{https://lmfit.github.io/lmfit-py/fitting.html}}. They minimized the quantity

\begin{center}
\begin{equation}
\chi^2 = \sum_{ij} (y_i - f_i ) V_{ij}^{-1} (y_j - f_j ),
\end{equation}
\end{center}
where $y_i$ are the measured data points, $f_i$ are the corresponding fit function values,
and $V_{ij}$ is the covariance matrix. The diagonal elements of $V_{ij}$
are the variances $\sigma_i^2$ of the data points $y_i$. The off-diagonal elements
of $V_{ij}$ are the covariances $\sigma_{ij}^2$ between the data points $y_i$, $y_j$.
Non-zero covariances were used in some analyses to handle correlations between
data points arising from the handling of cyclotron rotation, correction for
positron pileup, and construction of ratio histograms. The minimization of $\chi^2$
determines the optimal values of the model parameters of the fit function.

The nominal fit time ranges were 30.1 to 660.0 $\mu$s for the event-based analyses and
30.1 to 330.0~$\mu$s for the energy-based analyses.
The bin widths were \SI{149.2}{\nano\second} for the event-based analyses and \SI{150.0}{\nano\second}
for the energy-based analyses.
The $30.1$~$\mu$s start time is i) after the 
stabilization of beam scraping, and ii) as 
close as possible
to an $\omega_a$ anomalous precession node in order to minimize
any pull from miscalibration of the calorimeters (see Sec.~\ref{sss:gaincorrections}).

\subsubsection{$\omega_a^{m}$ fit model}
\label{sss:5parameterfit}

The fit function used for extracting $\omega_a^{m}$ from both the event-based and energy-based time distributions has the general form

\begin{eqnarray}
f( t ) &=& N_0 \cdot N_x(t) \cdot N_y(t) \cdot N_{xy}(t) \cdot \Lambda(t)  \cdot e^{ - t / \gamma \tau_{\mu} }  \nonumber \\
 & & ( 1 + A_0 \cdot A_x(t) ~ \cos( \omega_a^{m} t - (\phi_0 + \phi_x(t) )~ ) ~ ).\nonumber\\
\label{equation:omegaAfit}
\end{eqnarray}

The function incorporates the effects of muon decay and
anomalous precession through the time-dilated lifetime $\gamma \tau_{\mu}$, muon decay asymmetry $A_0$, anomalous precession frequency $\omega_a^{m}$, and
anomalous precession phase $\phi_0$.
$N_0$ is an overall normalization.
Note that the time-dependent terms $N_x$, $N_y$, $N_{xy}$, $A_x$, $\phi_x$, and $\Lambda$ are used to handle distortions from beam dynamics and muon losses~\footnote{In Ref.~\cite{Run1PRDomegaa} and~\cite{PhysRevLett.131.161802}, we used a simplified version of Eq.~\ref{equation:omegaAfit} with positive phase term $\phi_0$.}.
These distortions
are explained in detail in Secs.\
\ref{sss:beamdynamicsfit} and \ref{sss:muonlossfit}, respectively.

In addition,  we discuss in Sec.~\ref{sss:electronicsfit} an electronics ringing term
that was used in the energy-based analyses and in Sec.~\ref{sss:slowtermfit} a residual slow term  that was studied in the event-based analyses.

If $N_x$, $N_y$, $N_{xy}$, and $A_x$ are set to unity and $\phi_x$ is set to zero in Eq.~\eqref{equation:omegaAfit}, one obtains a five-parameter function involving $N_0$, $\gamma \tau_{\mu}$, $A_0$, $\omega_a^{m}$, and $\phi_0$. In subsequent sections,
we utilize the five-parameter fit residuals and their discrete Fourier transforms to illustrate the effects of beam dynamics.

\subsubsection{Beam dynamics distortions}
\label{sss:beamdynamicsfit}

In principle, the beam oscillations, in combination with detector acceptances introduced in Sec.~\ref{sec:beamdynamicsintro}, perturb the
overall normalization ($N_0$), decay asymmetry ($A_0$),
and precession phase ($\phi_0$), in the $\omega_a^{m}$ fit function.
In practice, we find  the large radial perturbations
require accounting  for beam distortions to $N_0$, $A_0$, and $\phi_0$
while the smaller vertical perturbations only require accounting
for distortions to $N_0$.

The time-dependent distortions from beam dynamics
were generally modelled by a sinusoidal oscillation
with an empirical decoherence envelope. For example,
leading effects of CBO perturbations
on the normalization $N_0$ could be modelled
by a term
\begin{equation}
N_x(t) = 1 + A_{\text{CBO}} ~ e^{-t/\tau_{\text{CBO}}} ~ \cos( \omega_{\text{CBO}} t + \phi_{\text{CBO}} ),
\end{equation}
where the associated parameters are the CBO amplitude, $A_{\text{CBO}}$,
CBO frequency, $\omega_{\text{CBO}}$, CBO phase, $\phi_{\text{CBO}}$, and CBO decoherence time constant $\tau_{\text{CBO}}$.
Similar functional forms were used for the
beam dynamics corrections  $N_y$, $N_{xy}$, $A_x$, and $\phi_x$.
Note that the term $N_{xy}(t)$, with a frequency $\omega_{\small \text{VW}} - \omega_{\small \text{CBO}}$, arises from a coupling between the dominant horizontal and vertical oscillations.

In practice, a number of monotonically decreasing functions, which involved combinations of exponential and
reciprocal functions, were used for modeling the decoherence envelope.
The envelope shape and time constant were found to differ
across the three datasets and the event-based and energy-based analyses. The $\omega_a^m$-sensitivity to the decoherence envelope is discussed in Sec.~\ref{sss:cbosystematic}.

In addition, an effective time variation of the CBO frequency was
identified in the time distributions of the individual calorimeters.
This effect was modelled through an exponentially decreasing time variation with a  10-20
$\mu$s time constant and a fitted
amplitude parameter. The $\omega_a^m$-sensitivity to the
frequency change is discussed in Sec.~\ref{sss:cbosystematic}.

\subsubsection{Muon loss distortions}
\label{sss:muonlossfit}

Muon losses, as described in Sec.~\ref{sec:beamdynamicsintro} and shown in Fig.~\ref{fig:muon_loss_changes_run_123}, reduce the number of stored muons and, consequently, the number of detected positrons.

As shown, such losses can be measured as a function of time $L (t)$ by muons traversing multiple calorimeters.
However, such measurements do not determine
the absolute rate of muon losses.
An absolute measurement of the muon loss rate would require modeling the calorimeter acceptance of aberrant trajectories to high precision. A data-driven approach was therefore employed.

Note that muon-loss effects on positron rates at time $t$ are determined by the integrated losses up to time $t$.
All $\omega_a^m$ fits therefore incorporate a muon loss term
\begin{equation}
\Lambda (t) = 1 - k_{\mathrm{loss}}  \int^t_0 L(t) e^{ t^{\prime} / \gamma \tau_{\mu}}  dt^{\prime},
\end{equation}
where $L(t)$ is the measured muon-loss time distribution and
$k_{\mathrm{loss}}$ is a fitted normalization parameter.

Figure~\ref{fig:muon_loss_changes_run_123} in Sec.\ \ref{sec:datasets}
compares the measured time distributions $L (t)$ for the different datasets. The changes made to the quadrupole and kicker settings
between the three datasets led to related changes in the loss rates and the time distributions. In \RunTwo the loss rates were significantly larger
as the field index was closer to beam resonances.

Another notable difference between the datasets was the appearance of a bump in the \RunTwo time distribution. The bump amplitude and bump time both varied around the storage ring and changed during \RunTwo operations.
Although the bump's cause is not fully understood, it was found to be correlated  with the magnet temperature and the vertical beam position.

Due to the \RunTwoThree differences in muon-loss time distributions, the procedures for fitting the losses differed between \RunTwo and \RunThree. These details are
summarized in Table~\ref{t:fitconfigs}.

\subsubsection{Electronics ringing distortions}
\label{sss:electronicsfit}

In the energy-based approach, the time distributions
are incremented with above-threshold, pedestal-subtracted
energies.
The pedestal is calculated from the rolling average of the
ADC samples in a window surrounding each above-threshold,  ADC sample.
Consequently, both drifts and oscillations of the baseline
during the fill can bias this calculation.

The largest bias arose from electronics ringing with a period of about \SI{600}{\nano\second} that resulted from the injection flash in the calorimeters.
To determine the effect on calculating the
pedestal, we computed the distribution of differences between
\begin{enumerate}
\item
ADC samples without
above-threshold signals, and
\item corresponding pedestal estimates
from the surrounding pedestal samples.
\end{enumerate}
This data-driven bias was then incorporated in the
fit function for the energy-based analyses
in a similar manner to the muon loss term.

\subsubsection{Residual slow effect}
\label{sss:slowtermfit}

Residual slow effects, a change in positron counts or integrated energy over the duration of the fill, have different sources.

One contribution arose in the local-fitting analysis from the handling of the single chopped islands with more than one positron cluster.
Such islands -- that are more probable at early times in the fill -- produced a time-dependent, energy-scale shift.

Another contribution stems from a remaining residual slow term that is common to both local and global fits. Possible sources of this effect include changes in gain, acceptance, or reconstruction over the duration of the fill. The introduction of either an ad hoc, time-dependent correction term or an ad hoc, time-dependent fit term is utilized to mitigate this residual effect. We noted that this term’s magnitude is highly correlated with analysis strategies that are applied to the fitting of other slow terms like the muon lifetime and the muon losses.
We chose not to apply the ad hoc, time-dependent fit term
in the extraction of the frequency $\omega_a^m$.

\subsection{Differences with respect to \RunOne}
The major differences between the \RunTwoThree analysis and the \RunOne analysis are listed below.

\begin{enumerate}
    \item 
    In \RunTwoThree we introduced a so-called kernel method for building ratio histograms. This method uses four identical copies of the time
distributions for the ratio construction.
It has the advantage of avoiding the statistical noise originating from the Run-1 randomization approach. It has the disadvantage of introducing bin-to-bin correlations in the ratio histograms.

\item 
In \RunTwoThree the ratio construction was additionally applied to the asymmetry-weighted positron time distributions and the integrated-energy time distributions.
Below we denote the original T-method ratio histograms by RT, the new A-method ratio histograms by RA, and the new Q--method ratio histograms by QR. 

\item In \RunTwoThree we introduced several improvements in the
local-fitting positron reconstruction. One improvement used the measured energy dependence of the SiPM time resolution \cite{KHAW2019162558}. It improved the separation of close-in-time clusters and reduced the positron pileup.
Another improvement by group I involved prioritizing the crystal hits with higher energies during clustering. It improved the positron time resolution.

\item In \RunTwoThree we improved the gain correction procedure by incorporating a temperature-dependent, short-term
gain correction.

\item In \RunTwoThree a new frequency corresponding to $\omega_{\mathrm{VW}} - \omega_{\mathrm{CBO}}$
was identified in the time distributions and incorporated in the $\omega_a$ fits.

\end{enumerate}

\subsection{Multi-parameter fits}
\label{ss:multiparameterfit}

\begin{figure}[hbt!]
\begin{center}
\includegraphics[width=\linewidth]{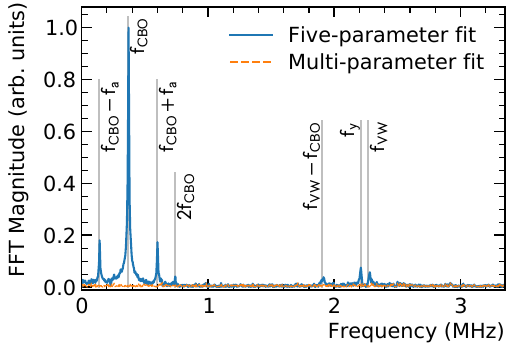}

\end{center}
\caption{Representative example of the discrete Fourier transform (FFT) of the fit residuals for a five-parameter fit (solid blue) and a
multi-parameter fit (dotted orange) to the \RunThreeB dataset. The five-parameter Fourier transform indicates the presence of perturbations due to beam dynamics, muon losses, {\it etc}.  The five-parameter fit shows peaks corresponding to radial beam oscillations ($f_{\text{CBO}}$, $2f_{\text{CBO}}$), vertical beam oscillations ($f_{\text{VW}}$, $f_y$), couplings between precession and radial frequencies ($f_{\text{CBO}} \pm f_{a}$), and radial and vertical frequencies ($f_{\text{VW}} - f_{\text{CBO}}$). Also evident at low frequencies are the effects of muon losses and other slow effects.}
\label{f:fft3b}
\end{figure}

Table~\ref{t:fitconfigs} summarizes
the analysis strategies and fitting choices that were made by the
seven groups in their multi-parameter $\omega_a^m$ fits.
The discrete Fourier transform
of the fit residuals for a representative multi-parameter fit
to the \RunThreeB dataset is shown in Fig.~\ref{f:fft3b}.

As discussed in detail in Sec.~\ref{ss:reconstructionapproaches}, the analyses span three distinct reconstructions:
the event-based, global-fitting reconstruction, the
event-based, local-fitting reconstruction, and the
energy-based reconstruction.
Positron pileup was corrected by three distinct, data-driven
approaches involving superimposing ADC waveforms, crystal hits,
or positron hits
(see Secs.~\ref{sss::PUcorrections} and \ref{sss::pileuphandling} for details).
The handling of cyclotron rotation involved
either randomizing the histogram  entries by times $\pm T_c / 2$
in event-based analyses or uniformly distributing
the histogram entries over times $\pm T_c / 2$ in energy-based analyses.
The time distributions themselves were constructed with
equally-weighted positron entries (T method), asymmetry-weighted positron entries (A method), and energy-weighted entries (Q method).
Ratio histograms for each weighting
were also constructed (TR, AR and QR methods).

In performing the fits, independent analysis groups used different
strategies for handling perturbations from beam dynamics,
muon losses, and residual slow effects. Choices included
the use of free, penalized, and fixed values for the
time-dilated muon lifetime $\gamma \tau_{\mu}$ \footnote{In fitting the muon lifetime, some analyses added a $\chi^2$ penalty term to  constrain the time-dilated lifetime to results from cyclotron rotation studies.}; the use of free,
fixed, or zero values for the muon loss parameter $k_{\mathrm{loss}}$;
and different handlings of the CBO envelope shape and
the CBO frequency time-dependence. The total number of free parameters varied with analysis choices
and histogramming methods and ranged from 14 parameters (in
one AR method fit) to 38 parameters (in the Q method fit). 

Note that two analysis groups (III and IV) used a randomization procedure similar to fast rotation randomization to handle the VW beam oscillation. This avoided the need for an associated fit term and reduced the number of fit parameters. 

The typical effects the aforementioned corrections have on the extraction of $\omega_{a}^{m}$ are $\mathcal{O}(\SI{1000}{ppb})$ for the beam dynamics,
$\mathcal{O}(\SI{10}{ppb})$ for the muon losses, $\mathcal{O}(\SI{100}{ppb})$ for the positron pileup,  and $\mathcal{O}(\SI{1}{ppb})$ for the cyclotron rotation.

\begin{table*}
\begin{center}
\caption{Summary of the fitting strategies of the seven analysis groups I-VII. Columns 1, 2 and 3 denote the groups, reconstruction and histogramming methods. Column 4 lists the total number of parameters varied in the fits to the datasets. Column 5 lists the strategy for handling the time-dilated muon lifetime. Columns 6 and 7 summarize the strategies for handling the muon-loss term in Runs 2 and 3, respectively. The $+$, $-$ denotes the sign of the muon-loss term in the wiggle fit (see Sec.\ \ref{sss:slowtermsystematic}). 
Columns 8-10 summarize the strategies for handling the various beam dynamics effects where the heading $f_{\text{CBO}}$(t)
denotes a time-dependent CBO frequency, the heading $e^{-t/\tau_{\text{CBO}}} + C$ denotes a CBO envelope with both an exponential and constant term, and the heading VW$-$CBO denotes the 1.9 MHz oscillation term.
An unlabeled check mark indicates the associated fit term was included in all datasets. A check mark with label `r3' or `r3b' indicates the associated fit term was included in the Run-3 or Run-3b datasets only.  Note in column 7, `fixed $\tau_d$' indicates the time constant of the CBO frequency change was not varied in the fit.
See text for details.}
\label{t:fitconfigs}
\begin{ruledtabular}
\begin{tabular}{cccccccccc}
 Group & Recon & Method & \# free & $\tau_{\mu}$ & Run-2 & Run-3 & $f_{\text{CBO}}$(t) & CBO env. & VW$-$CBO \\
 &  &  & parameters & handling & $k_{\mathrm{loss}}$ & $k_{\mathrm{loss}}$ & term & \small{$e^{-t/\tau}+C$} & term \\
 &  & & 2, 3a / 3b & & & & & & \\
\hline
I & local & A, T & 28 / 28 & free & free, $+$ &  free, $-$ &  & r3b $\checkmark$ & $\checkmark$ \\
II & local & A, T & 25 / 26 & free &  free, $+$ &  fix, $0$  & $\checkmark$ & r3b $\checkmark$ & $\checkmark$ \\
III & local & A, T & 28 / 28 & free & free, $+$ & free, $-$ & $\checkmark$, fixed $\tau_d$ & $\checkmark$ & $\checkmark$ \\
III & local & AR, TR & 14 / 14 & fix & free, $+$ & free, $-$ & $\checkmark$, fixed $\tau_d$ &  $\checkmark$  & \\
IV & local & A, T & 18 / 18 & free & free, $+$ & fix, $0$ & $\checkmark$, fixed $\tau_d$ & $\checkmark$ \\
IV & local & AR, TR & 15 / 15 & fix & fix, $+$ & fix, $0$ & $\checkmark$, fixed $\tau_d$ & $\checkmark$ & \\
V & global & A, T & 30 / 30 & free & free, $+$ & free, $-$ & $\checkmark$ & $\checkmark$ & $\checkmark$ \\
V & global & TR & 19 / 19 & fix & fix, $+$ & fix, $-$ & $\checkmark$ & $\checkmark$ &   \\
VI & global & A, T & 27 / 28 & penalize & free, $+$ &  free $-$ & $\checkmark$, fixed $\tau_d$ & r3b $\checkmark$ & $\checkmark$ \\
VII & energy & Q & 34 / 38 & free & free, $+$ & free, $-$ & $\checkmark$ & $\checkmark$ & r3 $\checkmark$ \\
VII & energy & QR & 26 / 24 & fix & fix, $+$ & fix, $-$ & $\checkmark$ & $\checkmark$ & r3 $\checkmark$ \\
\end{tabular}
\end{ruledtabular}
\end{center}

\end{table*}

\subsection{Commonly-blinded $\mathbf{\omega_a^m}$ results}
\label{ss:fitresults}

\begin{figure*}
\begin{center}
\includegraphics[width=\textwidth]{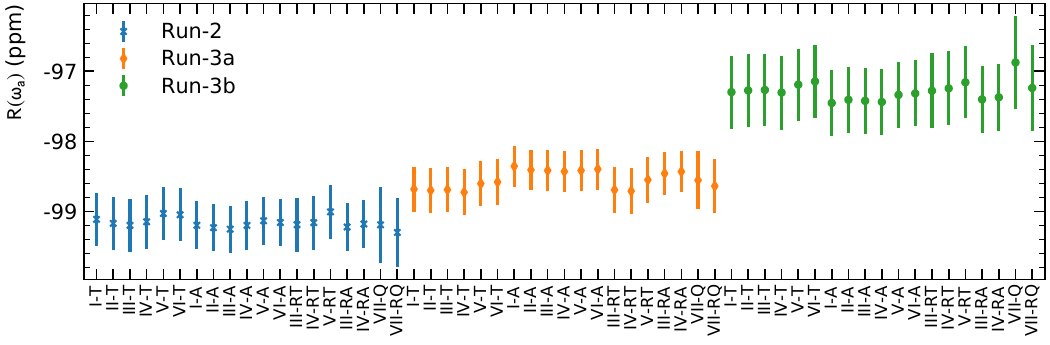}
\end{center}
\caption{Plot of the results for the 19 analyses of the three different
datasets. Note the muon-weighted magnetic field \ref{sec:field:muonWeighting} and beam dynamics corrections \ref{sec:bd:corr} are different for the three datasets. The plotted uncertainties are the statistical uncertainties from the multi-parameter fits to the
associated time distributions. The allowed statistical and systematic differences between the results for a given dataset
are discussed in \ref{ss:fitresults}.}
\label{f:comon-unblinded-R}
\end{figure*}

Table~\ref{t:comon-unblinded-R} and Fig.~\ref{f:comon-unblinded-R} list the commonly-blinded $\omega_a^m$ values and their statistical uncertainties for 19 distinct analyses covering the Run-2, Run-3a, and Run-3b datasets (the nineteen distinct analyses arise from the multiple histogramming techniques applied by the 7 analysis groups).
The results are expressed in terms of R[ppm] as
defined by Eq.~\eqref{e:omegaablindingequation} and described in Sec.~\ref{ss:softwareblinding}.
Across the datasets, the R-values may differ
due to dataset differences in the muon-averaged magnetic field \ref{sec:field:muonWeighting} and $\omega_a^m$ beam dynamics corrections \ref{sec:bd:corr}.

Within a given dataset the
R-values from different analyses are highly correlated. The R-values should agree within allowed statistical and systematic variations that account for the analysis-to-analysis correlations.

\begin{table*}
	\begin{center}
 \caption{R-values in units of ppm for the 19 distinct analyses of the three datasets.  Note the muon-weighted magnetic field \ref{sec:field:muonWeighting} and beam dynamics corrections \ref{sec:bd:corr} are different for the three datasets. Column 1 denotes the analysis group and column 2 denotes the histogramming method. The remaining columns give the commonly-blinded R-values and their statistical uncertainties for the Run-2, Run-3a, and Run-3b datasets, respectively. See text for the discussion of the allowed statistical differences between the different analyses.}
\label{t:comon-unblinded-R}
 \begin{ruledtabular}
\begin{tabular}{ccrrrrrr}
Group & Method & \multicolumn{2}{c}{Run-2} & \multicolumn{2}{c}{Run-3a} & \multicolumn{2}{c}{Run-3b} \\
 &  & R & $\sigma_{R}$ & R & $\sigma_{R}$ & R & $\sigma_{R}$\\
\midrule
I & T & -99.112 & 0.377 & -98.682 & 0.320 & -97.298 & 0.520\\
II & T & -99.171 & 0.376 & -98.700 & 0.323 & -97.274 & 0.519\\
III & T & -99.198 & 0.377 & -98.690 & 0.323 & -97.267 & 0.520\\
IV & T & -99.147 & 0.382 & -98.726 & 0.329 & -97.304 & 0.528\\
V & T & -99.029 & 0.378 & -98.603 & 0.325 & -97.191 & 0.513\\
VI & T & -99.047 & 0.378 & -98.581 & 0.325 & -97.145 & 0.522\\
\addlinespace
I & A & -99.197 & 0.339 & -98.355 & 0.290 & -97.453 & 0.468\\
II & A & -99.232 & 0.338 & -98.408 & 0.290 & -97.407 & 0.467\\
III & A & -99.253 & 0.337 & -98.416 & 0.291 & -97.422 & 0.468\\
IV & A & -99.199 & 0.344 & -98.430 & 0.295 & -97.438 & 0.476\\
V & A & -99.134 & 0.340 & -98.416 & 0.291 & -97.337 & 0.466\\
VI & A & -99.157 & 0.340 & -98.397 & 0.293 & -97.316 & 0.470\\
\addlinespace
III & RT & -99.189 & 0.383 & -98.693 & 0.334 & -97.279 & 0.533\\
IV & RT & -99.160 & 0.383 & -98.710 & 0.329 & -97.244 & 0.529\\
V & RT & -99.006 & 0.384 & -98.549 & 0.325 & -97.158 & 0.513\\
\addlinespace
III & RA & -99.222 & 0.345 & -98.458 & 0.301 & -97.402 & 0.480\\
IV & RA & -99.180 & 0.345 & -98.432 & 0.297 & -97.372 & 0.477\\
\addlinespace
VII & Q & -99.191 & 0.543 & -98.555 & 0.414 & -96.875 & 0.663\\
VII & RQ & -99.300 & 0.491 & -98.638 & 0.386 & -97.239 & 0.616\\
\end{tabular}
\end{ruledtabular}
\end{center}

\end{table*}

Various sources contribute to the allowed statistical variations
between the different analysis approaches. These sources  of statistical variations include:

\begin{itemize}

\item
differences between event-based and energy-based reconstructions
arise from different energy thresholds on
crystal pulses and positron candidates,
\item
differences between local-fitting and global-fitting reconstructions arise from different clustering of crystal hits into positron candidates,
\item 
differences between T-method and A-method histogramming arise from different thresholds and different weightings of positron candidates,
\item 
differences between ratio and non-ratio histogramming arise from the ratio-method time shifts and thereby differing data at the beginning and the end of the fit region.
  
\end{itemize}

Differing strategies for correcting for positron pileup, handling of beam dynamics, and compensating for muon losses, also introduce allowed differences in the systematic uncertainties for the different analyses.  Analysis groups also use different strategies in handling
slow effects.

One approach to estimating the analysis-to-analysis correlations uses
a Monte Carlo to generate positron candidates
and build time distributions.
The statistical correlation coefficients
between various approaches
are then determined by
running many Monte Carlo trials,
generating many time distributions,
and extracting $\omega_a^m$ variances
between different pairs of analysis approaches.

Another approach to estimating the analysis-to-analysis correlations involves
resampling of \RunTwoThree data into multiple subsets.
These subsets are then separately analyzed
using the different analysis approaches.
The statistical correlation coefficients between pairs of
analyses approaches are then extracted from the
measured variances of the $\omega_a^m$ differences
for the resampled subsets.

In Table~\ref{tab:correlations} in the appendix, we list the estimated correlations
between all 19 analyses.
The largest allowed differences are between
event-based analyses and energy-based analyses.
The analyses that employ either a common reconstruction approach or a common histogramming approach (the group of six A-method analyses or the group of
six T-method analyses) only allow much smaller differences.
Note in Table~\ref{t:comon-unblinded-R}, the apparent systematic differences between the A-method analyses and the T-method analyses  are consistent with the allowed differences
between these methods.

We define the pulls between pairs of $\omega_a^m$ determinations
as $( y_i - y_j ) / \sigma_{ij}$
where $y_i$, $y_j$ is the measurement pair and
$\sigma_{ij}$ is the corresponding allowed statistical
and systematic differences.
For each set of $19$ $\omega_a^m$-determinations,
there are $171$ analysis pairs and therefore a total $513$ comparisons across the three datasets.

\begin{figure}[hbt!]
\begin{center}
\includegraphics[width=\linewidth]{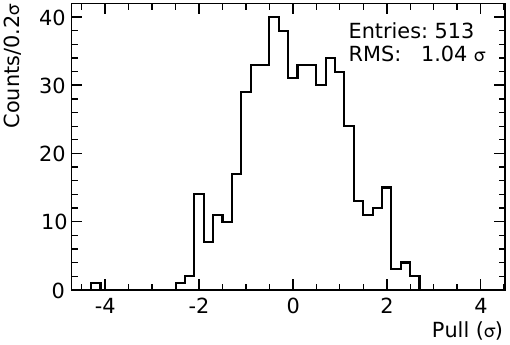}\\
\includegraphics[width=\linewidth]{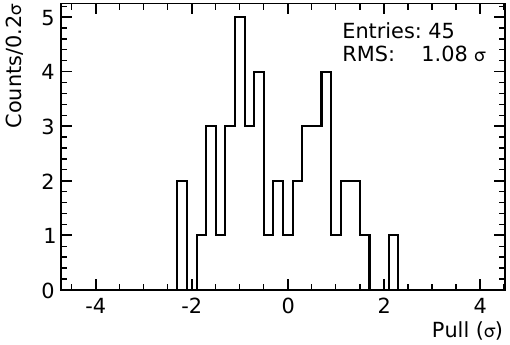}
\end{center}
\caption{Pulls between the 513 pairs of all $\omega_a$ measurements (top panel)
and 45 pairs of A- and RA-method measurements that are used in the $\omega_a$ averaging (bottom panel). The pulls are defined
as $( y_i - y_j ) / \sigma_{ij}$ where $y_i$, $y_j$ are the two measurements and  $\sigma_{ij}$ is the estimated
uncertainty on their difference. The values of $\sigma_{ij}$ are computed using the statistical and systematic uncertainties and their estimated correlations.}
\label{f:pulls}
\end{figure}

In Fig.~\ref{f:pulls}, we plot the
513 pulls for all $\omega_a^m$ measurements
and the 45 pulls from the eight A-method and RA-method measurements
that are most relevant to the $\omega_a^m$ averaging.
Their standard deviations are 1.04 and 1.08, respectively.

\subsection{Consistency checks}
\label{ss:crosschecks}

Beyond the fit $\chi^2$, fit residuals, and the discrete Fourier transform of the fit residuals, a number of checks were made on the robustness of the results for the frequency $\omega_a^m$ and other parameters.

\begin{figure}[hbt!]
\begin{center}
\includegraphics[width=\linewidth]{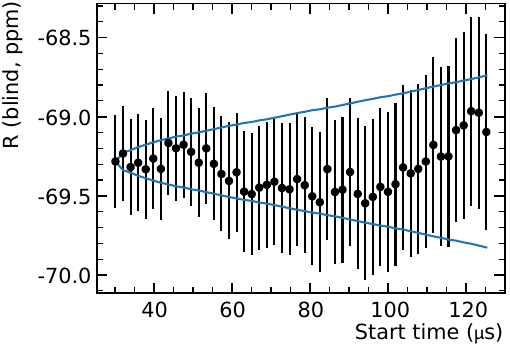}
\end{center}
\caption{A representative scan of the blinded R-value versus the fit start time for the Run-3a dataset and the asymmetry-weighted histogramming method. The black data points are the R-value fit results. The point-to-point values are highly correlated and the smooth blue curve is the $1$ allowed standard deviation band of any fit result from the canonical \SI{30.1}{\micro\second} fit start time. The allowed deviation band accounts for the statistical correlations between the \SI{30.1}{\micro\second} and $>\SI{30.1}{\micro\second}$ fit results. Note the vertical axis includes an analysis-dependent software blinding and cannot be compared to Fig.~\ref{f:comon-unblinded-R} and Table~\ref{t:comon-unblinded-R}.}
\label{f:3a-starttime}
\end{figure}

All analyses fit their time distributions
with incrementally increasing start times to 
probe the stability of the fit parameters.
A representative start time scan,
for an A-method analysis of the Run-3a dataset,
is shown in Fig.~\ref{f:3a-starttime}.
The start time scan dependence
of $\omega_a^m$ is sensitive to effects that vary
from early to late in fill such as cyclotron rotation,
positron pileup, and gain changes.
All analyses demonstrated
the start time scan stability of fitted $\omega_a^m$ values
within the allowed statistical deviations.

\begin{figure}[hbt!]
\begin{center}
\includegraphics[width=\linewidth]{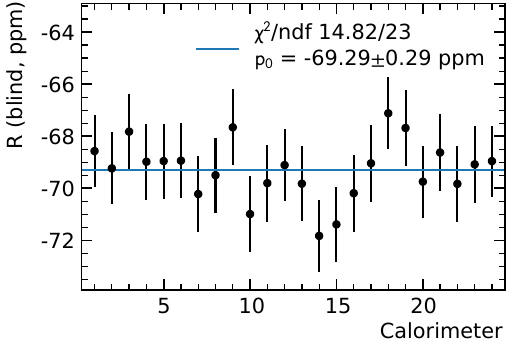}
\end{center}
\caption{A representative scan of the blinded R-value versus the calorimeter index for the \RunThreeA dataset and the asymmetry-weighted histogramming method. The black data points are the R-value fit results, and the solid blue line is a straight-line fit to the 24 individual calorimeter R-values.
Note the vertical axis includes an analysis-dependent software blinding and cannot be compared to Fig.~\ref{f:comon-unblinded-R} and Table~\ref{t:comon-unblinded-R}.}
\label{f:3a-calo}
\end{figure}

All analyses fit the 24 time distributions of the individual
calorimeters to perform calorimeter scans.
A representative calorimeter scan, for an A-method analysis
of the 3a dataset, is shown in Fig.~\ref{f:3a-calo}.
The calorimeter scan dependence of $\omega_a^m$ is
sensitive to effects from cyclotron rotation and CBO modulation
that are larger in the individual calorimeters
than the calorimeter sum (as a result of the
2$\pi$ phase advance of the cyclotron rotation
and the CBO modulation around the ring circumference).
All analyses demonstrated
the calorimeter scan stability of fitted $\omega_a^m$ values
within the allowed statistical deviations.

Fits as a function of the positron energy were also performed for the
event-based analyses. Such energy scans are sensitive to effects
of positron pileup and gain changes that vary with energy.
No evidence was found for $\omega_a^m$ variation with positron energy.

All analyses also reported the correlation coefficients between the fit
parameters in their $\omega_a^m$ fits. A large, known correlation exists
between the frequency $\omega_a^m$ and its phase $\phi$. A smaller, known correlation exists between the frequency $\omega_a^m$
and the frequency and phase parameters of the leading-order CBO term. 

\subsection{Systematic uncertainties}
\label{ss:systematicsomegaa}

The systematic uncertainties reflect the inevitable shortcomings in modeling the true behavior of beam dynamics and other effects.
Each analysis made reasonable choices
for the required modeling of the various effects in the data,
and each analysis made independent estimates of
systematic errors.  The reported errors are averaged
across the analysis groups with the same weightings
as the $\omega_a^m$ averages.

\begin{table}
\caption{Summary of the major systematic uncertainties
for the $\omega_a^m$ analysis of the three datasets.
The major systematic uncertainties arose from
the handling of CBO effects, the corrections for
gain changes and positron pileup, and the presence of
a residual slow effect. `Other systematics' refers
to the sum of all other systematic uncertainties.
}
\label{t:wasystematics}
\begin{ruledtabular}
\begin{tabular}{lrrrr}
Systematic uncertainty & \RunTwo & \RunThreeA & \RunThreeB & \RunTwoThree \\ 
& (ppb) & (ppb) & (ppb)  & (ppb) \\\hline
CBO handling & 22 & 18 & 28 & 21\\
Pileup corrections & 9 & 6 & 7 & 7\\
Gain corrections & 5 & 4 & 5 & 5\\
Residual slow effect & 5 & 14 & 10 & 10 \\
Other systematics & 2 & 5 & 3 & 4\\ \hline
Total & 25 & 24 & 31 &  25 \\
\end{tabular}
\end{ruledtabular}
\end{table}

The major sources of $\omega_a^m$ systematic uncertainties are summarized in Table~\ref{t:wasystematics}.
The treatment of the CBO distortions of the time distributions provides the largest source of systematic uncertainty.
The pileup and gain corrections (see Sec.~\ref{ss:wacorrections}) and presence of residual slow effects (see Sec.~\ref{sss:slowtermfit}) also yield significant systematic uncertainties.
The total systematic uncertainty for the three
datasets varies from 24 to \SI{31}{ppb}.

Each of the above systematic categories contains multiple contributions. In general, we assume that the contributions to a specific category may be correlated and are summed linearly \footnote{An exception to the policy of adding systematics linearly within a
systematics category is the CBO frequency drift and CBO decoherence envelope systematics. A dedicated study showed that the two systematic uncertainties are independent and therefore add in quadrature.}.  Conversely, we assume that systematics from different categories are not correlated and are summed quadratically.

The total systematic uncertainty for the $\omega_a^m$ analysis
is about two times smaller than \RunOne (\SI{56}{ppb}).
First, in \RunTwoThree, the CBO systematic was
reduced through studies that determined
that the contributions from the CBO decoherence envelope
and the CBO frequency change are uncorrelated and add in quadrature.
Second, in \RunTwoThree, the pileup systematic was
reduced through a combination of improved reconstruction algorithms, which yielded less pileup, and improved correction in more analyses. A pileup phase uncertainty was also shown
to be overestimated in the \RunOne analysis.
Third, in \RunTwoThree, the source of the residual slow effect became partially understood, thus reducing this systematic.

The following sub-sections discuss our procedures for estimating the CBO, pileup, slow term, gain and other systematics.

\subsubsection{CBO systematic}
\label{sss:cbosystematic}
Three significant uncertainties from beam dynamics were identified: uncertainty in the shape of the
CBO decoherence envelope, uncertainty in the drift of the CBO frequency, and uncertainty in the
lifetime of the CBO effects on the  precession asymmetry and its phase.

Note that the CBO envelope changed
from \RunThreeA to \RunThreeB as a result
of the increased kicker voltage.
For datasets \RunTwo and \RunThreeA, a simple exponential envelope
was sufficient to model the CBO decoherence.
For \RunThreeB, an additional constant term
was needed to model the CBO decoherence.

To estimate the systematic
associated with envelope shapes,
the analyses studied a variety of envelope functions.
The shapes incorporated constant, exponential, and reciprocal terms
and their combinations. The systematic
was estimated from the changes of the $\omega_a^m$ results
for all functions with an acceptable $\chi^2$ value. The average contribution of the CBO decoherence systematic across the  datasets and analyses in Table \ref{t:wasystematics} was about 16~ppb.

The \RunTwoThree CBO frequency drift was roughly ten times
smaller than the  \RunOne drift due to the repair of the ESQ resistors~\cite{Run1PRDomegaa}.
The \RunTwoThree drifts, attributed to the effects of quadrupole scraping and calorimeter acceptance, were modeled as an exponential relaxation of the CBO frequency.
The associated systematic uncertainty originates from the
poorly-known relaxation lifetime. The average contribution of the frequency-drift systematic across the datasets and analyses in Table \ref{t:wasystematics} was about 10~ppb.

Lastly, as discussed in Sec.~\ref{sss:beamdynamicsfit}, the CBO
also modulates the precession asymmetry $A_0$ and precession phase $\phi_0$.
These effects are similarly modeled by a sinusoidal oscillation
with a decoherence envelope. The effects on $A_0$ and $\phi_0$ are small and their impacts on determining $\omega_a$ are negligible compared to the CBO decoherence systematic and the CBO frequency-shift systematic. 

\subsubsection{Pileup systematic}

The procedures for correcting the time distribution for
pileup distortions are discussed in Sec.~\ref{sss::PUcorrections}.
The corrections involve superimposing either
digitized waveforms, crystal hits, or positron candidates. This pileup modeling is subject to inaccuracies in our knowledge of the detector response and the analysis reconstruction. Further systematics include errors in the pileup rate, errors in the pileup time distribution, and the truncation of the pileup correction at a finite order. Errors arising from unseen pileup -- pileup below the threshold for the reconstruction -- were also evaluated.

The two largest contributors to the pileup uncertainty are the accuracy of the pileup model, roughly \SI{2}{ppb},
and the error from the unseen pileup,
also roughly \SI{2}{ppb}. The various other sources of pileup systematic uncertainties were $\mathcal{O}$$(1~\mathrm{ppb)}$.

We note that the uncertainty in the overall normalization of the pileup correction is about 1\%.
This is determined by comparing the raw energy and reconstructed-pileup energy distributions in the region above 3.1~GeV (see Fig.~\ref{f:pileup}). This has a negligible contribution to the systematic uncertainty.

\subsubsection{Residual slow term systematic}
\label{sss:slowtermsystematic}

As already discussed, both \RunOne data and \RunTwoThree data indicated a
residual slow effect in the event-based time distributions.
Its handling is described in Sec.~\ref{sss:slowtermfit}.

In the local-fitting, event-based analyses, we identified
an energy-scale shift as a contribution to the residual slow
effect. The local-fitting analyses either explicitly corrected
their analyses for the energy-scale shift or treated the
effect as a systematic as in \RunOne.

The remaining effect -- about one-third of the size of the energy-scale shift -- has unknown origin(s).
To evaluate the associated systematic, we applied a `gain-like' correction
to accommodate the effect and evaluate its impact on $\omega_a^m$.
Two approaches for applying this correction were developed. One method utilized the $\chi^2$ of the fit, and another method equalized the muon-loss normalization across energy bins. Both methods were consistent, and the impact on $\omega_a^m$ was 5 to \SI{10}{ppb}.

Also included within this systematic category -- because it is highly correlated with the residual slow term -- is the uncertainty assigned to the fit preference for a non-physical, negative,
$k_{\mathrm{loss}}$ parameter in Run-3a and 3b.\footnote{A negative muon loss parameter would imply a gain of stored muons and therefore is considered nonphysical.}  This systematic is estimated from the $\omega_a^m$ shift required to return
to $k_{\mathrm{loss}} \geq 0$. The total systematic for this category was estimated at 5 to \SI{14}{ppb}.

\subsubsection{Gain systematic}

The procedures for correcting the time distributions for
gain changes are discussed in Sec.~\ref{sss:gaincorrections}.
The long-term gain correction has a negligible effect
on extracting $\omega_a^m$, since this correction is a time-independent factor for each muon fill.
The two other gain corrections, in-fill and short-term, do change with time in fill.

Both the in-fill gain change and short-term gain change were
modeled as exponential relaxations of gain sags.
The in-fill gain correction is larger and dominates the gain systematic.

The sensitivity to the in-fill gain
parameters is determined by scaling the correction
and observing the change in $\omega_a^m$.
This sensitivity is then
combined with the uncertainty on the parameters
obtained from the laser calibration system.
Uncertainties are conservatively assumed to be fully correlated across all calorimeter crystals.
The resulting in-fill gain systematic is roughly \SI{4}{ppb}.  The same procedure is applied in estimating the smaller short-term gain
systematic.

\subsubsection{Other systematics}

The remaining categories of systematic uncertainties considered are
the  timing calibration of the individual calorimeter channels, the time randomization for the fast rotation handling, the shape of the reconstructed muon loss time distribution, and the
requirement of a fixed muon lifetime and precession period in the ratio histogram construction.  The largest was the muon loss systematic, which contributed an uncertainty of 1 to \SI{5}{ppb}.

\subsection{Combination of $\mathbf{\omega_a^m}$ measurements}
\label{sec:omega_combination}

To define a single measured value of $\omega_a^m$ for each of datasets
Run-2, Run-3a, and Run-3b, we performed an equal-weighted average of the six measurements I-A, II-A, III-RA, IV-RA, V-A and VI-A where I-A,
{\it etc.}, denote the analysis group and histogram method.
This strategy combines two local-fitting A-method analyses, two global-fitting A-method analyses, and two ratio histogramming A-method analyses.
We did not include measurements using the T, RT, Q, or RQ methods because their statistical uncertainties are significantly larger, their systematic uncertainties are similar or larger, and their estimated correlations imply no appreciable reduction of the uncertainty of the average.

For each dataset, we conservatively assume that the statistical uncertainty and each systematic category uncertainty are fully correlated between the six averaged measurements.
In such circumstances, both the statistical uncertainty and the   individual systematic uncertainties of the dataset average, are the plain average of the six measurements. Each systematic category uncertainty is also conservatively assumed to be fully correlated across the three datasets.

As mentioned in Sec.~\ref{ss:fitresults}, we estimated the statistical correlations between the $\omega_a^m$ measurements within the same dataset (see Table~\ref{tab:correlations}). The statistical correlations between the six averaged analyses range from 0.993 to 1.000. The optimal linear combination of the six measurements in a $\chi^2$ fit using these correlations has an uncertainty that is only 1.5\% smaller than the plain average. 
Consequently, considering that the estimated correlations have significant uncertainties, we use the aforementioned plain average in computing $\omega_a^m$.
 
\section{Beam dynamics corrections\label{sec:bd}\label{sec:beamdynamicsection}}
\label{sec:bd:corr}

This section reviews the analysis and evaluation of the five beam dynamics corrections to $\omega^a_m$, introduced in Sec.~\ref{sec:intro}. 

\subsection{Electric-field correction\label{sec:bd:corr:Ce}}
The radial electric-field contribution from the ESQ to \oa in Eq.~\eqref{eq1} cancels only for magic-momentum muons.
The electric-field correction $C_e$ 
accounts for the spin precession in $\omega_a^m$ induced by 
the momentum spread of
the stored muon beam.

Expanding the second term in Eq.~\eqref{eq1} to the first order in the muon momentum offset from the magic momentum $p_0$, the shift relative to the ideal frequency is 
\begin{align}
  \frac{\Delta\omega_a}{\omega_a} &= - 2 \, \frac{\beta_0}{cB_0} \, \delta \, E_x,
\end{align}
where 
$\delta = (p - p_0)/p_0$, 
 $\beta_0$ is the magic-momentum  velocity, $B_0$ the vertical magnetic field, and $E_x$ the radial component of the ESQ electric field. For small radial displacements, $x$, from the center of the ESQ, the electric field is approximately linear
\begin{align}
  E_x \approx n \, \frac{\beta_0 cB_0}{R_0} \, x,
\end{align}
where $n \approx 0.108$ is the effective focusing field index (accounting for the finite lengths of the quadrupole sections) and $R_0$ is the magic-momentum bending radius. The muon-momentum offset can also be expressed in terms of the radial displacement from $R_0$, $x_e$, and the field index via the dispersion relation
\begin{align}
  \delta \approx \left( 1-n \right) \frac{x_e}{R_0}.
\end{align}
The electric-field correction averaged over all momenta is
\begin{align}
  C_e = - \left\langle \frac{\Delta\omega_a}{\omega_a} \right\rangle \approx 2n(1-n) \, \beta_0^2 \, \frac{\langle x_e^2 \rangle}{R_0^2}.
\label{eq:Ce_formula}
\end{align}
The following sections describe the two analyses used to evaluate the electric-field correction and the results.

\subsubsection{Fast-rotation analysis}\label{sss:cyclotronrotation}

Because the tangential speed, $\beta_0$, is constant to the {ppm-level} for the stored muons, the measured cyclotron angular frequency, $\omega_c$, determines the radial displacement $x_e$ through
\begin{align}
  \beta_0 \approx R\,\omega_c=(R_0 + x_e) \, \omega_c.
\end{align}

\begin{figure}[htbp!]
\begin{center}
         \centering
         \includegraphics[width=\linewidth]{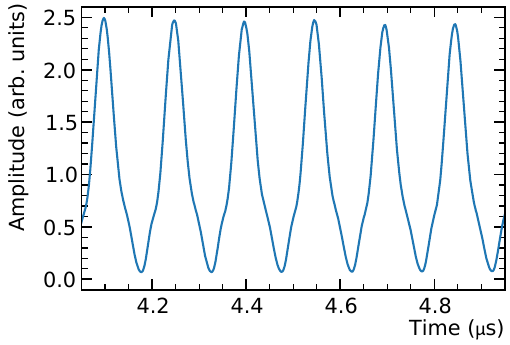}\\
         \includegraphics[width=\linewidth]{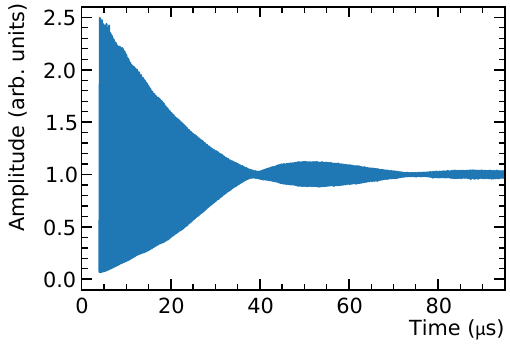}
\end{center}
\caption{Fast-rotation signal from Run-2 data, showing individual turns around the storage ring over short time scales (top) and broader decoherence envelope over long time scales (bottom).}
\label{Fig:FRSignal}
\end{figure}

The cyclotron frequency spread of the muons modulates the decay positron intensity detected by the calorimeters and is referred to as the fast-rotation signal. In the fast-rotation analysis, we use this signal to reconstruct the momentum distribution of the stored muons for the determination of $C_e$. 
At the start of a fill, the stored muons are tightly bunched. As the fill progresses, the muons spread out azimuthally over time due to the spread in their momenta.
This effect leads to decoherence of the fast-rotation signal shown in Fig.~\ref{Fig:FRSignal}.

The fast-rotation component of the positron intensity signal is isolated in two ways:
\begin{itemize}
  \item Smearing method: The pulses of the decay positron time spectrum are randomly split into two halves: a numerator and a denominator. Each detection time in the denominator is randomized by an amount uniformly distributed between $\pm T_c/2$, where $T_c$ is the revolution period. This randomization smears out the fast rotation in the denominator while slower features remain intact. Slowly changing features common to the numerator and denominator are eliminated in the ratio, leaving only the fast-rotation signal from the numerator.
  \item Fit method: The decay positron signal is binned at intervals of the expected revolution period, which approximately removes the fast rotation. The resulting histogram is then fit using a simplified version of the $\omega_a^m$ analysis fit model, which accounts for the most important features. The finely binned decay positron time spectrum is then divided by the fit function. As in the smearing method, the only prominent oscillation in the resulting ratio histogram is the fast rotation. Figure~\ref{Fig:FRSignal} shows an example of a fast-rotation signal from Run-2 isolated by the fit method.
\end{itemize}

The fast-rotation signal $S(t)$ can be modeled as a weighted combination of periodic impulse trains with frequencies $\omega$ and time offsets $\tau$, representing periodic detection of the circulating muon bunch, yielding
\begin{align}
  S(t) &= \int_{-\infty}^{\infty} \int_{-\infty}^{\infty} \sum_{m} \delta\!\left[ t - \left( \frac{2\pi m}{\omega} + \tau \right) \right] \rho(\omega, \tau) \, d\omega \, d\tau,
  \label{Eq:FRModel}
\end{align}
where $m$ is the turn index around the storage ring and $\rho(\omega, \tau)$ the joint distribution of revolution frequencies and injection times for stored muons. Analysis approaches,  based on Fourier analysis or a fit to the time-domain signal, are used to estimate the 
frequency distribution based on this model.

The Fourier analysis depends on the important assumption that $\rho(\omega,\tau)$ is separable. However, this is generally not true since the kicker pulse is not flat over the width of the injected pulse and preferentially stores different momenta in different time slices of the injected bunch. This ``momentum-time correlation'' causes a systematic distortion to the Fourier analysis, which depends on the kicker pulse shape. To rectify this feature, an alternative analysis, named the ``fast-rotation $\chi^2$ method'' and based on a method invented for the CERN
storage ring experiments, accounts for the momentum time correlation. The
results from this analysis can be used to correct the Fourier method. In the CERN method, the fast-rotation signal $S(t)$ is fit with a simple debunching model. Integrating Eq.~\eqref{Eq:FRModel} over narrow bins for $\omega$ and $\tau$, where the weight $\rho(\omega, \tau)$ is approximately constant for each bin, yields the contribution of each $(\omega, \tau)$ bin to the signal at time $t$. Denoting this component as $(\beta_{ij})_k$, where $i$ and $j$ label the $(\omega, \tau)$ bin, and $k$ labels the time bin of the fast-rotation signal, the overall signal $S_k$ may be expressed as a linear combination of these component signals, yielding
\begin{align}
  S_k &= \sum_{i,j} (\beta_{ij})_k \, \rho_{ij},
\end{align}
where $\rho_{ij}$ are the unknown weights of the discretized $\rho(\omega, \tau)$ distribution, treated here as fit parameters determined from the fits.

\begin{figure}[htbp!]
\begin{center}
         \centering
         \includegraphics[width=\linewidth]{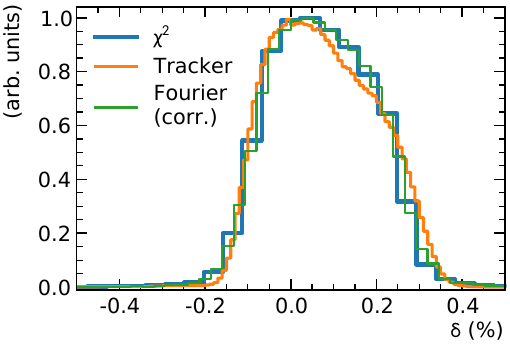}
\caption{Fractional momentum distributions from the fast-rotation $\chi^2$ method, the tracking analysis method (data from the straw tracking detector at $180^\circ$), and the corrected Fourier analysis for the data subset 3F.}
\label{Fig:MomDists}
\end{center}
\end{figure}

\begin{figure}[htbp!]
\begin{center}
         \centering
         \includegraphics[width=\linewidth]{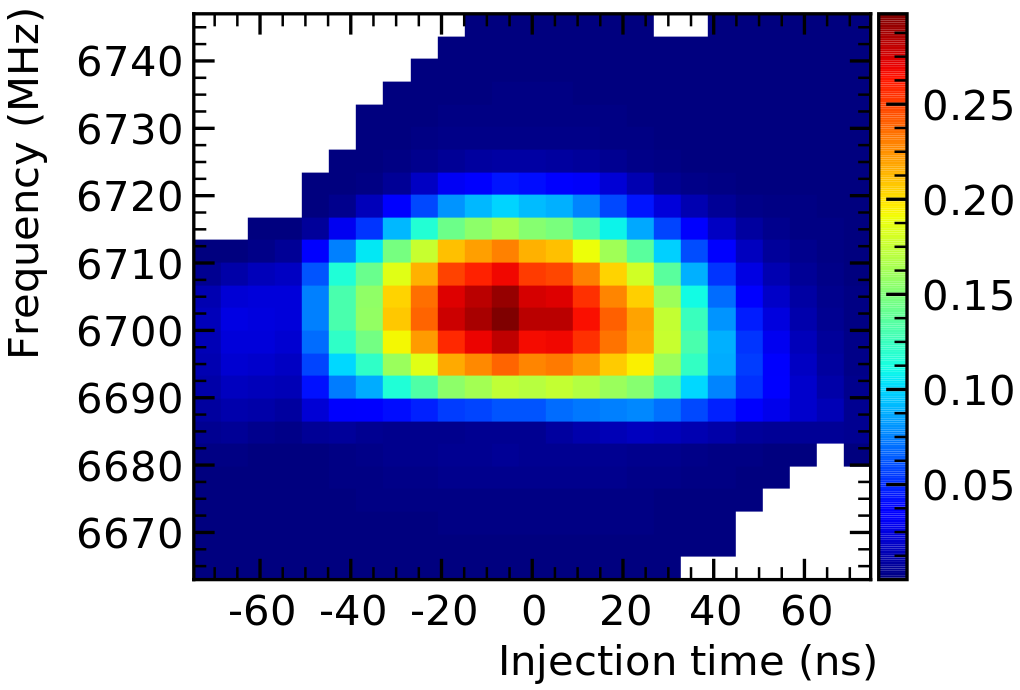}
\caption{Joint distribution from the fast-rotation $\chi^2$ method of revolution frequency and injection time determined by the direct fit method for the data subset 3N, first bunch in the beam pulse sequence.}
\label{Fig:FRDistributions}
\end{center}
\end{figure}

This prescription typically allows too many free parameters to obtain physically reliable fit results. To impose constraints, the frequency distribution in each injection time slice is assumed to have the same fundamental shape as in the central time slice, but with features of the three lowest moments (mean, standard deviation, and skew) varying smoothly as quartic polynomials over the injection time using the sinh-arcsinh transformation \cite{SinhArcsinh}. This modeling reduces the number of parameters to 62: one frequency distribution (25 bins), one overall injection time distribution (25 bins), and 12 polynomial coefficients, which describe the momentum-time correlation. Our $\chi^2$ minimization passes employed both the Davidon-Fletcher-Powell algorithm \cite{davidon1991variable} and refinements with simulated annealing. Each spectrum was fit multiple times from different starting parameters. Because of systematic shape variations in the beam pulses, fits were performed separately on time spectra for each of the bunches delivered by the Fermilab accelerator complex, as well as for the summed spectrum; see Fig.~\ref{Fig:MomDists} for a momentum distribution and Fig.~\ref{Fig:FRDistributions} for a joint distribution obtained in this manner for data subsets from Run-3a and Run-3b. 

We assessed the following systematic errors associated with the fast-rotation analysis methods: late start time, failure to remove stray frequencies from the signal, changes to the distribution created during scraping, and insufficient shape parameters.

With a quantitative description of the systematic distortions contributed by the correlation between $\omega$ and $\tau$, the Fourier analysis may then be corrected 
by evaluating the correlation-dependent parts using the correlation from the $\chi^2$ method as an external input (see Fig.~\ref{Fig:MomDists} for an example of the reconstructed momentum distribution obtained in this way). Thus, the corrected Fourier analysis is no longer completely independent from the fitting method, but it does enable a check for consistency between the two methods.

\subsubsection{Positron tracking analysis}

The stored beam exhibits a periodic pattern in which the initial narrow width imposed by passage through the inflector grows as the beam circulates due to the momentum dependence of the radial closed orbits. We developed a method for \RunTwo and \RunThree datasets to reconstruct the muon momentum distribution based on this behavior of the muons in the radial direction, $x$, which is directly observed by the positron tracking detectors until the betatron oscillations decohere. Figure~\ref{Fig:MomDists} includes a sample of a momentum distribution derived from this analysis.

The minimum and maximum radial spreads are apart by half of a betatron period, which appears in data from a detector located at a specific azimuth as the aliased coherent period (see Table~\ref{t:beamfrequencies}). The momentum-dependent magnetic rigidity $B_0 R=p_0(1+\delta)/e$ governs the amount of the spread. The linear matrix of an inhomogeneous magnet with field index $n$ \cite{berz2015introduction} well describes this spectrometric relation between the momentum and radial coordinates, which takes on a simple form for two states, $i$ and $f$, separated by a phase advance of $\pi/\sqrt{1-n}$ (or, equivalently, separated in time by ${\sim}T_{\text{CBO}}/2$ at a fixed detector):
\begin{align}
  \begin{pmatrix}
    x \\ x' \\ \delta
  \end{pmatrix}_{\!\!f}
  &=
  \begin{pmatrix}
    -1 & 0 & 2 \frac{R_0}{1-n} \\
    0 & -1 & 0 \\
    0 & 0 & 1
  \end{pmatrix}
  \begin{pmatrix}
    x \\ x' \\ \delta
  \end{pmatrix}_{\!\!i}.
  \label{BD:eq:MatInhDipole}
\end{align}

In Eq.~\eqref{BD:eq:MatInhDipole}, the variables $x$ and $x'$ represent the spatial and angular offsets in radial phase space. From the radial coordinate $x_f$ expressed in terms of the state-$i$ coordinates, the spectrometric relation is
\begin{align}
  \delta &= \frac{1-n}{2R_0} (x_i + x_f).
  \label{Eq:SpectrometerRelation}
\end{align}  

From Eq.~\eqref{Eq:SpectrometerRelation}, the radial distribution at state $f$ would equal the momentum distribution, shifted by $x_i$ and scaled by $(1-n)/2R_0$, if all the stored muons were to share the same coordinate $x_i$. For \RunTwo and \RunThree, the tracking detectors measured a radial beam that resembled this idealized scenario. Therefore, by defining $x_i$ as the radial mean of the stored beam when the radial width is minimal, we implemented Eq.~\eqref{Eq:SpectrometerRelation} to reconstruct the momentum spread from which $\langle \delta^2 \rangle$ is taken to calculate the electric-field correction via Eq.~\eqref{eq:Ce_formula}.

The method is validated with realistic beam-tracking simulations using the \texttt{gm2ringsim} package \cite{Run1PRAB}. The associated uncertainty is only significant for Run-3b, as shown in Table~\ref{tb:Ce_Tracker_Uncertainties}. 
\begin{table}[h]
    \centering
    \caption{Uncertainties of the electric-field correction from the tracking analysis.}
    \label{tb:Ce_Tracker_Uncertainties}
    \begin{ruledtabular}
    \begin{tabular}{lccc}
    Description &  \multicolumn{3}{c}{Uncertainty [ppb]} \\
    & \quad Run-2 & \quad Run-3a & Run-3b \\
    \hline
    \textit{Statistical}  & & & \\
    ~~Station 12  & 0.7 & 0.3 & 0.4 \\
    ~~Station 18  & 0.8 & 0.4 & 0.5 \\
    \hline
    \textit{Systematic}  & & & \\
    Method  & & & \\
    ~~Beam simulation  & 5.4 & 5.0 & 27.8 \\
    Detector effects  & & & \\
    ~~Tracker resolution  &  5.0 & 5.0 & 5.0 \\
    ~~Tracker acceptance  &  21.8 & 21.5 & 18.3 \\ 
    ~~Tracker alignment & 21.0 & 20.3 & 11.1 \\
    ~~Calorimeter acceptance & 2.0 & 2.0 & 2.0 \\
    Other effects & & & \\
    ~~Tracker station differences & 4.0 & 4.8 & 1.7 \\
    \hline
    Total & 31 & 31 & 35
    \end{tabular}
    \end{ruledtabular}
    \end{table}
In this dataset, the beam simulation shows a discrepancy between the truth and reconstructed momentum distributions using the tracking analysis. The discrepancy grows over time while the truth values stay stable, and the reconstructed value falls with time, which is not present in the Run-2 or Run-3a simulations. We see the same behavior in the data analysis of \RunThreeB, where the reconstructed value of $C_e$ steadily decreases over time, so we consider this behavior a real effect also present in the data. Hence, we apply a \SI{28}{ppb} correction to the results obtained for \RunThreeB, which comes directly from comparing truth and reconstruction in the simulation. Given the reliance on simulation, we apply a $100\%$ uncertainty \SI{28}{ppb} on this correction for the Run-3b dataset.

The uncertainties from the tracking analysis are dominated by acceptance correction, alignment, and simulation uncertainties.
The acceptance correction uncertainties are approximately \SI{20}{ppb} for all three datasets. This value comes from conservatively varying the shape of the known correction by $\pm 50\%$. 

The uncertainty in the analysis associated with tracker alignment emerges from the $\pm\SI{0.6}{mm}$ uncertainty of the detector radial locations, assumed as uncorrelated between the two tracker stations (its effect is thus reduced by a factor of $1/\sqrt{2}$). This uncertainty is smaller in Run-3b because the systematic bias resulting from an error in tracker alignment scales with the mean value of the muon momentum distribution. In Run-3b, 
the mean momentum relative to $p_0$, 
$\langle \delta \rangle$, is smaller than the width, $\sigma_{\delta}$,  due to increased kick strength, and thus, when we add the sum of squares to get 
\begin{align}
  C_e &= \frac{2n\beta_0^2}{1-n} \left( \langle \delta \rangle^2 + \sigma_{\delta}^2 \right),
\end{align}  
it is less significant.

The resolution uncertainty in this analysis assumes a detector resolution of $\sim$\SI{3.5}{mm} on the tracker reconstruction of the transverse muon coordinates.  Resolution studies at early times after injection indicate a $25\%$ uncertainty on this value, and we assess the associated systematic uncertainly by scaling the correction by $\pm 25\%$.
The sensitivity of the reconstructions to such resolution uncertainties has an upper limit of \SI{5}{ppb}, which we assign as a systematic uncertainty. The effect of mismatching calorimeter-vs-tracker acceptances is small, as shown in Table~\ref{tb:Ce_Tracker_Uncertainties}.

The last systematic error in this analysis arises from differences between $C_e$ reconstructions from the two tracker stations. Such difference potentially emerges from additional closed orbit distortions due to ESQ plate misalignments.

\subsubsection{Results}

\begin{figure}[htbp!]
\begin{center}
         \centering
         \includegraphics[width=\linewidth]{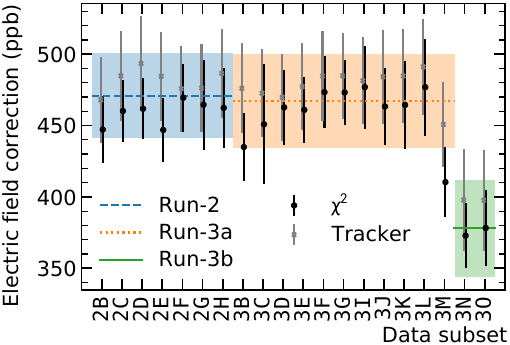}
\end{center}
\caption{Electric-field corrections $C_e$ by data subset obtained from the tracking analysis method and the fast-rotation $\chi^2$ method. The final values for Run-2, Run-3a, and Run-3b are shown in color, which come from the combination of the calorimeter and tracker-based analyses.}
\label{Fig:CaloTrackerResultsByDataset}
\end{figure}

Figure~\ref{Fig:CaloTrackerResultsByDataset} shows the electric-field correction from the fast-rotation fitting analysis, the positron tracking analysis, and the weighted average of the analyses. 

The tracking analysis is insensitive to the momentum-time correlation, whereas the fast-rotation fitting method was designed to incorporate momentum-time correlation, and the fast-rotation Fourier method is subject to significant distortions caused by momentum-time correlation. 

Results from the tracking analysis at the data-subset level are generally larger than the fast rotation by 16 -- \SI{31}{ppb}. The difference in the results from these independent methods is taken into account to estimate the systematic uncertainty of the electric-field correction.

The final results for $C_e$ 
are presented in Table~\ref{Tab:EFieldValues}.
The combined result is the weighted average, assuming the uncertainties for each are completely uncorrelated. 
The {electric-field} correction is significantly smaller for Run-3b due to the better-centered momentum distribution of the stored beam.
\begin{table}
\caption{\label{Tab:EFieldValues}Table of central values and uncertainties for $C_e$ (ppb) from the fast-rotation and tracking methods. Only the combined values are used for the full Run-2/3 dataset.}
\begin{ruledtabular}
\begin{tabular}{lcccccc}
\multirow{2}{*}{Dataset} & \multicolumn{2}{c}{Fast Rotation} & \multicolumn{2}{c}{Tracking} & \multicolumn{2}{c}{Combined} \\
 & Corr. & Unc. & Corr. & Unc. & Corr. & Unc. \\
\colrule
Run-2 & 459 & 24 & 485 & 31 & 469 & 30 \\
Run-3a & 459 & 28 & 475 & 31 & 466 & 32 \\
Run-3b & 367 & 27 & 398 & 35 & 378 & 33 \\
\end{tabular}
\end{ruledtabular}
\end{table}

A separate class of uncertainty in the final values of the combined result was evaluated, namely, the alignment and voltage errors of the ESQ stations, which correspond to an uncertainty of \SI{6}{ppb}. This error applies equally to the tracking- and fast-rotation-based analyses and is added in quadrature to the uncertainty of the combined result.  
We intend to conduct more extensive research to better understand the uncertainties associated with the recently developed techniques for determining the electric-field correction. For this reason, we increase the calculated uncertainties by a factor of 1.5.
The final uncertainty values are at the level of 30 -- \SI{33}{ppb}, as shown in Table~\ref{Tab:EFieldValues}.

\subsection{Pitch correction\label{sec:bd:corr:Cp}}
The electric field that keeps the beam confined in the vertical direction drives a radial component of the spin angular frequency \cite{PRAB_On_Pitch}, which biases $\omega_a$. The pitch correction 
\begin{equation} \label{eqCp}
C_p = \frac{1}{2}\langle \psi^2 \rangle,
\end{equation}
where $\psi=\frac{dy}{dz}$ is the  pitch angle, corrects this bias. This angle is calculated in accordance with sinusoidal vertical betatron motion: 
\begin{equation}
y = A \sin(kz + \phi) + \bar{y}, 
\end{equation}
where $z$ and $\bar{y}$ are the longitudinal coordinate and vertical mean position of muons in the storage ring, respectively. This expression allows Eq.~\eqref{eqCp} to be rewritten as
\begin{equation} \label{eqCpAmean}
C_p = \frac{n}{4 R_0^2}\langle A^2\rangle.
\end{equation}
Here, $A$ is the amplitude of the beam’s vertical oscillations, $n$ is the field index, and $R_0$ is the magic momentum radius. 

\begin{figure}[htbp!]
\begin{center}
         \centering
         \includegraphics[width=\linewidth]{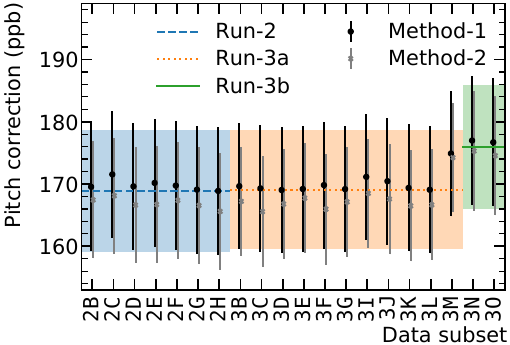}
\end{center}
\caption{Comparison between method-1 and method-2 of the pitch correction, $C_p$, results for all data subsets available in Run-2 and Run-3. The errors in the two methods are dominated by the tracking uncertainty.}
\label{fig:PitchComparison}
\end{figure}

Two independent analyses, ``method-1’’ and ``method-2,’’ determine $C_p$. Both start with the vertical decay distributions measured by the two straw tracking detectors located at 180$^\circ$ and 270$^\circ$, following equal selection criteria, but apply different corrections for tracker resolution and acceptance. The resulting tracker data is transformed into amplitude space, and $C_p$ is calculated using Eq.~\eqref{eqCpAmean}. Both methods then correct for the calorimeter acceptance. In this way, the calculated $C_p$ reflects the bias on $\omega^{m}_a$ for the muon population contributing to the calorimeter measurement.
The two methods calculate an average $C_p$ for each dataset, as seen in Fig.~\ref{fig:PitchComparison}. To make the switch to the amplitude space, method-1 derives a functional form, whereas method-2 uses a data-driven approach to estimate the amplitude distributions. In the end, results are within $\sim$\SI{2.5}{ppb} of each other, consistent with the statistical and systematical errors. Central values are calculated for each dataset, and we adopt the average of the final values from the two methods as the final $C_p$ result presented in Table~\ref{tbl:CentralPitch}. The $\sim 8\,\text{ppb}$ uncertainty from the tracking hardware and vertical coordinates reconstruction dominate the systematic uncertainties shown in Table~\ref{tbl:CentralPitch}, compared to other systematic errors from the amplitude fits, tracker acceptance and resolution correction, calorimeter acceptance, ESQ calibration, and tracker station differences.

\begin{table}[htbp!]
 \caption{Pitch correction values, $C_p$, and associated statistical/systematic uncertainties (ppb) for Run-2, Run-3a and Run-3b.}
  \label{tbl:CentralPitch}
\begin{center}
\begin{ruledtabular}
\begin{tabular}{l c c c}
 Dataset & Correction & Statistical Unc. & Systematic Unc.\\ 
 \hline
 Run-2 & 168.9 & 0.02 & 9.8 \\
 Run-3a & 169.1 & 0.01 & 9.5\\
 Run-3b &  175.9 & 0.02 & 10.0 \\
\end{tabular}
\end{ruledtabular}
\end{center}
\end{table}

\subsection{Muon-loss correction\label{sec:bd:corr:Cml}}

Muon losses, defined in Sec.~\ref{sec:beamdynamicintro:lm}, can bias the extraction of $\omega_a^m$ due mainly to the correlation between the \gm phase, $\phi_0$, and average momentum, $p$, of the lost muons distribution. The \gm phase is a single term in the parameter function to extract the anomalous precession frequency (see Sec.~\ref{sss:5parameterfit}), and it represents the ensemble-averaged spin phase referenced at the nominal injection time. Since the momentum of the stored beam could change over the data taking as muons are lost, we introduce the muon-loss correction, $C_{ml}$, to cancel out the resulting biasing on $\omega_a$, where
\begin{gather}
     C_{ml} = -\frac{\Delta\omega_a}{\omega_a} = \frac{1}{\omega_a}\frac{d \phi_0}{dt} = \frac{1}{\omega_a} \frac{d \phi_0}{dp} \left( \frac{dp}{dt} \right)_{ml}.
 \label{eq:dpdt}
 \end{gather}

The time dependence of the lost muons' momentum distribution, $\left({d p }/{dt}\right)_{ml}$, is directly proportional to both the momentum dependence of the loss probability and the overall rate of muon losses  \cite{Run1PRAB}. The mechanism in which the phase is correlated with momentum is described in Sec.~\ref{sec:bd:corr:Cbl}.

For \RunOne, $C_{ml}$ introduces a $\mathcal{O}(5-20\text{ppb})$ correction \cite{Run1PRAB}.
Post \RunOne, systematic studies show a momentum dependence of the muon losses for \RunTwoThree running conditions similar to \RunOne results;
meanwhile, the phase-momentum correlation $d \phi_0 / d  p $ at injection (which is denoted $t_0=0$) is increased in magnitude from $-10 \pm 1.6$ to $-13.5 \pm 1.4 ~\text{mrad}/(\% \delta)$. This increase is attributed to the addition of a momentum cooling wedge in the upstream beamline during \RunTwo \cite{prab_wedges}. The uncertainties of the measurements come from data fitting, magnetic field uncertainties, dataset differences, and gain changes. 

Despite these differences, the dominant factor in the determination of the muon loss correction is the order of magnitude reduction in losses from \RunTwo onward. Owing to this upgrade, the gradient $\left({d p }/{dt}\right)_{ml}$ and therefore $C_{ml}$ is reduced by an order of magnitude, reaching the sub-ppb level. $C_{ml}$ is calculated with a conservative uncertainty attached as \SI{3}{ppb}:
\begin{gather}
    C_{ml} = 0 \pm \SI{3}{ppb}.
    \end{gather}

\subsection{Differential decay correction} \label{sec:bd:corr:Cdd}
The differential decay correction, $C_{dd}$, accounts for the time dependence of the \gm phase $\phi_0$ (defined in Sec.~\ref{sec:bd:corr:Cml}) due to the spread of muon lifetimes in the beam.  We refer to this spread of decay rate as a function of beam particle momentum as ``differential decay.’’ The correction is thus expressed as
\begin{equation} \label{eq:Cdd}
C_{dd}=-\frac{\Delta\omega_a}{\omega_a}=\frac{1}{\omega_a}\frac{d\phi_0}{dt}=\frac{1}{\omega_a}\frac{d\phi_0}{dp}\left( \frac{d p}{dt}\right)_{dd},
\end{equation}
where $\left( d p/dt\right)_{dd}$ is the temporal variation of the beam-averaged momentum as muons decay in proportion to their time-dilated lifetimes, $\gamma(p)\tau_{\mu}$.
The evolution of the momentum distribution can be approximated by
\begin{equation} \label{eq:dpdt_sigmadp}
\left(\frac{ d p }{dt}\right)_{dd} \approx \frac{p_0}{\gamma_0 \tau_{\mu}}\sigma_{\delta}^2,
\end{equation}
where $\sigma_{\delta}^2$ is the variance of the fractional-momentum distribution. 

In addition to the initial $d\phi_0/dp$ from the upstream beamline (described in Sec.~\ref{sec:bd:corr:Cbl}), there is an additional correlation that develops from the non-symmetric kicker and longitudinal bunch structure during the injection process.
Because of differential decay, the ensemble average phase slightly evolves throughout a fill, interpreted as a slight shift in the value of $\omega_a^m$ from the precession data fits. On the basis of the orbital coordinates ${\bf r}=\{x,x',y,y',t_0\}$ (see Table~\ref{tab:table1}),\begin{table}[b]
\caption{\label{tab:table1}Orbital variables ${\bf r}=\{x,x',y,y',t_0\}$. All the coordinates are relative to the reference axis at injection.
}
\begin{ruledtabular}
\begin{tabular}{lldr}
$r_i$ & Definition \\
\colrule
$x,x'$ & Spatial and angular offsets in radial phase space \\
$y,y'$ & Spatial and angular offsets in vertical phase space \\
$t_0$ & Time relative to the nominal injection time.\\
\end{tabular}
\end{ruledtabular}
\end{table}
the linear momentum dependence of $\phi_0(x,x',y,y',t_0;p)$ is expanded as:
\begin{align*}
\frac{d\phi_0}{dp}&=\frac{\partial{\phi_0}}{\partial{x}}\frac{dx}{dp}+\frac{\partial{\phi_0}}{\partial{x'}}\frac{dx'}{dp}+\frac{\partial{\phi_0}}{\partial{y}}\frac{dy}{dp}\\
       &+\frac{\partial{\phi_0}}{\partial{y'}}\frac{dy'}{dp}+\frac{\partial{\phi_0}}{\partial{t_0}}\frac{dt_0}{dp}+\frac{\partial{\phi_0}}{\partial{p}}.      \stepcounter{equation}\tag{\theequation}\label{eq:dphidp_exp}\\
\end{align*}
Beam tracking studies of the stored muons at injection from \texttt{gm2ringsim} simulations confirm the validity of this equality. From Eqs.~\eqref{eq:Cdd} and \eqref{eq:dphidp_exp}, we divide the $C_{dd}$ correction into three independent contributions based on their physical origins, namely: the \textit{beamline}, \textit{p\text{-}x} \textit{correlation}, and $p\text{-}t_0$ \textit{correlation} effects.

\subsubsection{Beamline effect} \label{sec:bd:corr:Cbl}

The direct correlation between the $g\text{-}2$ phase and momentum drives the beamline effect:
\begin{equation}
\label{eq:Cdd_Beamline}
C_{dd}^{bl}=\frac{1}{\omega_a} \frac{\partial \phi_0}{\partial p} \frac{d p}{d t} \approx \frac{\sigma_\delta^2}{\omega_a \gamma_0 \tau_\mu} \frac{\partial \phi_0}{\partial \delta}.
\end{equation}
After four revolutions of the muon beam around the Delivery Ring (DR) at Fermilab \cite{beam-delivery}, the magnetic field of the bending dipole magnets contribute to a momentum-dependent angle advance between the muon spin and momentum by $\Delta \phi \approx 8\pi a_{\mu} \gamma$, which leads to $|\Delta \phi/\Delta \delta| =8.6\,\mathrm{mrad}/(\%\delta)$ \cite{Run1PRAB}. For Run-1, beam tracking simulations and direct measurements of the correlation determined $|\partial \phi_0/\partial \delta|$ at beam injection to be $10\pm1.6\,\mathrm{mrad}/(\%\delta)$; a result in agreement with the DR-only contribution $|\Delta \phi/\Delta \delta|$.

The first step to calculate $C_{dd}^{bl}$ is to recreate the joint distribution for $\phi_0\text{-}\delta$ of the stored muons at $t=0$ for each data subset from a bivariate normal distribution. The correlation is defined from the $\partial \phi_0/\partial \delta$ measurements and the momentum projection is scaled with the corresponding momentum distributions, determined in the electric-field correction analysis. Then, a Monte Carlo signal with a simplified five-parameter version of Eq.~\eqref{equation:omegaAfit} is prepared out of the $\phi_0\text{-}\delta$ distribution, where the differential decay $e^{-\frac{t}{\gamma(t) \tau}}$ transforms the distribution over time. Finally, we fit the Monte Carlo signal to extract the shift in $\omega_a^m$ due to differential decay.

The difference between the results from the steps described above and Eq.~\eqref{eq:Cdd_Beamline} is negligible. The main purpose of the step-by-step procedure is to test the sensitivity of $C_{dd}^{bl}$ to two possible systematic effects: correlations of $\gamma$ and $\phi_0$ with the muon-momentum dependence of (a) the asymmetry, $A$, and (b) emitted positrons, $N$, based on the leading-order Michel spectrum.  Because these effects produce systematic uncertainties below $2\,\mathrm{ppb}$, we assign a conservative upper limit of $3\,\mathrm{ppb}$ to the differential-decay beamline correction. Table~\ref{tab:Cdd} summarizes the evaluation of $C_{dd}^{bl}$ for all the datasets based on the weighted results of the procedure for each data subset.
The larger $\phi_0\text{-}\delta$ correlation induced by the cooling wedge increases the beamline effect in Run-3a and Run-3b. 

\subsubsection{$p\text{-}x$ effect} \label{sec:bd:corr:Cpx}

At the exit of the inflector, the Muon Campus delivers a muon beam where the only sizable momentum-phase correlation is the one that is measured for the differential-decay beamline effect (i.e., $\partial \phi_0/\partial \delta$). This specific feature of the injected beam, which tracking simulations corroborate, is perturbed due to momentum-orbit correlations that develop during beam injection, where the radial and vertical phase-space coordinates $x,x',y\,\mathrm{and }\,y'$ are the ``orbit'' coordinates in this context (see Table~\ref{tab:table1}). 

The beam injection is optimized to accommodate the radial beam within the storage ring admittance. The process introduces correlations between the radial phase coordinates and momentum, $dx'/d\delta$ and $dx/d\delta$, of the stored muons at injection time ($t=0$). The resulting differential-decay contribution from injection is hence expressed as
\begin{equation}
\label{eq:Cdd_px}
C_{dd}^{p\text{-}x}=\frac{\sigma_{\delta}^2}{\omega_a\gamma_0\tau_{\mu}}\left(\frac{\partial \phi_0}{\partial x} \frac{dx}{d\delta}+\frac{\partial \phi_0}{\partial x^{\prime}} \frac{dx^{\prime}}{d\delta}\right).
\end{equation}

While the pion beam decays into muons as it is transported down the muon-production beamline, the angle $\phi$ between each muon's momentum in the lab frame and its spin direction depends on the parental pion momentum, $p_{\pi}$, as
\begin{equation}
\label{eq:phi_ppi}
\sin\left(\phi\right)\approx \frac{2m_{\mu}}{m_{\pi}^2-m_{\mu}^2}\frac{p_{\pi}}{c}\sin\theta,
\end{equation}
where $\theta$ is the angle between the muon momentum and the pion direction in the lab frame. In our case, as muons are  emitted in the lab frame in a forward cone of semi-angle $\theta_{max}\approx 12.7\,\mathrm{mrad}$, Eq.~\eqref{eq:phi_ppi} is further simplified to
\begin{equation}
\label{eq:phi_ppi_simplified}
\sin\left(\phi\right)\approx 78.8 x_0',
\end{equation}
where $x_0'$ is the phase-space coordinate of the muon's trajectory at birth. Therefore, a nonzero correlation $\partial \phi_0/\partial x_0'$ exists, which yields nonzero $\phi_0\text{-}x$ and $\phi_0\text{-}x'$ correlations in Eq.~\eqref{eq:Cdd_px} as muons subsequently execute betatron oscillations and cross bending magnets along the muon-production beamline. As shown in Eq.~\eqref{eq:Cdd_px}, these spin-orbit correlations couple with $dx/d\delta$ and $dx'/d\delta$ to alter the original phase-momentum relationship before injection.

\begin{figure}[htbp!]
\begin{center}
         \centering
         \includegraphics[width=\linewidth]{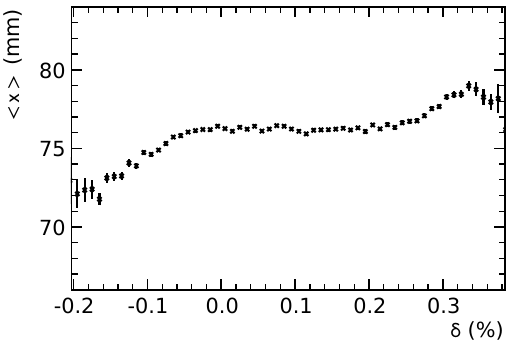}
\end{center}
\caption{Average radial coordinate $\langle x\rangle$ of the beam distribution per momentum offset at injection, from a \texttt{gm2ringsim} tracking simulation of stored muons. In this example, a nominal configuration of the injection parameters is implemented in the simulation. The $dx/d\delta$ correlations to quantify $C_{dd}^{p\text{-}x}$ are obtained from these tracking simulation results.}
\label{fig:DD_x_p}
\end{figure}

With beam tracking simulations using the \texttt{BMAD} and \texttt{gm2ringsim} injection models \cite{Run1PRAB}, we calculate the beam correlations necessary to determine the differential-decay $p\text{-}x$ effect. Figure~\ref{fig:DD_x_p} shows the radial coordinate versus fractional momentum of the stored muons at injection, which is the dominant momentum-orbit correlation in $C_{dd}^{p\text{-}x}$.
With Eq.~\eqref{eq:Cdd_px} and the simulation results, the $p\text{-}x$-effect contribution to the differential-decay correction for Runs-2/3 is
\begin{equation}
C_{dd}^{p\text{-}x}=-5\pm 6\,\mathrm{ppb}.
\end{equation}
The uncertainty accounts for several simulation configurations in view of injection parameter configurations within operational ranges (i.e., inflector current, beam distributions at the inflector exit, and injection kicker strengths, pulse shapes, and relative timings). 

\subsubsection{$p\text{-}t_0\,$\textit{effect}} \label{sec:bd:corr:Cpt}

\begin{figure}[htbp!]
\begin{center}
         \centering
         \includegraphics[width=\linewidth]{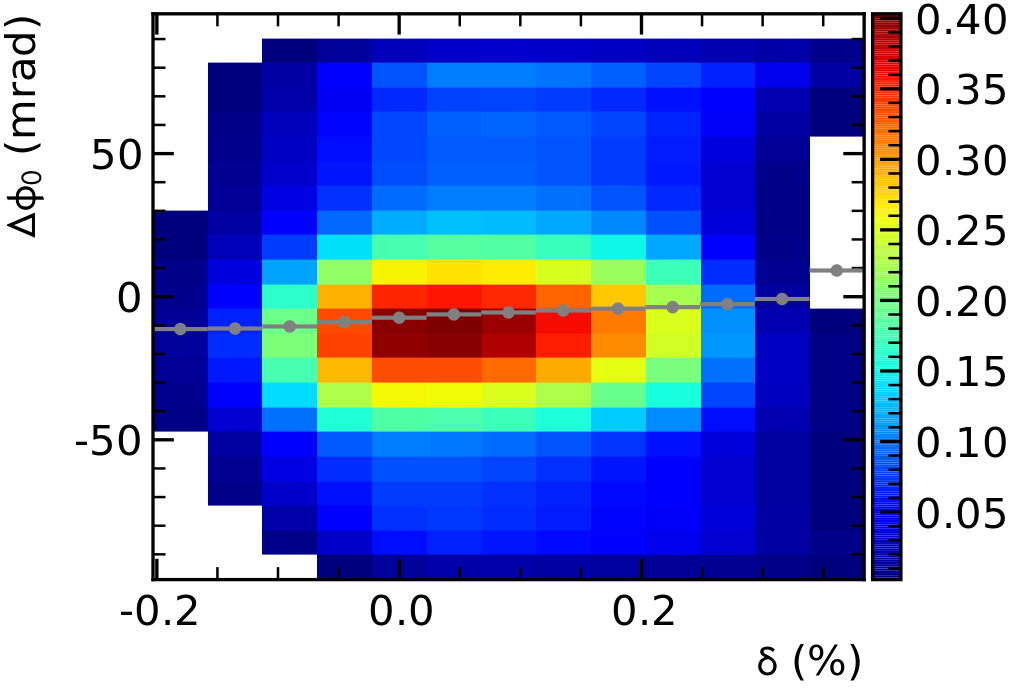}
\end{center}
\caption{Momentum-phase distribution from the momentum-time distribution for one bunch in data subset 2C. The gray markers are the averaged relative spin phases per fractional momentum, exhibiting the correlation that drives $C_{dd}^{p\mathrm{-}t_0}$.}
\label{fig:DD_dphi_dp}
\end{figure}

\begin{figure}[htbp!]
\begin{center}
         \centering
         \includegraphics[width=\linewidth]{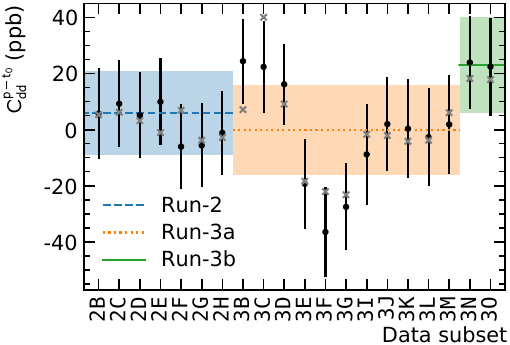}
\end{center}
\caption{Momentum-time differential decay correction $C_{dd}^{p\text{-}t_0}$ per data subset (black). In gray crosses, correction predictions where the ratio between $p\text{-}t_{0}$ correlations and kicker timing offsets relative to beam injection, based on \texttt{gm2ringsim} beam tracking simulations, is scaled in proportion to the per-data-subset kicker timing offsets.}
\label{fig:DD_pt_central}
\end{figure}

A muon's spin starts to precess as soon as it enters the storage ring. Typical muon bunches are $120\,\mathrm{ns}$ long; the spin of muons at the head of the bunch accumulates an additional precession $\Delta \phi\approx (120\,\mathrm{ns})\omega_a $ relative to muons at the tail while they enter the ring. This longitudinal phase variation across the bunch, together with the $t_0$-dependent momentum acceptance induced by the time dependence of the injection kicker, produce the momentum-time effect:
\begin{equation}
C_{dd}^{p\text{-}t_0}=\frac{1}{\omega_a} \frac{\partial \phi_0}{\partial t_0} \frac{d t_0}{d p} \frac{d p}{d t} \approx \frac{\sigma_{\delta}^2}{\gamma_0\tau_\mu}\frac{dt_0}{d\delta}.
\end{equation}
The method to evaluate $C_{dd}^{p\text{-}t_0}$ is similar to the procedure used for the differential-decay beamline effect explained in Sec.~\ref{sec:bd:corr:Cbl}, except for the first step where the muon distributions are prepared from the momentum-time distributions of the electric-field correction analysis; the time coordinates are transformed to relative spin phase advance via $\Delta\phi_0=\omega_at_0$ (Fig.~\ref{fig:DD_dphi_dp} shows one example).
The $C_{dd}^{p\text{-}t_0}$ is evaluated at the bunch level because each of the bunches in a sequence has characteristically different longitudinal intensity profiles. The results are then combined to obtain the corrections per data subset, as shown in Fig.~\ref{fig:DD_pt_central}.
The final momentum-time corrections per Run are summarized in Table~\ref{tab:Cdd}. 
The effect in Run-2 and Run-3a is consistent with zero, whereas a more constant timing offset between the kicker pulse and injection time leads to the non-zero correction for Run-3b.

To assess the uncertainties in this correction, we prepare 100 momentum-time distributions, each seeded by different initial conditions in the fitting method for the electric-field correction. The $C_{dd}^{p\text{-}t_0}$ correction is thereafter calculated for each seed, where the standard deviation for each set of bunches is treated as the uncertainty. The uncertainties per data subset are the correlated combination of the uncertainty from each bunch. An additional uncertainty, added in quadrature with the previously explained errors, is assigned from the RMS of all the mean-subtracted data subsets to account for the intrinsic ambiguity in the momentum-time distributions used to calculate the $p\text{-}t_0$ effect.

\subsubsection{Total effect}

The total differential decay correction is the combination of the beamline, $p\text{-}x$, and $p\text{-}t_0$ effects:
\begin{equation}
C_{dd}=C_{dd}^{bl}+C_{dd}^{p\text{-}x}+C_{dd}^{p\text{-}t_0},
\end{equation}
summarized in Table~\ref{tab:Cdd}.
To first order, these are uncorrelated; their physical origin is independent of each other. Therefore, the errors of each individual differential-decay effect are added in quadrature.

\begin{table}
\caption{\label{tab:Cdd} Differential decay corrections (ppb) for Run-2, Run-3a and Run-3b. The corresponding uncertainties (ppb) are enclosed in parentheses.}
\begin{ruledtabular}
\begin{tabular}{lcccc}
$C_{dd}$ & Run-2 & Run-3a & Run-3b & Section \\ \hline
Beamline & -12(3) & -17(3) & -20(3) & \ref{sec:bd:corr:Cbl} \\
$p\text{-}x$ & -5(6) & -5(6) & -5(6) & \ref{sec:bd:corr:Cpx} \\
$p\text{-}t_0$ & 6(15) & 0(16) & 23(17) & \ref{sec:bd:corr:Cpt} \\ \hline
Total & -11(16) & -22(17) & -2(18) & \ref{sec:bd:corr:Cdd} \\
\end{tabular}
\end{ruledtabular}
\end{table}

\subsection{Phase acceptance correction\label{sec:bd:corr:Cpa}}

The detected \gm phase, as measured by the calorimeter detectors, varies over time as a function of the transverse beam coordinates of the muons $(x,y)$. The beam transverse distribution changes with time and creates in-fill variations of the detected phase that could affect the fit model for $\omega_a^m$, where the phase is expected to be time-independent. For this detector-acceptance effect, we introduce the phase acceptance correction, $C_{pa}$.

\begin{figure}[htbp!]
\centering
\includegraphics[width=0.45\textwidth]{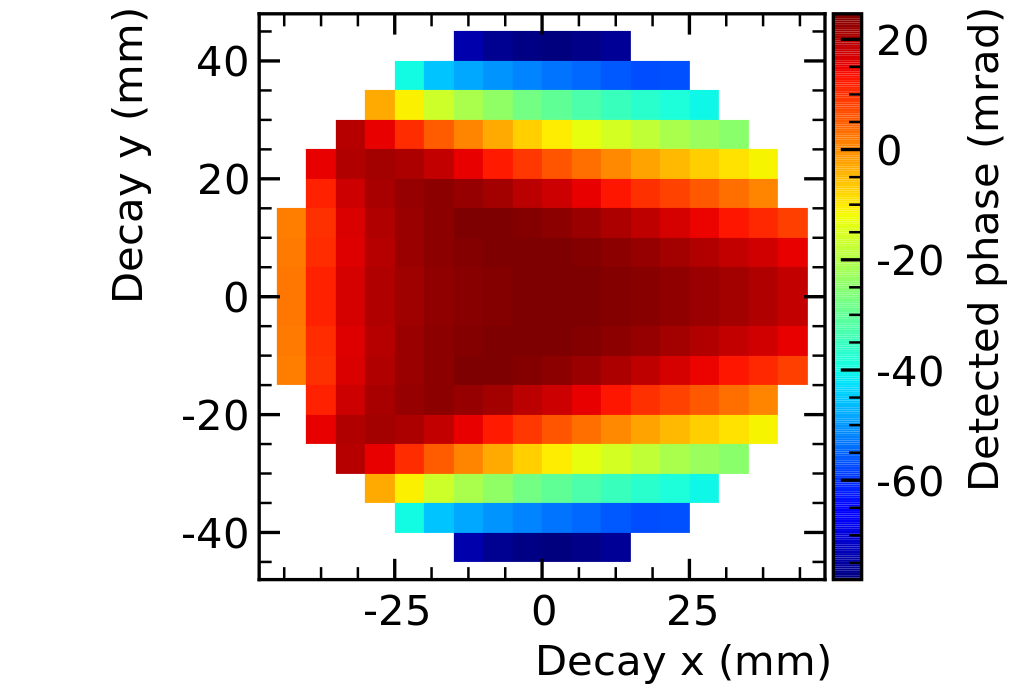}
\caption{Simulated azimuthally averaged phase maps for the asymmetry-weighted analysis. The coupling between the overall quadratic-like detected phase acceptance in the vertical direction and the in-fill reduction in vertical beam width is the most significant effect on $C_{pa}$.}
\label{fig:phase_map}
\end{figure}

The time-dependent phase $\phi_{\mathrm{pa}}(t)$ is computed by averaging the measured phase as a function of transverse coordinates ($x$,$y$) that are obtained from \texttt{gm2ringsim}. 
The time dependence of the transverse beam coordinates is extracted from tracker beam profiles $M^T(x,y,t)$, which generates a time-dependent phase by virtue of the correlation between the phase and the beam transverse distribution.
Figure~\ref{fig:phase_map} is a transverse map of $\phi_{\mathrm{pa}}(x,y)$ averaged over the azimuth, obtained by fitting the asymmetry-weighted histogram used to extract $\omega_a^m$ (see Sec.~\ref{ss:overviewomegaa}).

The tracker stations measure the $M^T(x,y,t)$ distribution at two locations around the ring, but the extraction of the measured $\omega_a^m$ is performed by calorimeters at 24 azimuthal locations. Therefore, we extrapolate the $M^T(x,y,t)$ profiles around the ring using \texttt{gm2ringsim} and \texttt{COSY INFINITY} beam dynamics simulations. 
Vertical ($y(\varphi,t)$) and radial ($x(\varphi,t)$) muon coordinates at any given azimuthal position $\varphi$ are calculated by scaling the transverse coordinates from tracker measurements with the mean and width values from simulated beam distributions as
\begin{equation}\label{eq:pa:yrms}
    y(\varphi, t)=y_{\mathrm{trk}}(t)\frac{y^{\mathrm{rms}}(\varphi,t)}{y_{\mathrm{trk}}^{\mathrm{rms}}(t)},
\end{equation}
for the vertical width, and
\begin{equation}\label{eq:pa:xrms}
\begin{aligned}
    x(\varphi, t) = & \frac{x^{\mathrm{rms}}(\varphi, t)}{x^{\mathrm{rms}}_{\mathrm{trk}}( t)}\cdot [x_{trk}(t) - \bar{x}_{\mathrm{trk}}(t)]+\bar{x}(\varphi, t),
\end{aligned}
\end{equation}
for the radial motion of the beam, where $(x^{\mathrm{rms}},y^{\mathrm{rms}})$ are the root mean squares of the transverse beam distributions and $\bar{x}$ is the radial distribution average. The quantities from simulated distributions on the right-hand side in Eq.~\eqref{eq:pa:xrms} and Eq.~\eqref{eq:pa:yrms} do not have subscripts, whereas tracker-based values are denoted with the subscript ``trk.’’ By modifying the $M^T(x,y,t)$ distribution using Eq.~\eqref{eq:pa:xrms} and Eq.~\eqref{eq:pa:yrms}, we obtain the spatial and time distribution of the muons $M^c(x,y,t)$ at each calorimeter location. Combining the simulated maps with the muon distributions, a time-dependent phase $\phi^c_{pa}(t)$ can be computed for each calorimeter using the following weighted sum:

\begin{equation}
\begin{aligned}
\phi_{\mathrm{pa}}^c(t) = \mathrm{arctan} & \left[ \frac{\sum_{ij}M^c(x_i,y_j,t)\cdot \varepsilon^c(x_i,y_j)}{\sum_{ij}M^c(x_i,y_j,t)\cdot \varepsilon^c(x_i,y_j)} \right. \\
& \left. \frac{\cdot A^c(x_i,y_j)\cdot \sin[\phi^c_{\mathrm{pa}}(x_i,y_j)]}{\cdot A^c(x_i,y_j)\cdot \cos[\phi^c_{\mathrm{pa}}(x_i,y_j)]} \right],
\end{aligned}
\label{eq:master_formula}
\end{equation}
where acceptance, asymmetry and phase maps for a calorimeter ``$c$'' are represented by $\varepsilon^c$, $A^c$ and $\phi_{\mathrm{pa}}$, respectively.

The calculation of the phase acceptance correction is done by comparing  $\omega_a^m$ to the fit of the simulated data. A histogram is generated for each calorimeter and for each parameter of the $\omega_a^m$ fit, including the modified ${g-2}$ phase obtained by fitting $\phi_{\mathrm{pa}}^c(t)$.
Simulated data (produced using values extracted from histograms) are fitted with a constant phase. The difference between $\omega_a^m$ and the fit result determines $C_{pa}$ for a given calorimeter.

\begin{figure}[htbp!]
\centering
\includegraphics[width=\linewidth]{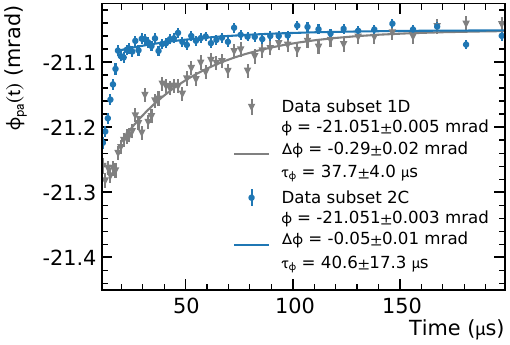}
\caption{Calculation of $\phi_{\mathrm{pa}}$ for calorimeter 13 in data subset 1D (gray) and data subset 2C (blue) using data from the tracker station at 180$^{\circ}$. The shown fit function is of the form $\phi + \Delta\phi \cdot e^{{(-t/\tau_{\phi})}}$.}
\label{fig:exponential}
\end{figure}

Figure \ref{fig:exponential} shows the $\phi_{\mathrm{pa}}$ time evolution for a Run-2 data subset, superimposed with one from Run-1d for comparison. After replacing the damaged resistors of the ESQ system from Run-1, the variation of the phase is highly reduced during Run-2/3, and the $C_{\mathrm{pa}}$ is hence smaller. The central values of the correction are calculated by taking the average of the results from all calorimeters. The central values are shown in Table~\ref{tab:final_value}, where further improvement on the effect is observable in Run-3  with respect to Run-2. This outcome is due to the improved stability of the beam motion thanks to more optimized kicker settings and a better temperature stability of the main magnet. The evaluations of the statistical and systematic uncertainties are also reported in Table~\ref{tab:final_value}. The statistical uncertainty, which ranges from 2.0 to \SI{7.8}{ppb}, originates from the limited number of tracks from the $M^T(x,y,t)$ collected by tracker stations. The sources of systematic uncertainty can be divided into three main groups. The first one stems from imperfect knowledge of the straw trackers’ alignment, resolution, and acceptance, which directly affects the measured distribution $M^T(x,y,t)$. Next are the uncertainties associated with the estimation of the phase, asymmetry, and acceptance maps in Eq.~\eqref{eq:master_formula} estimated using \texttt{gm2ringsim}. Lastly, the calculation utilizes beam dynamics functions obtained by simulation to extract the calorimeter $M^c(x,y,t)$ distribution from the tracker-based $M^T(x,y,t)$. Uncertainties are estimated by calculating $C_{pa}$ while varying the beta functions and magnetic field within expected deviations based on the measurements.
\begin{table}[!htbp]
\centering
\caption{Values of the phase-acceptance correction $C_{\mathrm{pa}}$ (ppb) and their statistical, systematic, and total uncertainties (ppb) for each of the Run-2/3 datasets.}\label{tab:final_value}
\begin{ruledtabular}
\begin{tabular}{lccc}
Quantity           & Run-2 & Run-3a & Run-3b \\ \hline
\textit{Correction} & -50   & -16    & -13    \\ \hline
\textit{Statistical Unc.} & 9   & 2    & 3   \\ 
\textit{Systematic Unc.} & & & \\
\, Tracker and CBO   &   13    &   8    &  7      \\
\, Phase maps       &   13    &    3    &    3    \\
\, Beam dynamics    &   5    &   3     &   2     \\ 
\hline
Total uncertainty & 21    & 9      & 8      \\ 

\end{tabular}

\end{ruledtabular}
\end{table}

\subsection{Summary}

The beam dynamics corrections and their uncertainties for \RunTwoThree are listed in Table~\ref{tab:C_BD}. 
\begin{table}
\caption{\label{tab:C_BD} Values and uncertainties of the beam dynamics corrections (ppb) for \RunTwoThree.}
\begin{ruledtabular}
\begin{tabular}{lcc}
Quantity & Correction & Uncertainty \\ \hline
$C_e$ & 451 & 32 \\
$C_p$ & 170 & 10 \\
$C_{ml}$ & 0 & 3 \\
$C_{dd}$ & -15 & 17 \\
$C_{pa}$ & -27 & 13  \\
\hline
Total & 580 & 40 \\
\end{tabular}
\end{ruledtabular}
\end{table}

Each individual correction is highly correlated for different datasets, and therefore, the per-dataset combination of the uncertainties is fully correlated. To obtain the total beam dynamics correction uncertainty, we add the uncertainties of all the individual corrections in quadrature because they are uncorrelated.

A combination of improvements in the experimental setup (listed in Sec.~\ref{sec:conditions}) and analysis reduced both the beam dynamics correction magnitudes and uncertainties in \RunTwoThree compared to \RunOne. The replacement of the ESQ high-voltage resistors damaged in \RunOne leads to a smaller and more precise determination of $C_{pa}$. The muon loss correction is negligible thanks to the significantly reduced mechanical muon loss rates. With the stronger injection kickers in \RunThreeB, the more symmetric momentum distribution requires a lower electric-field correction, whereas the determination of the momentum-time beam correlations at injection, as well as an independent reconstruction of the momentum distribution based on the tracker detector data, reduce the uncertainty of $C_e$. While the differential decay correction was not included in \RunOne, the momentum-time correlations analysis for the electric-field correction allowed us to fully quantify this correction in \RunTwoThree.

\section{Magnetic field measurement}
\label{sec:field}

In Eq.~\eqref{eq:omega_pB}, $\tilde B$, the magnetic field averaged over space and time by the muons, is expressed as the precession frequency of protons in a spherical water sample at a reference temperature: $\opprimetilde(T_r)$.
In this notation, the tilde indicates the muon weighting, and the prime indicates that the proton magnetic moment is shielded in H$_2$O. The reference temperature is $T_r=\SI{34.7}{\celsius}$, the temperature at which the shielded proton magnetic moment was measured relative to the bound-state electron in hydrogen~\cite{phillips_magnetic_1977}. This section describes the measurements and analyses leading to $\tilde\omega_p^\prime$, which follows from the general approach of \RunOne~\cite{Run1PRAField}.

\subsection{Magnetic field measurement principle\label{sec:field:principle}}

The muon-weighted magnetic field is derived from 
time-dependent maps of the magnetic field in the muon storage region $\omega_p^\prime(x,y,\phi,t)$. 
The maps are derived from measurements by a set of NMR probes in a trolley that is pulled through the storage ring every two to three days and maps the full circumference in about 70 minutes. The field is mapped at the 17 NMR-probe positions  ($x$, $y$)  ($x=0$ at $r=R_0$) and about 9000 azimuthal positions $\phi$. 
Corrections for differences of the physical ring configuration and from magnetic field transients from the kickers and ESQs, which are not operating during the trolley measurements, are discussed in section~\ref{sec:field:transients}.

The trolley's NMR probes, described in~\cite{Run1PRAField}, contain samples of proton-rich petroleum jelly (petrolatum). 
The  trolley probes are calibrated to account for the sample and the different magnetic environment 
due to magnetic perturbations from the aluminum shell, the wheels of the trolley,  the other probes, and other trolley components, including the electronics, cables, etc. 
A dedicated calibration magnetometer was used to correct each probe to the frequency that would be measured with a spherical water sample at temperature $T_r$. 
The details of this calibration procedure are described in sections~\ref{sec:field:absoluteCalibration} and~\ref{sec:field:trolleyCalibration}.

The time-dependent trolley maps are parameterized  as 
\begin{equation}
    \omega_p^\prime(x,y,\phi,t) =
    \sum_{i=1}^{N_\mathrm{max}} m_i(\phi,t) f_i(r,\theta)
    \, ,    \label{eq:field:MomentExpansion}
\end{equation}
where
\begin{equation}
    f_i(r,\theta) = \left\{
    \begin{array}{ll}
    1 & \text{for } i=1, \\
    \left(\frac{r}{r_0}\right)^{\frac{i}{2}}\cos\left(\frac{i}{2}\theta\right) & \text{for even } i > 1, \\
    \left(\frac{r}{r_0}\right)^{\frac{i-1}{2}}\sin\left(\frac{i-1}{2}\theta\right) & \text{for odd } i > 1. \\
    \end{array}
    \right.
    \label{eq:MomentDefs}
\end{equation}

Here $r_0=\SI{4.5}{cm}$ is a reference radius, $x=r\cos(\theta)$, $y=r\sin{(\theta)}$. The $\cos(\theta)$ and $\sin(\theta)$ terms are referred to as normal and skew moments, and $t$ is the time of the measurement.
The moments $m_i(\phi,t)$ are determined from fits of the 17 trolley-probe frequencies at the time $t$ when the trolley is at the position $\phi$. 
The parameterization in Eq.~\eqref{eq:field:MomentExpansion} is motivated by solutions to a 2-D Laplace equation and is analogous to a 2-D Taylor expansion around $(x,y)=(0,0)$ with constraints. The 2-D Laplace-equation solution is strictly valid only if $B$ has no azimuthal dependence; the impact and validation of this parameterization and the effect of truncating the parameterization at $N_\mathrm{max}$
are discussed in Sec.~\ref{sc:2Dvs3D}. 

The time-dependence of the moments $m_n(\phi,t)$ between trolley runs is estimated by interpolation 
making use of a set of 378 NMR magnetometers (fixed probes) mounted on the outside of the vacuum chambers
at 72 azimuthal positions, called stations.  Each fixed probe is read out with a rate of ${\sim}\SI{0.5}{Hz}$. 
Each station has either four or six NMR probes, half above and half below the storage region, and can interpolate the magnetic field moments up to $i=4$ or $i=5$, respectively. 
As a trolley run proceeds, the moments calculated from the fixed probes at the stations near the trolley are set equal to the corresponding moments calculated from the trolley probes at that time, which we call ``tying''. Moments up to $n=4,5$ are tracked with the fixed probes by interpolating in time between two trolley runs, and higher-order moments are interpolated assuming linear time dependence. The limitation of this interpolation results in ``tracking errors'' that are estimated from the difference between the moments predicted by the fixed probes and the moments actually measured by the subsequent trolley run. Studies with different intervals between trolley runs and at different times after the magnet was ramped to the nominal operating field were used to reduce the tracking errors and uncertainties.

The muon-weighted field is
\begin{equation} 
\tilde \omega_p^\prime = \frac{\int \omega_p^\prime(x,y,\phi,t)M(x,y,\phi,t)\, \mathrm{d}x\, \mathrm{d}y\, \mathrm{d}\phi\, \mathrm{d}t}{\int M(x,y,\phi,t)\, \mathrm{d}x\, \mathrm{d}y\, \mathrm{d}\phi\, \mathrm{d}t},
\label{eq:muonweightedfield2}
\end{equation}
with the muon distribution $M(x,y,\phi,t)$ determined by a combination of measurements with the trackers and modeling of beam dynamics (Sec.~\ref{sec:bd:muonDistribution}).
Expanding $M(x,y,\phi,t)$
in the basis introduced in Eq.~\eqref{eq:field:MomentExpansion}, 
the muon weighted azimuth- and time-dependent magnetic field is
\begin{equation}
    \tilde\omega_p^\prime(\phi,t) = \sum_{i} m_i(\phi,t)k_i(\phi,t)\,,
\label{eq:field:time_azimuth_dependent_muon_weighted_field}
\end{equation}
where 
\begin{equation}
    \label{eq:multipole_projections}
    k_i(\phi,t) = \frac{ \int_{} M(x,y,\phi,t) f_i(x,y) dx\, dy }{\int_{} M(x,y,\phi,t) dx\, dy}\,.
\end{equation}
The time-dependent azimuthally averaged field is
\begin{equation}
    \Tilde{\omega}_p^\prime(t) = \frac{1}{2\pi}\int_{0}^{2\pi} \Tilde{\omega}_p^\prime(\phi, t) \,\mathrm{d}\phi, 
    \label{eq:field:time_dependent_muon_weighted_field}
\end{equation}
which is weighted by the number of detected muon decays and  time averaged over few day intervals.

\subsection{Absolute calibration with a high-purity water probe\label{sec:field:absoluteCalibration}}
\label{sec:CalibrationProbe}\label{sec:field:absoluteCalibration:H2O}

Each trolley probe reading is corrected for the field perturbations caused by the trolley components to the NMR frequency expected from a bare spherical water sample at \SI{34.7}{\celsius}. This is done using an H$_2$O absolute calibration probe installed in the \gm storage ring.
The calibration probe for \RunTwo and \RunThree was similar to that described in detail in~\cite{FlayPP,Run1PRAField}.

Corrections must be applied to the measured calibration probe NMR frequencies to those expected from a bare spherical water sample at $T_r$. 
Corrections to the measured calibration frequency are listed in Table~\ref{tb:PP2Corrections} and described below. These corrections were cross-checked with respect to a $^3$He magnetometer in a dedicated high uniform \SI{1.45}{T} solenoid and with simulations. 
All corrections are expressed as fractions of the measured NMR frequency, i.e., $\omega^\text{corr}=\omega^\text{meas}(1+\delta)$, where $\omega^\text{corr}$ is the frequency corrected for the effect $\delta$. For corrections $\ll$1 (the largest is \SI{1.5}{ppm}), the combination of two corrections is $(1+\delta_a)(1+\delta_b)\approx (1+\delta_a+\delta_b+\mathcal{O}(\delta^2))$; only the first-order corrections are applied.

\noindent
{\bf Sample-shape correction $\mathbf{\delta^b}$} The calibration probe consists of a cylindrical sample filled with high-purity water. The temperature-dependent correction to a spherical sample is 
    \begin{equation}
        \delta^b(T_n)=\chi(T_n)(\epsilon-1/3),
    \end{equation}
    where $\chi(T_n)$ is the susceptibility at the temperature of the calibration probe for calibration of probe $n$, and 
    $\epsilon=0.4999(0,-0.0003)$
    for the finite cylindrical sample, which was calculated in closed form from~\cite{HoffmanShapeCorrection} and confirmed by numerical simulation ($\epsilon=1/2$ for an infinite cylinder). 
    
    The temperature-dependent volume susceptibility is 
    \begin{equation}
        \chi_{V}(T) = \chi_{V}(\SI{22}{\celsius}) \times \left[ \frac{\chi_{m}(T)}{\chi_{m}(\SI{22}{\celsius})} \right] \times \left[ \frac{\rho(T)}{\rho(\SI{22}{\celsius})} \right],
    \end{equation}
    where $\chi_{V}(\SI{22}{\celsius})=-9.056\times  10^{-6}$ is the value recommended by CODATA~\cite{CODATA1998} with $3\times 10^{-8}$ uncertainty due to additional measurements at unspecified temperatures~\cite{Blott1993}.
We use the ratio of mass susceptibilities from~\cite{philo1980}:
\begin{eqnarray}
    \frac{\chi_{m}(T)}{\chi_{m}(\SI{22}{\celsius})} &=& \frac{\chi_{m}(T)}{\chi_{m}(\SI{20}{\celsius})} \frac{\chi_{m}(\SI{20}{\celsius})}{\chi_{m}(\SI{22}{\celsius})}\nonumber \\
    \approx 1 &+& 1.3881 \times (T-\SI{22}{\celsius})\frac{10^{-4}}{\SI{}{\celsius}} \nonumber\\
    &+& \mathcal{O}\left(\left((T-\SI{20}{\celsius})\frac{10^{-4}}{\SI{}{\celsius}}\right)^2\right). 
\end{eqnarray}
The temperature-dependent density $\rho(T)$ from \cite{Kell1967} 
was used, because that is what was used in the analysis by~\cite{philo1980}.

\noindent
{\bf Material effects $\mathbf{{\delta^s}}$} The calibration probe consists of the sample contained in a glass cylinder NMR sample tube, a concentric glass cylinder holding the NMR coil wires, a concentric aluminum cylinder shell, end caps, the temperature sensor, tuning capacitors, connectors, and mounting fixtures. 

Due to their finite magnetic susceptibility, each of these components becomes magnetized by the external \SI{1.45}{T} field, and the resulting magnetization contributes to the field measured by the probe. The contribution depends on the orientation (roll and pitch) of the probe with respect to the vertical magnetic field. The approximate cylindrical symmetry of the probe construction mitigates these effects, and a combination of direct measurements of intrinsic-probe effects $\delta^s$, and simulations specific to the configuration in the \gm storage ring are used to determine the remaining material corrections.
Additionally, the high-permeability pole pieces of the storage-ring magnet act as magnetic mirrors that create images of the magnetized calibration-probe components, leading to a correction $\delta^{s,\text{img}}$ that depends on the probe position. 

\noindent
{\bf Sample (im)purity $\mathbf{\delta^P}$ }
Potential impurities, in particular, dissolved paramagnetic O$_2$ and salts, in the water sample could lead to a shift of the NMR frequency.
Degassed ultra-pure (ASTM Type-1) water from several vendors was used, with no observed variation within an uncertainty of \SI{2}{ppb}. A variety of additional tests were performed in which the glass water sample tube was rotated, and different sample tubes were used. No systematic shifts were observed.

\noindent
{\bf Magnetization dependent effects $\mathbf{\delta^{RD}}$ and $\mathbf{\delta^d}$}
The sample magnetization $\vec M=\chi_{\rm H_2O}\vec B$ can lead to two shifts. Radiation damping is the result of the oscillating current in the NMR coil that rotates the magnetization toward the external magnetic field. This leads to a time-dependent precession frequency shift $\delta^{RD}$ that depends on the magnetization along the magnetic field, the detuning of the NMR coil, and the coupling between the coil and the precessing spins (filling factor) \cite{Vlassenbroek1995RadiationDI}.
A second, shape-dependent frequency shift is caused by the dipolar field from the precessing protons, $\delta^d$. Both effects are estimated as in \RunOne~\cite{Run1PRAField}.

\noindent
{\bf Calibration probe temperature dependence $\mathbf{\delta^T}$}
The gyromagnetic ratio of protons diamagnetically shielded in a spherical sample of water was measured at \SI{34.7}{\celsius}~ ~\cite{phillips_magnetic_1977}. This diamagnetic shielding is temperature-dependent~\cite{PetleyH2Otemp}. The correction from $T^\text{cp}_n$, the calibration-probe temperature for calibration of trolley probe $n$, to $T_r$, is $\delta_n^T=(-10.36\pm 0.30)\times \frac{10^{-9}}{\SI{}{\celsius}} (T_r-T^\text{cp}_n)$. The calibration probe temperature was measured with a platinum resistive temperature device (PT1000 RTD) with an accuracy of \SI{0.5}{\celsius}, and a different correction per probe was applied to account for the calibration-probe and trolley temperature during the calibration of each probe as discussed in the next section. 

\noindent
{\bf Corrections dependent on the calibration-probe environment}
As noted in the discussion of material effects, the magnetized components of the calibration probe contribute to the measured magnitude of the magnetic field that depends on the orientation with respect to $\vec B$ and due to magnetic images. Additional corrections for the calibration configuration vary with the individual trolley probe being calibrated and are discussed in Sec.~\ref{sec:field:trolleyCalibration}.

\subsubsection*{Calibration-probe cross checks}\label{sec:field:absoluteCalibration:3He}

Work is underway to cross-check the intrinsic corrections applied to the calibration probe, i.e., corrections not dependent on the environment ($\delta^b$, $\delta^s$, $\delta^{RD}$, $\delta^d$, and $\delta^P$), using $^3$He magnetometry and a separate H$_2$O probe based on continuous wave (CW) NMR.
The Mark-I $^3$He absolute magnetometer provided an indirect \SI{42}{ppb} cross-check on the calibration probe~\cite{Farooq,FlayPP,Run1PRAField}. A Mark-II $^3$He probe was designed and constructed with much smaller intrinsic corrections, and a campaign is underway to directly calibrate the muon $g-2$ calibration probes for \RunOne and Runs 3-6. Preliminary analysis confirms agreement with uncertainties less than \SI{20}{ppb}.
The calibration probes were also compared to the CW H$_2$O NMR probe under development for JPARC's MuSEUM and g-2/EDM (E34) experiments~\cite{yamaguchi_development_2019}. Cross-checks with earlier CW prototypes at \SI{1.4}{T} and \SI{1.7}{T} showed a tension on the $\sim$\SI{50}{ppb} level with a precision around \SI{15}{ppb}. The same cross-check, with newer probe versions, performed at \SI{3}{T}, is in good agreement with an uncertainty of \SI{10}{ppb}. The discrepancy with the earlier version is not yet understood; additional work is ongoing.

\begin{table}[ht]
    \centering
    \caption{Calibration probe intrinsic corrections and uncertainties. Shape corrections are temperature dependent and hence different for each trolley probe. Thus, the range of all probes is given.}
    \label{tb:PP2Corrections}
    \begin{ruledtabular}
    \begin{tabular}{llcc}
        Description & & Corr. (ppb) & Unc. (ppb) \\
        \hline
        Shape, susceptibility  & $\delta^{b}$ &  -1508.7 to -1507.4 & 6.0 \\
        Material effects & $\delta^{s}$ & 10.3& 5.0 \\
        Radiation damping & $\delta^{RD}$ & 0 & 3.0\\
        Proton dipolar field & $\delta^{d}$ &0 & 2.5 \\
        Sample purity & $\delta^{P}$ & 0& 2.0 \\
        \hline
        Subtotal &  & & 8.9 \\
    \end{tabular}
    \end{ruledtabular}
\end{table}

\subsection{Trolley-probe calibration \label{sec:field:trolleyCalibration}}

Trolley-probe calibration provides a set of corrections to the frequencies $\omega_n^{tr}$ measured by each trolley probe 
\begin{equation} \label{eq:field:calib}
    \omega^\prime_n  = \omega_n^\text{tr}(1+\delta_n^{calib}),
\end{equation}
where $\omega_n^\prime$ is the field that would be measured by a spherical water sample at $T_r=\SI{34.7}{\celsius}$ at the position of probe $n$. Corrections for the temperature dependence of the vaseline-filled trolley probes are discussed in Sec.~\ref{sec:field:trolleyMaps}. 

Calibration campaigns before the start of \RunTwo and after \RunThree took place in vacuum in a dedicated region of the storage ring magnet using the calibration probe described in Sec.~\ref{sec:CalibrationProbe}. Magnetic field gradients applied in all three directions were used to place the effective volumes of the calibration probe and each trolley probe within \SI{0.5}{mm} of the same position, and the magnetic field in the calibration region was carefully mapped and shimmed. 

The calibration correction was determined from a sequence of measurements swapping the trolley and calibration probe into the calibration position. During this swapping, the magnetic field was tracked with fixed probes to mitigate the effect of drifts. 
Additionally, the \RunTwoThree calibration campaigns 
and the \RunOne calibration campaign provided data on the stability of 
the trolley-probe calibrations over a three-year period. 

Uncertainties from the calibration procedure are listed in Table~\ref{tb:AbsCorrections}. These include 
uncertainties due to mis-alignment of the calibration probe and trolley probe, temperature corrections of the diamagnetic shielding $\delta^{T}$, the variance between the calibration constants of different measurement campaigns and analyzer $\delta^{var}$, the difference between the active volume of the calibration probe and trolley probe $\delta^{av}$, the influence of the trolley and calibration probe's materials on the
the magnetic environment of the other, called magnetic footprint $\delta^{fp}$ and $\delta^{cp}$ , the frequency extraction $\delta^{f}$ and the material effects including the magnetic image in the pole pieces $\delta^{img}$.
The per-probe calibration constants with a graphical representation is given in Table~\ref{tb:PPCorrections} and Fig.~\ref{fig:calibrationConstants}, in the appendix.

\begin{table}[H]
    \centering
    \caption{Uncertainties from the calibration procedure on the muon-weighted field. The uncertainties for the individual probes are shown in Table~\ref{tb:PPCorrections}. The probe individual corrections due to temperature dependence of the diamagnetic shielding range from \SI{-126.3}{ppb} to \SI{-59.1}{ppb}.}
    \label{tb:AbsCorrections}
    \begin{ruledtabular}
    \begin{tabular}{ll|c}
    Description & &  Uncertainty [ppb] \\
    \hline
    Swapping and misalignment & $\delta^{tr}$ & 1.6 \\
    Temperature of diamag. shielding & $\delta^{T}$ & 5.2 \\
    Variance & $\delta^{var}$ & 11.0 \\
    Active volume & $\delta^{av}$ & 1.7 \\
    Footprint trolley & $\delta^{fp}$ & 8.0 \\
    Footprint CP & $\delta^{cp}$ & 4.0 \\
    Frequency extraction CP & $\delta^{freq (cp)}$ & 1.0\\
    Material and mag. image & $\delta^{img}$ & 9.0 \\
    \hline
    Subtotal & & 17.8
    \end{tabular}
    \end{ruledtabular}
\end{table}

\subsection{Magnetic field maps\label{sec:field:trolleyMaps}}

In this section, we describe the detailed extraction of the field maps  $\omega_p^{\prime\, \text{tr}}(x,y,\phi, t)$ (Eq.~\eqref{eq:field:MomentExpansion}). The transverse positions are fixed by the probe locations, while the trolley position is radially constrained by the trolley rails. The trolley azimuthal position is determined by reading the barcodes etched into the bottom of the vacuum chambers.
Encoders that measure the length of the trolley cables are a backup, however, the encoder precision is inferior compared to the barcode due to tension variations in the cables. 
The 17 trolley NMR probes are triggered in sequence every ${\sim}\SI{30}{\milli\second}$, resulting in a ${\sim}\SI{2}{Hz}$ sampling rate for each probe. The corrected frequencies  are interpolated to a grid of azimuthal positions  $\phi_k(t)$.
Different interpolation schemes were tested and agreed within \SI{1}{ppb}.

\begin{figure*}
    \centering
    \includegraphics[width=\linewidth]{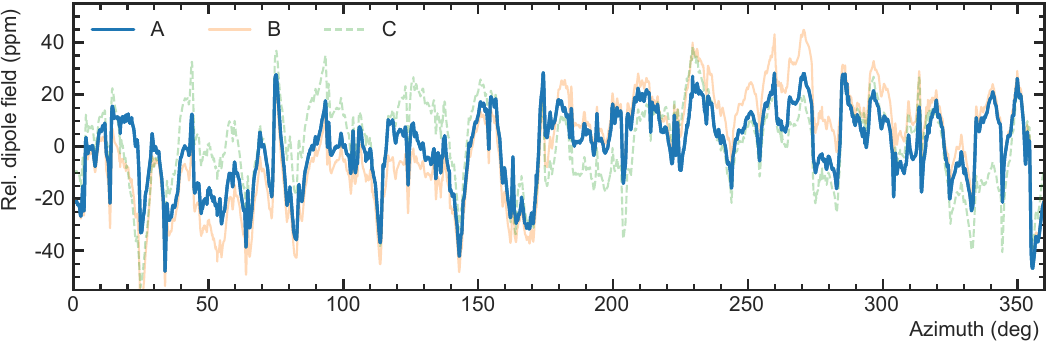}
    \caption{The relative (Rel.) dipole $m_1$ coefficient as a function of azimuth for three field maps with respect to its azimuthal average. A) is from April \nth{8} 2019, the beginning of \RunTwo, B) is from June \nth{20}, 2019, the end of \RunTwo, and C) from March \nth{11}, 2020, the end of \RunThree. The peak-to-peak amplitudes are \SI{76}{ppm}, \SI{108}{ppm}, and \SI{93}{ppm}, respectively, with RMSs of \SI{14.6}{ppm}, \SI{20.5}{ppm}, and \SI{15.8}{ppm}.}
    \label{fig:field:fieldMap}
\end{figure*}

The multipole coefficients $m_i(\phi_k(t_\text{tr}))$ 
are determined 
for each $\phi_k$ by fitting the corrected frequencies to  Eq.~\ref{eq:MomentDefs}, where $t_{\text{tr}}$ is the time when the trolley is at $\phi_k$.
A lower bound on $N_\mathrm{max}$ is derived from azimuthal averaged fit residuals, which show a transverse dependence if $N_\mathrm{max}$ is chosen too small. 
An upper bound comes from degeneracies of the multipoles with our trolley probe configuration. 
The truncation at $N_\mathrm{max}=12$ of the parametrization in  Eq.~\eqref{eq:field:time_azimuth_dependent_muon_weighted_field} is used.
The difference between using different minimization algorithms to extract the multipole coefficients is negligible.
Representative field maps $m_1(\phi)$ for three different trolley runs are shown in Fig.~\ref{fig:field:fieldMap}.

Corrections and uncertainties to the trolley multipole coefficients are presented in Table~\ref{tb:field:maps} and summarized here.

\textbf{Trolley motion effects ($\mathbf{\delta^{motion}}$):} The trolley motion in a nonuniform magnetic field generates eddy currents in the conducting components, most significantly the aluminum shell. We use the \RunOne correction for $\delta^\text{motion}=\SI[separate-uncertainty = true]{-15(18)}{ppb}$ from \RunOne analysis~\cite{Run1PRAField} estimated from the comparison of standard continuous motion trolley runs with stop-and-go runs and from the comparison for clockwise and counter-clockwise trolley runs. 

\textbf{Difference in configuration ($\mathbf{\delta^{config}}$):} During the trolley runs, the collimators that radially constrain the stored-muon distribution are retracted, and the trolley rails are in a different position than when the muons are stored. 
The effect of these two configuration changes is estimated from calculations of the magnetic field produced by the diamagnetic copper and paramagnetic aluminum in the respective configurations. The uncertainty of the \RunOne correction of $\delta^\text{config}=\SI[separate-uncertainty = true]{-7(22)}{ppb}$~\footnote{Note that in Ref.~\cite{Run1PRAField}, \SI{-5}{ppb} is used for the central value of the effect from the garage alone.} is dominated by a discrepancy in the calculation and what a local fixed probe measures. The same value is used for \RunTwoThree. 
The effect from the collimators on the azimuthally  averaged field is smaller than \SI{1}{ppb}.

\textbf{Trolley frequency extraction ($\mathbf{\delta^{freq}}$):} Trolley NMR-probe FID analysis is described in~\cite{field_FID_frequency_extraction}.  
Briefly, the phase function (phase vs time) for the free-induction-decay (FID) signals is extracted from in-phase and  Hilbert-transform quadrature signals. The phase functions are fit to polynomials of varying order from two to six and for a varying time ranging from 0.20 to 0.75 of $T_2^*$ (the FIDs are not exponential, so in this case, we refer to the time for the FID amplitude to reach $1/e$ of the maximum).
The frequency-extraction correction $\mathbf{\delta^{freq}}$  on $m_1$ is below \SI{12}{ppb}. Potential effects from incorrect $t_\text{FID}=0$ on the $\SI{100}{\micro\second}$ level are shown to be negligible. 
Temperature changes affect the phase function of FIDs. This effect on the extracted precession frequencies is included in the correction below.
The uncertainty due to correcting from the $\approx\SI{25}{^\circ C}$ trolley temperature during field mapping  to around $\approx\SI{33}{^\circ C}$ temperature during calibration is \SI{5}{ppb}.

The total uncertainty from the frequency extraction, taking the \RunTwoThree beam shapes and correlations between the multipoles into account, is shown in Table~\ref{tb:field:maps}. 
In \RunOne, this correction had a different meaning because every trolley NMR position was treated as an independent point with  frequency extraction uncertainty of \SI{10}{ppb}. In fact the NMR sample active volume is ${\sim}\SI{1.8}{cm}$, while the measurements are separated by ${\sim}\SI{0.5}{cm}$ leading to oversampling.

\textbf{Trolley temperature dependence ($\mathbf{\delta^{temp}}$):} 
A dedicated study in the Argonne National Laboratory magnet facility with two temperature-controlled probes to track magnet drifts revealed a temperature dependence of the vaseline frequency of \SI[separate-uncertainty = true]{-0.8(2)}{ppb/C}. However, a conservative uncertainty of \SI{2}{ppb/\celsius} is used, since the uncertainty is dominated by the frequency extraction uncertainties discussed above.

The trolley-probe NMR frequencies are not actively temperature corrected, rather,
we apply a correction and uncertainty $\delta^\text{tr,temp}$. The temperature difference of the trolley probes with respect to the mean temperature during the calibration (\SI{33.1}{\celsius}) range from \SI{-8.0}{\celsius} to \SI{-1.9}{\celsius}. 
The temperature-dependent frequency correction is calculated using the temperature dependence of  \SI[separate-uncertainty = true]{-0.8(20)}{ppb/C}. 
The muon weighted corrections for the three datasets are \SI{-3.6}{ppb}, \SI{-5.5}{ppb}, and \SI{-6.0}{ppb}, respectively.
In addition, the temperature spread during one field map is \SI[separate-uncertainty = true]{1.8(3)}{\celsius} and an uncertainty of $\SI{1}{^\circ C}$ on the temperature sensor is used. The resulting uncertainties for \RunTwo, \RunThreeA, and \RunThreeB are listed in Table~\ref{tb:field:maps}. 

\textbf{Trolley transverse and azimuthal position ($\mathbf{\delta^{xy}}$, $\mathbf{\delta^{azi}}$):} The trolley position is constrained in the transverse plane by the rails. 
A laser tracker was used to estimate rail distortions before the vacuum chambers were installed. The effect in the transverse plane $\delta^\text{xy}$ is evaluated by taking the \RunTwo and \RunThree beam shapes into account by running one of the analysis chains with and without incorporating rail distortions. The observed difference of \SI{11.8}{ppb} (\RunTwo), \SI{4.1}{ppb} (\RunThreeA), and \SI{1.8}{ppb} (\RunThreeB) are used to correct the other analysis. The corresponding uncertainties are listed in Table~\ref{tb:field:maps}. For \RunTwoThree the corrections are smaller than for \RunOne due to the smaller higher-order multipole moments. 
    
The azimuthal trolley position is determined using the barcode except for small gaps between adjacent vacuum chambers and for barcode errors, where cable-length encoders are used. 
A conservative estimate of the azimuthal position resolution of \SI{2}{mm} leads to a systematic uncertainty of $\delta^\mathrm{azi}=\SI{4}{ppb}$ on the average dipole field.

\textbf{Parametrization ($\mathbf{\delta^{param}}$) and Azimuthal averaging ($\mathbf{\delta^{avg}}$):} The finite number of measurements and the parametrization of Eq.~\eqref{eq:field:MomentExpansion} lead to additional uncertainty with three contributions: A. an uncertainty due to the truncation $N_\text{max}$ in Eq.~\eqref{eq:field:MomentExpansion}, B. uncertainty due to interpolation between the finite number of azimuthal slices and C. the use of 2D multipole expansion, which is only valid if there is no azimuthal magnetic field dependence\label{sc:2Dvs3D}.
The uncertainty due to the choice of $N_\text{max}$  is estimated from the residuals of the fits to Eq.~\eqref{eq:field:MomentExpansion} 
weighted by the azimuthally averaged beam distribution within $\Delta l = \SIrange{1}{10}{\milli\meter}$ around each probe.

The uncertainty due to the interpolation between these finite azimuthal slices was determined by interpolating with linear, quadratic, and cubic splines. To estimate the effect of 2D multipole expansion, the averaged magnetic fields following the above analysis approach were compared to an analytic azimuthal average using simulated magnetic fields based on a toroidal 3D multipole-based field description. The observed differences from such comparisons are $<\SI{1}{ppb}$.

\begin{table}[H]
    \centering
    \caption{Corrections and uncertainties from the spatial field maps. A single value per line indicates the same value for all datasets.}
    \label{tb:field:maps}
    \begin{tabular}{ll|c|ccc}
    \hline\hline
    Description & &  Corr. [ppb] & \multicolumn{3}{c}{Uncertainty [ppb]}\\
    & & &  Run-2 & Run-3a & Run-3b \\
    \hline
Motion effects & $\delta^{motion}$ & -15.0 & \multicolumn{3}{c}{18.0} \\
Configuration & $\delta^{config}$ & -7.0 & \multicolumn{3}{c}{22.0} \\
\hline
Freq. extraction & $\delta^{freq (tr)}$ & - & 19 & 18 & 16 \\
Temperature         & $\delta^{temp}$ & - & 9.2 &	13.8 & 15.2 \\
Transverse pos. & $\delta^{xy}$ & - &	10.0 & 9.9 & 9.0 \\
Azimuthal pos.  & $\delta^{azi}$ & - & \multicolumn{3}{c}{4.0} \\
Parameterization    & $\delta^{param}$ & - & 3.4 & 6.3  &	7.6 \\
Azi. averaging      & $\delta^{avg}$ & - & 0.8 & 1.4 & 1.7 \\
\hline
Subtotal & & &  37.2 & 38.5 & 38.1 \\
\hline\hline
    \end{tabular}
\end{table}

\subsection{Magnetic field tracking\label{sec:field:tracking}\label{sec:field:synchronization}}

The fixed probes track the magnetic field between trolley runs (see Sec.~\ref{sec:field:trolleyMaps}) for moments up to $i=5$. For higher-order moments, we use linear interpolation in time.  Fixed-probe tracking entails the following steps:
1) extracting fixed-probe moments defined in Eq.~\eqref{eq:field:MomentExpansion}; 
2) tying the fixed-probe moments to the trolley-map moments; 
3)  parameterizing the moments as a function of azimuth and time.

\subsubsection{Fixed probe moments
\label{sec:field:tracking:fxp_moments}}

Linear combinations of measurements from the four or six fixed probes at each station provide fixed-probe moments
$m_i^\text{fp}(\phi_s,t)$ following the procedure described in~\cite{Run1PRAField}.

To reduce the effect of probe noise, the $m_5^\text{fp}(\phi_s,t)$ moment is first tied to the measured $m_5$ from the trolley run pair (see Sec.~\ref{sec:field:tracking:synchronization}) before the change of moment basis.

Fixed probes in three stations close to the inflector experience large gradients 
resulting in very short FIDs and increased frequency uncertainty (noise). Two additional probes with a PEEK housing are installed inside the vacuum chamber at the position of one of the stations. These additional measurements verified that linear interpolation of the moments from neighboring stations gives a better estimate than the determination from the noisy fixed probe frequencies. Therefore, the multipole moments for these three stations are linear interpolations from their neighboring stations.

The relative fixed probe frequency extraction is very robust and the uncertainty from the fixed probe frequency extraction $\delta^\text{freq(fp)}$ is ${\sim} \SI{1}{ppb}$, consistent with \RunOne~\cite{Run1PRAField}. Non-linear temperature changes of the yoke and thus the fixed probes are on the $\SI{0.06}{^\circ C}$ level, and thus the uncertainty due to fixed probe temperature is negligible. 
Linear components are canceled by tracking between two subsequent field maps. 

Fixed probe data are subject to general data quality cuts (Sec.~\ref{sec:datasets}). Additionally, events with FID amplitudes or FID power more than seven standard deviations from the probe's mean  amplitude and power are removed.

\subsubsection{Tying fixed probe to trolley-map moments\label{sec:field:tracking:synchronization}}

The  change of the magnetic field at a fixed-probe station  before or after $t_{s}^\text{tr}$, the time the  trolley passes the station at $\phi_s$ during a trolley run, is
\begin{align}
\Delta m_i^{\text{fp}}(\phi_s,t) = m_i^{\text{fp}}(\phi_s,t)-m_i^\text{fp}(\phi_s,t_{s}^\text{tr}),
&
    \label{eq:field:fixed_probe_drift}
\end{align}
where $m_i^\text{fp}(\phi_s,t_{s}^\text{tr})$ is the moment measured using the fixed probes within station $s$ averaged around the time the trolley passes by that station.

To determine $t_{s}^\text{tr}$, we make use of the fact that the material effects of the trolley and its onboard electronics produce a characteristic field perturbation (footprint) that is measured by the fixed probes when the trolley passes. The time of the largest field perturbation sets $t_{s}^\text{tr}$ and the trolley's azimuthal location sets $\phi_s$. 
Varying the station positions $\phi_s$ by ${\sim}\SI{0.25}{deg}$ has an effect less than $\SI{1}{ppb}$.

The 
field perturbation due to the trolley when passing a fixed probe station  
is removed from the fixed-probe data and replaced with a linear interpolation 
of $m_i(\phi_s,t)^\text{fp}$ based on the \SI{30}{\second}  before and after $t_{s}^\text{tr}$. 
The effect of the trolley footprint replacement is tested on data in regions without footprint by comparing the field estimated by the replacement algorithm and the actual measured data. The uncertainty is listed in Table~\ref{tb:field:tracking} and is similar to \RunOne, as described in~\cite{Run1PRAField}.

\subsubsection{Fixed-probe tracking \label{sec:field:tracking:tracking}}

For azimuth $\phi$ and time $t$ for one or more trolley runs at $t_k$ the fixed-probe tracked moments are 
\begin{widetext}
\begin{align}
   m_i(\phi,t) = &\sum_k W_k(t) \left(m_i^\text{tr}(\phi,t_k) + \sum_{s} W_s(\phi) 
    \sum_j J_{ij}(\phi_s)\Delta m_j^{\text{fp}}(\phi_s,t)\right)
    \label{eq:field:tracking}
\end{align}
\end{widetext}
where $k$ labels the trolley runs, and  
$W_k(t)$ is the weighting of each trolley run at time $t$. The azimuthal weighting factor $W_s(\phi)$ interpolates between stations on either side of $\phi$,  $J_{ij}(\phi_s)=\frac{\partial m_i^\text{tr}(\phi_s)}{\partial m_j^\text{fp}(\phi_s)}$ is the Jacobian that relates small changes of the fixed probe moments to changes of the trolley moments for station $s$, and
  $\Delta m_i^\text{fp}({\phi_s},t)$ is defined in  Eq.~\eqref{eq:field:fixed_probe_drift}. 

  Ideally, magnetic field tracking uses two consecutive trolley runs, {\it e.g.} $k=1,2; 2,3$ {\it etc.}. Occasional unplanned magnet incidents, such as the loss of magnet power allow tracking only from the trolley run before the incident, in which case
$W_k(t)=1$.

Field changes not tracked by the fixed probes lead to errors of the $m_i(\phi,t)$ that is a maximum at the midpoint between the two paired trolley runs. To quantify this, tracking from a single trolley run is used to predict the field moments at the  later trolley run. The difference between the predicted and measured field moments for the second trolley run is called the \textit{tracking offset}. The tracking offset can be modeled as a random walk process caused by changes in the magnet shape. For tracking using a pair of consecutive trolley runs, the random walk becomes a Brownian bridge 
that uses a linear interpolation between the first and second trolley run (see Ref.~\cite{Run1PRAField} for details). A single parameter $M$ parametrizes the rate of the process.

The distribution of the azimuthally--averaged tracking offsets can be used to account for potential correlations between different stations. In order to reduce the statistical error, 
the random-walk parameters are determined from 
the azimuthally--averaged tracking offsets for all of \RunTwoThree. 
We determine 
$M =\SI{0.018}{Hz/\sqrt{s}}$ for the $m_1$ coefficient. Similar rate of change parameters are determined for each multipole moment. 
The resulting uncertainties, taking the muon-weighted corrections for the different datasets and the correlations between the different multipole moments into account,  are summarized in Table~\ref{tb:field:tracking}. Note that this uncertainty is
statistically independent and hence reduces if multiple datasets are combined.

\begin{figure}
    \centering
    \includegraphics[width=\columnwidth]{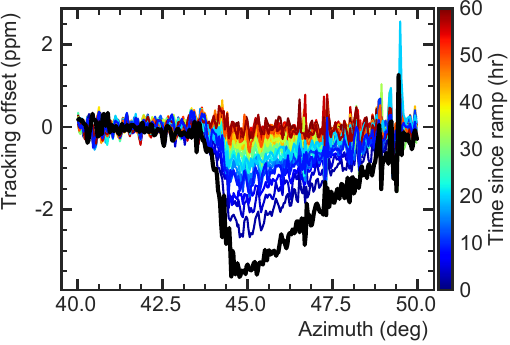}
    \includegraphics[width=\columnwidth]{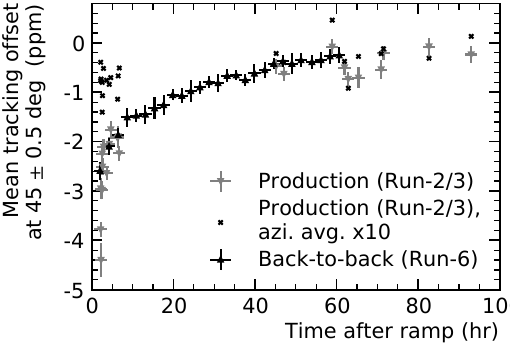}
    \caption{Top: Tracking offset (inability to track field) as a function of azimuth (azi.) around a yoke boundary. Different colors indicate different times after the magnet ramp. Bottom: Amplitude of effect at \SI{45}{\degree} as a function of time after magnet ramp. The x show the azimuthally averaged values scaled up by a factor of x10. A dedicated campaign of back-to-back trolley runs was performed in Run-6 to study this effect.}
    \label{fig:field:time_after_ramp}
\end{figure}

We observe that the tracking offset depends on the time after the magnet was ramped up and shows a characteristic azimuthal dependence that is largest at magnet yoke boundaries as shown in see Fig.~\ref{fig:field:time_after_ramp}. A dedicated measurement was performed, repeatedly measuring the field with the trolley for \SI{60}{h} after the magnet was ramped. 
We use the azimuthally averaged tracking offset  
to estimate the bias. We model the effect by an exponential function with amplitude and time constants as parameters. 
The amplitude and time constant may depend on the history of the magnet before the ramp. Therefore, we determine a correction and uncertainty conservatively; the result is an  initial amplitude of $(100\pm100)$\ ppb 
and a relaxation time constant of \SI{12}{h}. The correction and uncertainty depend on the time periods relative to the magnet ramp time in which muon data have been taken. The resulting correction and uncertainties are listed in Table~\ref{tb:field:tracking}.

A detailed comparison between interpolation analyses from two groups was performed to identify inconsistencies and bugs in the analysis, while the individual groups had individual software blinds. Comparisons  performed on the azimuthal averaged field and on a station-by-station basis agree within a few ppb after relative unblinding. The difference in analysis results due to different analysis choices is added as additional uncertainty and listed in Table~\ref{tb:field:tracking}.

\begin{table}[H]
    \centering
    \caption{Corrections and uncertainties (in parenthesis) from magnetic field tracking. A single value per line indicates the same value for all datasets. All values are given in units of ppb.}
    \label{tb:field:tracking}
    \begin{ruledtabular}
    \begin{tabular}{l|ccc}
    Description &  \multicolumn{3}{c}{ Correction (Uncertainty)}\\
    & Run-2 & Run-3a & Run-3b \\
    \hline
    Tying & & & \\
    ~~Trolley footprint     & \multicolumn{3}{c}{(7.0)} \\
    ~~Fixed probe resolution &  \multicolumn{3}{c}{(1.0)} \\
    \hline
    Tracking & & & \\
    ~~Brownian bridge  & (15.4) & (10.7) & (16.0) \\
    ~~Magnet ramp effect & -3.0 (3.0) & -10.0 (10.0) & -3.0 (3.0) \\
    ~~Fixed probe temperature & 
    (0) & (0) & (0) \\
    \hline
    Analysis choices  & (1.8) & (2.5) & (1.5) \\
    \hline
    Subtotal & (17.3) & (16.5) & (17.8)  \\
    \end{tabular}
    \end{ruledtabular}
\end{table}

The  multipole moments averaged over azimuth and weighted by the detected muons (including DQC) $\langle m_i \rangle_{\phi,t}$ are listed in Table~\ref{tab:field:multipoles} for all three datasets. The lowest order $\langle m_i\rangle_{\phi,t}$, the normal and skew quadrupoles, are shown as a function of time over the full dataset in Fig.~\ref{fig:field:overview}.

\begin{table}[h]
    \centering
    \caption{Field multipole moments in ppb (see Eq.~\eqref{eq:field:MomentExpansion}) averaged over azimuth and time (including DQC) per dataset. The \RunThree the experiment hall temperature was more stable than  \RunTwo  due to a climate-control upgrade.  }
    \label{tab:field:multipoles}
    \begin{ruledtabular}
    \begin{tabular}{l|rrr}
         Multipole  & Run-2 & Run-3a & Run-3b \\
         \hline
$m_2/m_1$   &  331 &   -113 &   -14 \\
$m_3/m_1$    &  611 &     -6 &   -43 \\
$m_4/m_1$    & -310 &     23 &    17 \\
$m_5/m_1$    &  383 &     40 &    35 \\
$m_6/m_1$    &   94 &     -9 &   -20 \\
$m_7/m_1$    &  217 &    127 &   127 \\
$m_8/m_1$    &  -24 &    -22 &   -21 \\
$m_9/m_1$    &   23 &     15 &    12 \\
$m_{10}/m_1$ & -697 &   -725 &  -727 \\
$m_{11}/m_1$ & -167 &   -203 &  -215 \\
$m_{12}/m_1$ &-1068 &  -1056 & -1057 
    \end{tabular}
    \end{ruledtabular}
\end{table}

\subsection{Muon weighted magnetic field\label{sec:field:muonWeighting}}

\subsubsection{Muon beam distribution}
\label{sec:field:muonDistribution}
\label{sec:bd:muonDistribution}

The muon beam distribution $M(x, y, \phi)$ 
is reconstructed from measured positron tracker profiles combined with beam-dynamics calculations of the azimuthal dependence of the muon distribution around the ring. Two trackers 
provide well-localized muon beam distributions with an azimuthal sensitivity with an RMS of \SI{4.9}{\degree} and \SI{4.8}{\degree}, respectively. Following Eq.~\eqref{eq:muonweightedfield2}, the mapped magnetic field is weighted by the muon distribution to determine the magnetic field seen by the muons.

Tracker profiles $M^\text{T}_{\text{i}}(x,y)$ for the muon-weighted magnetic field are accumulated in time intervals of $T_\text{interval}=$\SIrange[]{2}{3}{h} and corrected for detector resolution and acceptance. Only positrons with decay times between the analysis start time $t_\text{start} = \SI{30.2876}{\micro\second}$ and end time $t_\text{end} = \SI{650.0644}{\micro\second}$ enter the tracker profiles. The time intervals $T_\text{interval}$ are chosen to contain more than $\SI{6e5}{}$ total tracks, avoid gaps $>\SI{6}{h}$, stay within a trolley-run pair and contain entire $\omega_a$ DAQ runs.

The measured beam profiles at azimuthal locations where the tracker detectors do not provide beam diagnostics are reconstructed from tracker profiles by shifting the mean and scaling the transverse widths of the distribution relative to the tracker station using
\begin{align}
    \langle x \rangle (\phi) & = x_\text{COD}(\phi) + D_x(\phi) \langle \delta \rangle \label{eq:field:muon_weighting_extrapolation_1},\\
    \langle y \rangle (\phi) & = 0,\label{eq:field:muon_weighting_extrapolation_2}\\
    x_\text{RMS}(\phi) & = \bigg[\frac{\beta_x(\phi)}{\beta_x(\phi_\text{tkr})}\left(x_\text{RMS}^2(\phi_\text{tkr}) - D_x^2(\phi_\text{tkr})\delta_\text{RMS}^2\right) \nonumber \\
    & + D_x^2(\phi)\delta_\text{RMS}^2\bigg]^{1/2},\label{eq:field:muon_weighting_extrapolation_3}\\
    y           _\text{RMS}(\phi) & = \left[\frac{\beta_y(\phi)}{\beta_y(\phi_\text{tkr})} y_\text{RMS}^2(\phi_\text{tkr})\right]^{1/2} \label{eq:field:muon_weighting_extrapolation_4}.
\end{align}
The beam widths $x_\text{RMS}$ and $y_\text{RMS}$ at the azimuth of the tracker stations $\phi_{\text{trk}}$ are extracted from the tracker profiles $M^\text{T}_{\text{i}}(x,y)$.
The beta functions $\beta_x(\phi)$, $\beta_y(\phi)$, and radial dispersion function $D_x(\phi)$ are determined from the optical lattice calculated with the \texttt{COSY INFINITY}-based model of the storage ring. The mean and RMS fractional momentum $\langle\delta\rangle$ and $\delta_\text{RMS}$ are extracted from the fast-rotation analysis discussed in Sec.~\ref{sec:bd:corr:Ce}.  The average fractional momentum is $\sim$\SI{0.07}{\percent} except for \RunThreeB, which is lower ($\sim$\SI{0.01}{\percent}) owing to stronger injection kickers, whereas the RMS of the distribution is $\sim$\SI{0.1}{\percent}. The field indices are listed in Table~\ref{tab:datasets}.

Closed orbit distortions (COD) shift the ideally circular closed orbit away from the equilibrium position. Azimuthal variation in the vertical dipole component of the magnetic field causes a radial COD
\begin{align}
    x_\text{COD}(\phi) \approx \frac{R_0}{n}\frac{b_1(m_1)}{B_0} \cos\left(\phi-\phi_1(m_1)\right),
\end{align}
where $R_0$
is the nominal radius, $B_0$ is the nominal field,  $n$ is the effective field index given in Table~\ref{tab:datasets}, and $b_1(m_1)$ and $\phi_1(m_1)$ are the $N=1$ Fourier amplitude and phase of $m_1(\phi)$. The Fourier components are extracted with an FFT from field maps in each $T_\text{interval}$, and $x_\text{COD}$ is calculated for each individual $T_\text{interval}$.
The amplitudes of the radial COD range from \SIrange{0.6}{1.5}{\milli\meter} and \SIrange{0.2}{0.4}{\milli\meter} for \RunTwo and \RunThree, respectively.

An azimuthally varying radial magnetic field would cause a vertical COD. Because the radial field dependence on azimuth is not measured during the experiment, $y_\text{COD}$ is set to zero and considered separately as a systematic. Misalignments of the electric quadrupole plates also cause radial and vertical CODs by steering the beam. These are considered separately as a systematic.

Each tracker station is extrapolated separately, and the reconstructed distributions from both stations are averaged to get the nominal beam distribution.

Figure~\ref{fig:bd:beamShape} in Sec.~\ref{sec:beamdynamicintro:beamdistribution} illustrates azimuthally averaged muon beam distributions based on the beam extrapolation around the ring of tracker measurements.

\subsubsection{Muon weighting}

Following Eq.~\eqref{eq:muonweightedfield2}, the reconstructed muon beam distribution $M(x,y,\phi,t)$ (see Sec.~\ref{sec:bd:muonDistribution}) is projected onto the moments used to describe the magnetic field for time intervals $T_\text{interval}$ and evaluated every \SI{5}{\degree} because the azimuthal variation of the beam moments is small. Since the tracker profiles and thus the beam moments are only determined every $\SIrange{2}{3}{h}$, the field moments $m_i(t,\phi)$ are averaged in time, weighted by the number of muons in the storage region $N_\mu(t)$. 
Eq.~\eqref{eq:field:time_azimuth_dependent_muon_weighted_field} is used to calculate the muon-weighted field per $T_\text{interval}$ and azimuthal bin $\phi_i$. Additional averaging over all azimuthal bins and thus implementing Eq.~\eqref{eq:field:time_dependent_muon_weighted_field} yields the muon-weighed field per time interval $T_\text{interval}$. Averaging all time intervals within a dataset, weighting by $N_\mu(t)$ and accounting for DQC cuts, yields the muon weighted magnetic field $\tilde\omega_p^\prime$ per dataset defined in Eq.~\eqref{eq:muonweightedfield2},
listed in Table~\ref{t:Rmu-results} for each dataset. 

\begin{figure}[h]
    \centering
    \includegraphics[width=\linewidth]{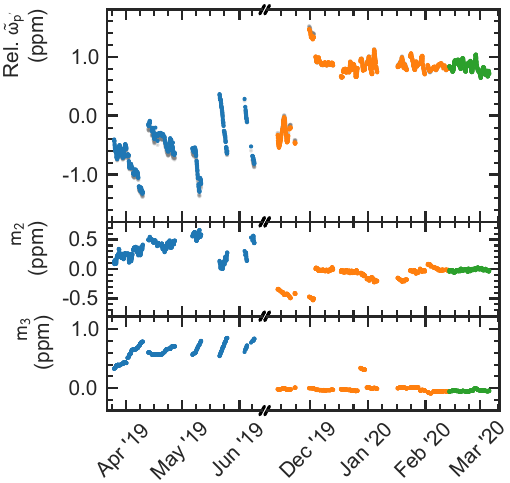}
    \caption{The relative muon-weighted magnetic field ($\tilde\omega_{p'}$) as a function of time for the \RunTwo (left side) and \RunThreeA and \RunThreeB (right side). The dipole $m_1$ contribution alone is shown in gray below. On this scale, they barely differ.  The lower two plots show the tracked $m_2$ and $m_3$ moments.}
    \label{fig:field:overview}
\end{figure}

The improvement in the kick for dataset Run-3b reduces the $k_2$ and $k_5$ parameters (see Eq.~\eqref{eq:multipole_projections}) since the muon distribution is more centered. This has the effect that weighted moments $m_i, i>1$ are reduced, and thus systematic uncertainties that only couple through moments with $m_i, i>1$ are reduced as well. The beam multipole projections averaged over azimuth over the times when muons are stored to 
extract $\omega_a$ ($\langle k_i\rangle_{\phi,t}$) are listed in Table~\ref{tab:field:ks} for all three datasets. Figure~\ref{fig:field:overview} provides an overview of the muon-weighted field as a function of time. 

\begin{table}[h]
    \centering
    \caption{Average beam multipole projections in each dataset, including DQC. Projections are normalized to beam profile intensity and are unitless.}
    \label{tab:field:ks}
    \begin{ruledtabular}
    \begin{tabular}{l|rrr}
         Beam Projection &  \RunTwo & \RunThreeA & \RunThreeB \\
         \hline
         $k_1$    &  1.000 &  1.000 &  1.000 \\
         $k_2$    &  0.139 &  0.136 &  0.073 \\
         $k_3$    & -0.001 & -0.006 & -0.005 \\
         $k_4$    &  0.001 & -0.001 &  0.000 \\
         $k_5$    &  0.081 &  0.076 &  0.046 \\
         $k_6$    &  0.000 & -0.001 &  0.000 \\
         $k_7$    & -0.001 & -0.001 & -0.006 \\
         $k_8$    & -0.002 & -0.001 &  0.003 \\
         $k_9$    &  0.001 &  0.001 &  0.000 \\
         $k_{10}$ & -0.004 & -0.003 &  0.001 \\
         $k_{11}$ &  0.000 &  0.000 &  0.000 \\
         $k_{12}$ & -0.001 & -0.001 &  0.001 \\
    \end{tabular}
    \end{ruledtabular}
\end{table}

\subsubsection{Systematics}

Tracker-specific systematics cause uncertainties in the beam distribution, which lead to uncertainties in $\tilde\omega_p^\prime$. The relevant uncertainties for muon weighting are tracker resolution $\delta^\text{reso,tkr}$, acceptance $\delta^\text{accept,tkr}$, and alignment $\delta^\text{align,x,tkr}$, $\delta^\text{align,y,tkr}$. 
These systematics are evaluated by varying each parameter by $\SI{1}{\sigma}$, producing corresponding beam distributions in the usual time intervals $T_\text{interval}$ and evaluating the effect on $\tilde\omega_p^\prime$  averaged over each dataset. 
The resulting uncertainties are listed in
Table~\ref{tb:field:muonWeighting}.

The tracker acceptance uncertainty is $\leq \SI{2}{ppb}$ from changing the acceptance function by $\pm20\%$, and the resolution uncertainty is $<\SI{1}{ppb}$ by changing the radial and vertical resolution by $\pm\SI{0.5}{mm}$. Changing the tracker alignment in $x$ and $y$ by $\pm\SI{0.6}{mm}$ yields uncertainty on the size of \SI{1}{ppb}.
The uncertainty due to
tracker profile statistics are insignificant. 

The muon-weighted field should be calculated for muons that enter the $\omega_a$ determination and thus are seen by the calorimeters. Because the spatial acceptance from tracker and calorimeters is different, the muon distribution from the tracker would have to be corrected for calorimeter acceptance. However, the effect is small and thus is only treated as an uncertainty. 

As discussed above, an azimuthal radial magnetic field variation can contribute to $y_\text{COD}$. Since the radial magnetic field was only measured in pre-\RunOne while no vacuum chambers were installed, the effect is estimated by assuming an amplitude of $\SI{0.5}{mm}$, which is a factor two larger than the pre-\RunOne measured value, for the $N=1$ COD and the worst case phase. 

Misalignments of the electric quadrupole plates cause an $x_\text{COD}$ or $y_\text{COD}$ by steering the beam. The expected COD calculations use the central displacements of the electric quadrupole plates measured in a survey. Survey uncertainties cause uncertainties in the CODs. These effects were evaluated using the same method from \RunOne~\cite{Run1PRAField}, resulting in a correction and uncertainty listed in Table~\ref{tb:field:muonWeighting}.

The momentum deviation $\delta$ used in the beam reconstruction procedure in Eq.~\eqref{eq:field:muon_weighting_extrapolation_1} and Eq.~\eqref{eq:field:muon_weighting_extrapolation_3} slightly differ from different analyzing teams in Sec.~\ref{sec:oa}. The related systematic uncertainty is determined by varying $\langle \delta \rangle$ and $\delta_\text{RMS}$ by $\pm0.0001$.

A changing muon distribution over time in a fill can be caused by magnetic field transient effects from the electric quadrupoles and kicker eddy currents. 
Tracker profiles are reconstructed for different times in a fill. Studies show that the related uncertainties are negligible in \RunTwoThree.

\begin{table}[H]
    \centering
    \caption{Corrections and uncertainties (in parenthesis) due to spatial muon weighting of the magnetic field. }
    \label{tb:field:muonWeighting}
    \begin{ruledtabular}
    \begin{tabular}{l|ccc}
    Description &  \multicolumn{3}{c}{Correction (Uncertainty) (ppb)} \\
    & \RunTwo & \RunThreeA & \RunThreeB \\
    \hline
    Detector effects  & & & \\
    ~~Tracker acceptance  &  (2.1) & (1.1) & (0.1) \\
    ~~Tracker resolution  &  (0.1) & (0.1) & (0.1) \\ 
    ~~Tracker y-alignment & (10.7) & (0.6) & (0.4) \\
    ~~Tracker x-alignment &  (4.5) & (1.3) & (0.3) \\
    ~~Calorimeter acceptance & (1.0) & (0.2) & (0.2) \\
    Closed Orbit Distortion & & & \\
    and azimuthal effects & & &  \\
    ~~yCOD (radial B) & (1.8) & (3.7) & (2.9) \\
    ~~xCOD (quad misalig.) & +1.3 (5.9) & +2.7 (6.7) & +2.5 (6.3) \\
    ~~yCOD (quad misalig.) & -0.9 (0.1) & -0.5 (0.2) & -0.3 (0.2) \\
    ~~Mean momentum offset & (0.2)  & (0) & (0) \\
    \hline
    Subtotal & (13.4) & (7.9) & (6.9)
    \end{tabular}
    \end{ruledtabular}
\end{table}

\subsection{Transient magnetic fields\label{sec:field:transients}}

The fixed probe system measures the magnetic field at intervals of \SIrange{1.2}{1.4}{s} asynchronous to beam injection. Thus, any time-dependent, \SI{}{\micro\second}-timescale magnetic field transient that is synchronized with beam injection is not accounted for in $\tilde\omega_p^\prime$. In addition, the skin-depth effect in the aluminum of the vacuum chambers reduces the effects on high-frequency magnetic field transients.  
Transient magnetic fields synchronized with beam injection are caused by eddy currents in the kicker and time-varying fields caused by the pulsing of ESQs. Both effects lead to corrections on the muon-weighted magnetic field and are improved compared to \RunOne by additional measurements. Additional transient effects related to magnetic fields in the booster are $<\SI{7}{ppb}$ as determined for the \RunOne analysis~\cite{Run1PRAField}.

\subsubsection{Transient magnetic fields from kickers}
\label{sec:field:transients:kicker}
The magnetic field kick of \SI{22}{mT} to store muons on the stable orbit is a fast transient field (${\sim}\SI{150}{ns}$) that introduces eddy currents in the region of the kicker magnets that lasts longer than the initial kick. NMR magnetometers are too slow to measure the effect on the magnetic field. The transient magnetic field has been measured with two magnetometers based on Faraday rotation using terbium gallium garnet (TGG) crystals~\cite{Run1PRAField}. For Run 2/3, additional measurements with improved setups have been performed using the same magnetometers.

One of the magnetometers utilizes fibers to guide the light from the laser source, which is housed in the center of the storage ring magnet, to the 3D printed magnetometer where the laser light is polarized and sent through two 14.5-mm-long TGG crystals. A polarization-sensitive splitter divides the laser beam into two returning fibers. The two beam intensities are measured by PIN diodes; the polarization is reconstructed from the difference. This differential readout scheme reduced the sensitivity on laser instabilities. The magnetometer base consists of a glass block with small Sorbothane legs, lowering the magnetometer's center of mass and reducing mechanical vibrations. 

The measurements in \RunOne~\cite{Run1PRAField} were limited by noise picked up from mechanical vibrations of the kicker cage through the magnetometer and the fibers themselves. To reduce the noise in the measurements, 
a PEEK bridge was machined with Sorbothane legs that allow the magnetometer to be anchored to the vacuum chamber instead of the cage that holds the kicker plates. In addition, the returning fibers are routed on top of silicon bands that dampen out potential vibrations.

Two measurement campaigns in summer 2021 and summer 2022 have been performed. To calibrate the magnetometer, the magnetic field of the main magnet was ramped up and down at a constant rate to \SI{1.4513}{T}. 
The calibration constants change from ramp to ramp due to temperature changes affecting the Verdet constant of the TGG crystal and small tilt angles changing the effective length of the crystal.

Since the laser was operated in constant current mode, the calibration factor changed over time, which was tracked by measuring the $\SI{12}{\micro\tesla}$ magnetic field transient from charging the kicker plates prior to the kick. 

\begin{figure}
    \centering
    \includegraphics[width=\linewidth]{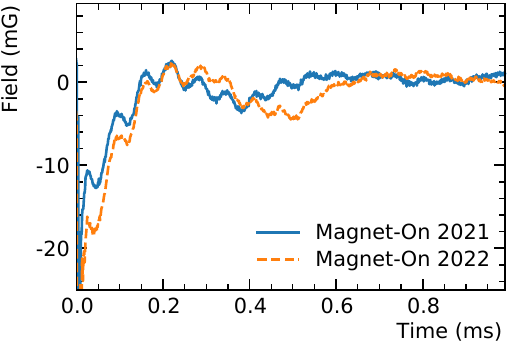}
    \caption{Magnetic field transient induced by kicker magnets measured by the optical fiber magnetometer in summer 2021 and summer 2022.}
    \label{fig:field:kicker_transient}
\end{figure}

The measured transient field is shown in Fig.~\ref{fig:field:kicker_transient} for two measurement campaigns one year apart. The average of the two campaigns is used to estimate the effect of the measured field perturbations.
The effect on \oa is estimated by integrating the effect of the transient over the muon lifetime. A five-parameter fit is used to estimate the overall correction. The corrections are estimated based on measurements in the first of the three kickers with upgraded kicker cables and operated at nominal kicker setting of \SI{53.1}{kV} as present during Run-3b. The results from this measurement are scaled to the other kickers, which operate at slightly different operation voltages (\SI{53.1}{kV}, \SI{53.0}{kV}, and \SI{55.0}{kV}), and to the conditions in Run-2 and Run-3a, during which the kickers were operated at lower voltages (\SI{47.7}{kV}, \SI{47.1}{kV}, and \SI{47.1}{kV}). Azimuthally, the kicker transient is treated as uniform within the regions occupied by the kicker plates. The steep fall-off at the edges was modeled and confirmed by measurements outside the kicker plates, resulting in a suppression for the azimuthal average of $0.085$. Overall, this results in corrections to \opprimetilde of \SI{-21.1}{ppb} and \SI{-22.5}{ppb} for Run-2/3a and Run-3b, respectively. The associated uncertainties are summarized in Table~\ref{tb:field:transient:kicker} and described briefly below.

The effect of residual vibrations in the measured signal is estimated by comparing results with the main magnet powered and not powered. 
The origin of the perturbations with a time scale of about \SI{1}{ms} and amplitude of a few \SI{0.1}{\micro\tesla} remains ambiguous.  The measurement data cannot distinguish between an actual change in the total magnetic field and mechanical vibrations of the fibers or the crystal. This ambiguity contributes to the leading systematic uncertainty on the transient measurement.
The observed differences between the two campaigns is not fully understood and might indicate local variations of the effect. This ambiguity is accounted for by assigning the observed difference as a ``transient variance'' uncertainty. Further contributions to the uncertainty come from the azimuthal and transverse modeling, as well as from the above-mentioned calibration procedure and baseline determination. Like the total effect, the uncertainties are scaled to the different run conditions in the Run-2 and Run-3a datasets. The scaling and potential differences in pulse shapes due to using different cables lead to additional uncertainties for these datasets. 

\begin{table}[H]
    \centering
    \caption{Uncertainties to \opprimetilde due to transient magnetic fields from eddy currents in the kicker system. The uncertainties from the two campaigns in 2021 and 2022 are combined for the Run-3b dataset. The values are scaled for the Run-2 and Run-3a datasets accounting for the different run conditions.}
    \label{tb:field:transient:kicker}
    \begin{ruledtabular}
    \begin{tabular}{lrrrr}
    Description & \multicolumn{4}{c}{Uncertainty (ppb)}\\
    & 2021 & 2022 & Run-3b & Run-2/3a \\
    \hline
    Vibration ambiguity & 8.3 & 12.8 & 10.5 & 9.9 \\
    Transient variance  &     &      &  4.2 & 3.9 \\
    Azimuthal           & 3.1 &  4.7 &  3.9 & 3.7 \\
    Transverse          & 4.4 &  6.8 &  5.6 & 5.3 \\
    Calibration         & 0.3 &  0.2 &  0.3 & 0.3 \\
    Baseline            & 2.5 &  0.2 &  1.3 & 1.2 \\
    Scaling             &     &      &      & 1.7 \\
    Pulse shape diff.   &     &      &      & 4.2 \\
\hline
Subtotal & & & 13.3 & 13.3 \\
    \end{tabular}
    \end{ruledtabular}
\end{table}

\subsubsection{Transient magnetic fields from ESQs}
\label{sec:field:transients:esq}

\begin{figure}
    \centering
    \includegraphics[width=\columnwidth]{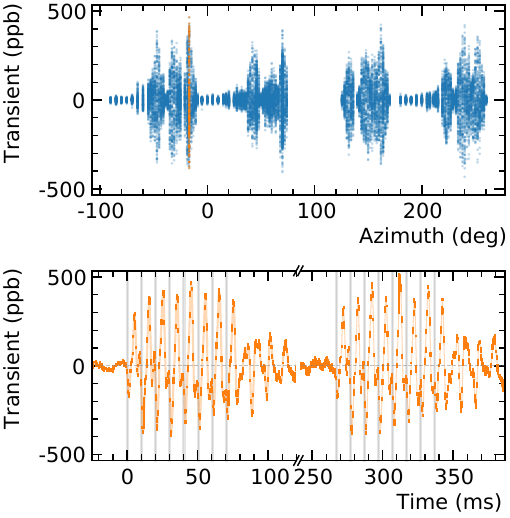}
    \caption{Top) The transient magnetic field from the vibration caused by the ESQ pulsing for all times as a function of azimuth in the storage ring. Bottom) The transient magnetic field as a function of time at one specific location (\SI{-17}{deg}). The times during which muons are stored are highlighted by gray bands. The shown field transients are scaled up to the ESQ operation voltage. }
    \label{fig:field:esq}
\end{figure}

The beam-synchronous pulsing of the ESQ plates causes time-dependent magnetic field changes on the \SI{}{\micro\second}-timescale. 
These fast synchronous changes are not captured by the field maps nor tracked by the fixed probe system. 
Besides the asynchronous operations of the fixed probes with respect to beam injection times, skin depth effects in the aluminum walls of the vacuum chambers suppress field transients on that time scale. In-situ measurements are required.
While the exact mechanism creating this magnetic field transition is not fully understood, the effect is associated with the ESQ plates' and support structure's mechanical vibrations. 
The injection of muons and associated pulsing of the ESQ plates every \SI{10}{\milli\second} for $8$ bunches drives an oscillation around \SI{100}{Hz}, close to the system's intrinsic frequencies around \SI{50}{Hz}. 
The bottom plot in Fig.~\ref{fig:field:esq} shows an example of this effect as a function of time at one fixed location. 
A second train of eight bunches is injected after \SI{266.7}{\milli\second}, a gap long enough for the vibration to mostly ring down.
This pattern repeats every \SI{1.4}{\second} or \SI{1.2}{\second}. 
Since this field changes during the time muons are stored and are not reflected in the direct measurement of \opprimetilde, this transient results in a correction term $B_Q$. 

In Run-1, the transient fields from ESQs were measured in a dedicated measurement campaign with a set of trolley NMR probes sealed inside plastic tubes for vacuum compatibility, held in place in the center of the storage volume on static legs sitting on the trolley rails. 

The ESQs span \SI{43.3}{\percent} of the ring and are grouped into four stations, each consisting of a short and a long section. 
The azimuthal dependence was mapped coarsely for one such section. Significant differences in the oscillation pattern were observed as a function of azimuth. The long sections were approximated with two short ones. Due to the static nature of the used probes, only one measurement per section was feasible for most sections. The total shift of the magnetic field during the times the muons are stored averaged around the ring was determined from these spatially sparse measurements, leading to the dominant systematic uncertainty of the Run-1 result~\cite{Run1PRL}. 

In dedicated measurement campaigns, the identical sealed NMR probes were mounted on a frame that can be moved around the ring using the trolley infrastructure. The NMR probes were pulsed and read out in the same scheme used in \RunOne through a dedicated multiplexer of the fixed probe systems, now through the $\sim$\SI{50}{\meter} long trolley cable.  
This scheme allows mapping of the effect with finer resolution, significantly improving the precision.  
In the summer of 2020, a quarter of the ring was mapped, and in summer 2021, the full ring was mapped.
The top plot in Fig.~\ref{fig:field:esq} shows the transients for all times as a function of azimuth around the ring.
 The measurements were performed at a reduced ESQ voltage of \SI{14}{kV}. The confirmed voltage-squared dependence was used to scale the measurement to the nominal ESQ operations voltage of \SI{18.2}{kV}.

The effect of the magnetic field perturbations on \oa in a particular fill at a particular azimuthal position is estimated by a linear fit of the magnetic field transient over the muon storage time of around \SI{700}{\micro\second} of this fill. The effect accumulates over the muon lifetime in the storage ring
~\cite{Run1PRAField}.
The azimuthally resolved effect from the different measurement positions is averaged around the ring, accounting for the different azimuthal spacings between the measurements. Segments outside vacuum chambers containing ESQs and where no time-dependent field perturbations are observed don't contribute.  
Table~\ref{tb:field:transient:ESQ} shows the total correction $B_Q$=\SI[separate-uncertainty = true]{-21.0(19.5)}{ppb} due to transient magnetic fields from the ESQ and lists the corresponding uncertainties, which are discussed in more detail below.   

The frequency extraction from NMR FID signals requires a minimal length of more than $\sim$\SI{0.5}{\milli\second} for the required resolution. The time scale of the observed transient changes the field within an FID. Hence, magnetic field perturbations from outside the fit window of the transient effect leak into the frequency. Alternatively, the phase function from multiple FIDs with different delays with respect to the muon injection time can be combined and fitted directly in the relevant time window. 
The NMR probes have a 0.5-mm-thick aluminum shell, and the corresponding skin depth suppresses higher-frequency components. 
This effect was evaluated in a dedicated measurement.
The transient caused by the ESQ was mapped partially one year after \RunThree and around the full ring the year afterward. In addition, starting mid-\RunThree, periodic measurements at static positions were taken. The different measurements over time are in good agreement. In addition, the fixed probe system is used to monitor the effect of the transient from outside of the vacuum chambers parasitically during data taking.

All the measurements are point estimates, and the values in between the measurement points are unknown, resulting in uncertainty in the azimuthal averaging. 
In addition, the mapping was performed in the center of the storage volume. The radial dependence of the transient was measured on the diagonal along the ESQ \SI{0}{\volt}-line at one location. A flat dependence was found up to \SI{2}{\centi\meter}, where most of the muon beam is located, and variations up to \SI{25}{\percent} were observed at a radius of \SI{4}{\centi\meter}, at the edge of the storage volume. 
As mentioned above, the ESQ can only be operated consistently at \SI{14}{kV} with the mapper device present. Perturbations of the electric field from the mapping device itself might modify the local forces on the ESQ plates and change the mechanical oscillation of the system.
Other sources for uncertainties are fill-by-fill intensity variations not accounted for the averaging between the 16 fills and small changes in the time structures in the second eight bunches between running conditions and the measurements. 

\begin{table}[H]
    \centering
    \caption{Correction and associated uncertainties to \opprimetilde due to transient magnetic fields caused by the pulsing of the ESQ system. }
    \label{tb:field:transient:ESQ}
    \begin{ruledtabular}
    \begin{tabular}{lcc}
    Description & Correction (ppb) & Uncertainty (ppb)\\
    \hline
    frequency extraction    && 5\\
    skin depth              && 2\\
    stability over time     && 8\\
    azimuthal averaging     && 11\\
    transverse dependence   && 5.3\\
    measurement apparatus   && 10.5\\
    fill-by-fill variations && 2\\
    second bunch train      && 5\\
    \hline
    Subtotal &-21.0 & 19.5
    \end{tabular}
    \end{ruledtabular}
\end{table}

\subsection{Summary and differences with respect to \RunOne\label{sec:field:summary}}

The dataset averaged $\tilde\omega_p^\prime$ are listed in Table~\ref{t:Rmu-results}. All non-negligible uncertainties are summarized in Table~\ref{tb:field:uncertainty}. For uncertainties that have been determined on a probe-by-probe basis, the uncertainties are translated to multipole moments and further to $\tilde\omega_p^\prime$ taking the correlation between moments and the spatial and temporal muon distribution into account. 
Uncertainties are highly correlated and thus treated as fully correlated, except the Brownian bridge-based tracking uncertainty, which is random in nature and reduced by combining datasets. 
Calibration constants and corrections are taken into account in the final $\tilde\omega_p^\prime$ and are not listed individually.
The total uncertainty on the muon-weighted magnetic field, including corrections from magnetic field transients, is $\leq\SI{52}{ppb}$, a factor of ${\sim}2$ improvement compared to the \RunOne analysis~\cite{Run1PRAField}. The main reason is the improved understanding of the electrostatic quadrupole transient due to additional measurements. Overall, the current uncertainty budget is well below the systematic uncertainty goal from the technical design report of $<\SI{70}{ppb}$. 

\begin{table}[H]
    \centering
    \caption{Summary of uncertainties on $\tilde \omega_p^\prime$ for each step in the analysis. A detailed breakdown of each contribution is given in the corresponding section. A single value per line indicates the same value for all datasets. All contributions are assumed to be fully correlated, except the Brownian bridge uncertainty in the Tracking section, which is treated as statistical uncertainty.}
    \label{tb:field:uncertainty}
    \begin{ruledtabular}
    \begin{tabular}{lcccl}
    Description &  \multicolumn{3}{c}{Uncertainty (ppb)} & Section \\
    & \RunTwo & \RunThreeA & \RunThreeB & \\
    \hline
Calibration probe   & \multicolumn{3}{c}{8.9}  & \ref{sec:field:absoluteCalibration} \\
    Trolley calibration & \multicolumn{3}{c}{17.8} & \ref{sec:field:trolleyCalibration}\\
    Spatial Field Maps & 37.2 & 38.5 & 38.1 & \ref{sec:field:trolleyMaps} \\
    Tracking & 17.3 & 16.5 & 17.8 & \ref{sec:field:tracking} \\
    Muon Weighting & 13.4 & 7.9 & 6.9 & \ref{sec:field:muonWeighting}\\
    Transient Booster & \multicolumn{3}{c}{7} & \ref{sec:field:transients}\\ 
    Transient Kicker & \multicolumn{3}{c}{13.3} & \ref{sec:field:transients:kicker}\\
    Transient ESQ & \multicolumn{3}{c}{19.5} & \ref{sec:field:transients:esq}\\
\hline
Sub total uncorrelated   & 15.4& 10.7& 16.0&   \\
Sub total correlated     & 51.3& 52.0& 50.6& 
    \end{tabular}
    \end{ruledtabular}
\end{table}

The major differences in the \RunTwoThree analysis of \opprimetilde with respect to the \RunOne analysis are listed below:

\begin{itemize}
    \item In \RunOne, the transverse multipole expansion was truncated at $N_\mathrm{max}=9$, for \RunTwoThree, $N_\mathrm{max}=12$ was used.

    \item In the frequency extraction of the trolley FIDs, in \RunTwoThree, slightly earlier times in the phase function fits were used compared to \RunOne.

    \item While in \RunOne only one of the barcode readers was used to determine the azimuthal position, in \RunTwo and \RunThree the second barcode reader is used as a cross-check, increasing reliability. This has the advantage that measurements in the small gaps between adjacent vacuum chamber positions can still be reconstructed even though one of the barcode readers fails.  In addition, better timing alignment of the barcode and encoder systems is possible due to additional timing information in the raw data of both systems. These two developments led to improved reliability of the position determination. 

    \item For \RunTwoThree, the trolley calibration procedures were improved with respect to \RunOne. The improvements include the following:
1) moving the trolley further from the calibration position during measurements with the calibration probe; 
2) revised corrections to the calibration-probe mounting configuration; 
3) inclusion of improved magnetic image measurements described in Sec.~\ref{sec:CalibrationProbe}; 
4) Corrections for second-order gradients near the calibration position due to the different effective sample volumes of the trolley probe and calibration probe.

\item A ground loop issue that was present in \RunOne was removed between \RunOne and \RunTwo.

\item Higher-order multipole moments are smaller in \RunTwoThree than in \RunOne. They were shimmed out better after \RunOne due to the availability of trolley calibration constants. This reduces the uncertainty from the rail misalignments, as well as from muon weighting.
     
\item The temperature dependence of the trolley NMR probes was measured more precisely for \RunTwoThree. It was evaluated as \SI[separate-uncertainty = true]{-0.8(20)}{ppb/\celsius}. In \RunOne, a temperature dependence of \SI[separate-uncertainty = true]{-0(5)}{ppb/\celsius} was used.

\item The rate of change parameter $M$ used for the uncertainty evaluation of the field tracking with a random walk or Brownian bridge model was evaluated in \RunOne station-by-station, manually including observed correlations. This approach was chosen due to the statistics of field periods. In \RunTwoThree, $M$ is evaluated directly from azimuthal averages, which intrinsically includes correlations.  

\item Additional measurements with a dedicated magnetometer with significantly reduced vibrations lowered the uncertainty on the measurements of transient magnetic fields from the kickers.

\item An extensive azimuthal mapping of the transient magnetic field from the ESQ system reduced the corresponding uncertainty significantly. 
    
\end{itemize}

 \section{Overall \textbf{$\mathbf{\omega_{a}/\tilde{\omega}^{\prime}_{p}}$} consistency checks}
\label{sec:sliceanddice}
The $R^{\prime}_{\mu}$ ratio values have been investigated for any inconsistencies and unexpected correlations to external parameters. These external parameters are representative of the conditions that the experiment \RunTwoThree data had been collected in. Eight external parameters had been identified for these checks, namely, average temperature of the muon storage ring, average vacuum pressure of the muon storage ring, magnet current, inflector current, time of data collection since last magnet ramp up, time of data collection (day or night), amplitude of CBO and $k_{loss}$. 

\subsection{Methodology}
In order to perform these checks the data were split into five slices based on the external parameter values, for each of the three Run sets. The $\omega_{a}$ and $\omega_{p}$ values with their respective uncertainties are subsequently extracted from each of the fifteen data slices. These in turn are used to calculate the $R^{\prime}_{\mu}$ ratio and its uncertainty for each of the data slices. It should be noted that for this study the beam dynamics and magnetic field transient corrections are assumed to be constant within the \RunTwo, \RunThreeA, and \RunThreeB datasets.  
These checks were performed on relatively unblinded but overall still blinded data, and repeated eventually on unblinded data.

For the purposes of these tests, we perform a $\chi^{2}$ minimization on the calculated $R^{\prime}_{\mu}$ ratios and their uncertainties in order to evaluate the overall optimal error weighted $R^{\prime}_{\mu}$ ratio value for each external variable studied. Thereafter, the p-value for the sliced $R^{\prime}_{\mu}$ ratios against the optimal $R^{\prime}_{\mu}$ ratio is extracted.

Furthermore, the sliced $R^{\prime}_{\mu}$ ratio values are plotted against the external parameter values for each of the slices and fitted against a constant.
The pull histograms for these plots are then evaluated for any skewness in order to identify dependencies on the external parameters at hand.

\subsection{Results}
The p-values for all the different external parameter cross-checks performed using the methodology described above are summarised in Table~\ref{Tab: slice and dice table}.
In the \RunTwo, \RunThreeA, and \RunThreeB overall consistency study, none of the sliced $R^{\prime}_{\mu}$ ratio values show any direct dependency on the eight investigated external parameters, with p-values within nominal ranges. Moreover, the pull histograms for each of the external parameter slicing fits show a Gaussian distribution of the data centered around $0.0\pm0.2$.

\begin{table}[H]
    \centering
    \caption{$R^{\prime}_{\mu}$ ratio vs. external parameter value with optimal $R^{\prime}_{\mu}$ ratio fit p-values, for combined Run-2, -3a and -3b slicings.}
    \label{Tab: slice and dice table}
    \begin{ruledtabular}
    \begin{tabular}{lc}
    External variable & p-value   \\ \hline
        Average ring temperature & 0.43\\
        Inflector current & 0.75   \\
        Magnet current & 0.13   \\
        Time since magnet ramp up & 0.91   \\
        Day/Night split & 0.70 \\
        Average vacuum pressure & 0.75  \\ 
        Amplitude of CBO & 0.77  \\ 
        $k_{loss}$ & 0.93 \\
    \end{tabular}
    \end{ruledtabular}
\end{table}

The magnet current slicing has a relatively small p-value due to a pull from the slices containing data from runs 2F, 2H, and 3N.  Detailed analysis cross-checks have been made for datasets 2F, 2H and 3N, by $\omega_{a}^m$ and $\tilde{\omega}^{\prime}_{p}$ analyzers. In these cross-checks no extraordinary anomaly was discovered by the analyzers, consequently, the datasets remain valid datasets with statistical fluctuation.

\begin{figure}[h]
\centering
\includegraphics[width=\linewidth]{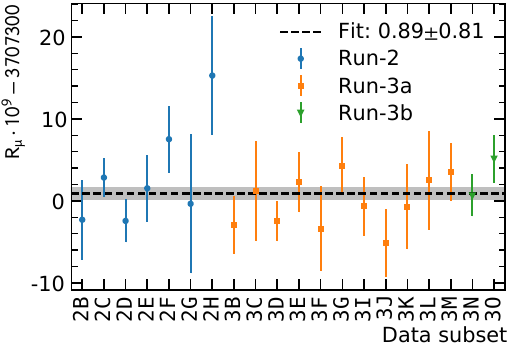}
\caption{$R^{\prime}_{\mu}(T_{r})$ versus data subset. The fit line has a $\chi^2$/ndf$ = 19.31/19$ with a p-value of \SI{44}{\percent}.}\label{fig: data_set_slice_and_dice}
\end{figure}

Additionally, a slicing over different datasets was also performed in order to examine the consistency of the extracted $R^{\prime}_{\mu}$ ratio values over different datasets and time. The results for this data splitting can be visualized in Fig.~\ref{fig: data_set_slice_and_dice}. There are no observed inconsistencies for the $R^{\prime}_{\mu}$ ratio values extracted for different datasets.

\section{Calculation of $\mathbf{a_{\mu}}$} \label{sec:result}

Following Eq.~\ref{eq:R}, for each dataset, the measured $\omega_a^{m}$ is corrected by adding the beam dynamics corrections, and the ratio $R^\prime_\mu(T_r) = \omega_a / \Tilde{\omega}'_p(T_r)$ is computed. Table~\ref{tab:uncertainties} provides an overview of all contributions. All uncertainty contributions to $\omega_a^{m}$, to the beam dynamics corrections and to $\Tilde{\omega}'_p(T_r)$, are propagated to $R^\prime_\mu(T_r)$.  

\begin{table}
\caption{Values and uncertainties of the $\mathcal{R}_\mu'$ terms in Eq.~\eqref{eq:R} and uncertainties due to the external parameters in Eq.~\eqref{eq:final} for $a_{\mu}$. Positive $C_i$ increase $a_{\mu}$; positive $B_i$ decrease $a_{\mu}$. The $\omega_{a}^{m}$ uncertainties are decomposed into statistical and systematic contributions.}
\label{tab:uncertainties}
\begin{center}
\begin{ruledtabular}
\begin{tabular}{lrrr}
Quantity & Correction & Uncertainty & Section\\
          & (ppb) & (ppb) \\
\hline
$\omega_a^{m}$ statistical & - & 201 & \ref{sec:omega_combination} \\
$\omega_a^{m}$ systematic & - & 25 & \ref{ss:systematicsomegaa} \\
$C_{e}$ & 451 & 32 & \ref{sec:bd:corr:Ce}\\
$C_{p}$ & 170 & 10 & \ref{sec:bd:corr:Cp} \\
$C_{ml}$ & 0 &  3 & \ref{sec:bd:corr:Cml}\\
$C_{dd}$ & -15 & 17 & \ref{sec:bd:corr:Cdd}\\
$C_{pa}$ & -27 & 13 & \ref{sec:bd:corr:Cpa} \\
\hline
$\langle \omega_p^\prime\times M\rangle$ & - & 46 & \ref{sec:field:summary}\\
$B_{K}$ & -21 & 13 & \ref{sec:field:transients:kicker}\\
$B_{Q}$ & -21 & 20 & \ref{sec:field:transients:esq}\\
\hline
$\mu_{p}'(\SI{34.7}{\celsius})/\mu_{e}$ & - & 11 & \cite{phillips_magnetic_1977} \cite{Karshenboim:2003qv}\\
$m_{\mu}/m_{e}$ & - & 22 & \cite{RevModPhys.93.025010}\\
$g_{e}/2$ & - & 0 & \cite{ParticleDataGroup:2022pth}\\
\hline
Total systematic & - & 70\\
Total external parameters & - & 25\\
Totals & 622 & 215 \\
\end{tabular}
\end{ruledtabular}
\end{center}
\end{table}

Uncertainty contributions that are assumed to be fully correlated between different \RunTwoThree datasets and also between different measurements by the Fermilab Muon \gm (E989) collaboration are tracked separately from the statistical uncertainties and the other uncertainty contributions that can be considered uncorrelated: the magnetic field uncorrelated uncertainty. The correlation matrix between the ratios is reported in Table~\ref{t:Rmu-ds-corr}. The three $R^\prime_\mu(T_r)$ values are found to be statistically consistent and are fit to obtain the measured $R^\prime_\mu(T_r)$ for the \RunTwoThree sample. The fit $\chi^2$ probability is about 20\%. The results are summarized in Table~\ref{t:Rmu-results}.
\begin{table}
\caption{Correlation matrix of the \RunTwoThree datasets measurements of $R^\prime_\mu(T_r)$.}
\label{t:Rmu-ds-corr}
\begin{center}
\begin{ruledtabular}
\begin{tabular}{lrrr}
$R^\prime_\mu(T_r)$ & \RunTwo & \RunThreeA & \RunThreeB\\
\hline
Run-2 & 1.00 & 0.05 & 0.03\\
Run-3a & 0.05 & 1.00 & 0.03\\
Run-3b & 0.03 & 0.03 & 1.00\\
\end{tabular}
\end{ruledtabular}
\end{center}
\end{table}

\begin{table}
\begin{center}
\caption{\RunTwoThree datasets measurements of $\omega_a$, $\Tilde{\omega}'_p(T_r)$, and their ratios $R^\prime_\mu(T_r)$ multiplied by 1000.}
\label{t:Rmu-results}

\begin{ruledtabular}
\begin{tabular}{lrrr}
 Dataset & $\omega_a/2\pi$\,(Hz) & $\Tilde{\omega}'_p(T_r)/2\pi$\,(Hz) & $R^\prime_\mu(T_r) \times 1000$\\
\hline
\RunTwo & 229077.408(79) & 61790875.0(3.3) & 3.7073016(13)\\
\RunThreeA & 229077.591(68) & 61790957.5(3.3)  & 3.7072996(11)\\
\RunThreeB & 229077.81(11) & 61790962.3(3.3) & 3.7073029(18)\\
\RunTwoThree  &  &  & 3.70730088(79)\\
\end{tabular}
\end{ruledtabular}
\end{center}
\end{table}

Over the course of this analysis, three small errors in the Run-1 analysis~\cite{Run1PRL} were identified. The total shift in the previous result due to these errors is \SI{+28}{ppb}, resulting in
$R^\prime_\mu(T_r)_{\text{Run-1}} = 0.0037073004(16)(6)$.
The measured $R^\prime_\mu(T_r)_{\text{Run-2/3}} = 0.00370730088(75)(26)$ is combined with the Run-1 result~\cite{Run1PRL}, assuming that the systematic uncertainties are fully correlated, to obtain the Fermilab experimental measurement, $R^\prime_\mu(T_r)_{\text{Run-1/2/3}} = 0.00370730082(68)(31)$. This value is combined with the BNL measurement of $R_\mu$ for free protons in vacuum~\cite{PhysRevD.73.072003},  $R_\mu = 0.0037072063(20)$, after converting it using the measured diamagnetic shielding correction $\sigma_{p'}(T_r)$~\cite{phillips_magnetic_1977}:
\begin{align}
R^\prime_\mu(T_r) =
\frac{R_\mu}{1 - \sigma_{p'}(T_r)} = 0.0037073019(20)~.
\end{align}
We compared the systematic uncertainties for the BNL and FNAL measurements and, due to the significant changes in the beam characteristics and detectors between the experiments, concluded that those uncertainties were largely uncorrelated between the two experiments.
The resulting experimental average is $R^\prime_\mu(T_r)_{\text{Exp}} = 0.00370730095(70)$.

The muon magnetic anomaly is computed 
from
\begin{equation}
\label{eq:final}
a_{\mu} = R^\prime_\mu(T_r) \frac{\mu'_{p}(T_r)}{\mu_e(H)} \frac{\mu_e(H)}{\mu_e} \frac{m_{\mu}}{m_{e}} \frac{g_e}{2}~.
\end{equation}
Here $\mu^{\prime}_p(T_r) / \mu_e(H)$ is the ratio of the magnetic moment of the proton in a spherical water sample at 34.7\,\si{\celsius}
and the magnetic moment of the electron in a hydrogen atom~\cite{phillips_magnetic_1977} (\SI{10.5}{ppb}). ${\mu_e(H)}/{\mu_e}$ is the ratio of the magnetic moment of the electron in a hydrogen atom and the magnetic moment of the free electron in vacuum, obtained with a theory QED calculation~\cite{Karshenboim:2003qv}, whose precision is limited to \SI{100}{ppt} by the number of reported digits. ${m_\mu}/{m_e}$ is the ratio of the muon and electron masses (\SI{22}{ppb}), taken from the CODATA 2018 fit~\cite{RevModPhys.93.025010}, primarily driven by the LAMPF 1999 measurements of muonium hyperfine splitting~\cite{PhysRevLett.82.711}. ${g_e}$ is the electron gyromagnetic factor, computed from the electron anomaly $a_e = (g{-}2)/2$ world average~\cite{ParticleDataGroup:2022pth} (\SI{100}{ppt}), dominated by~\cite{PhysRevLett.130.071801}.

\begin{figure}[h]
\centering
\includegraphics[width=\linewidth]{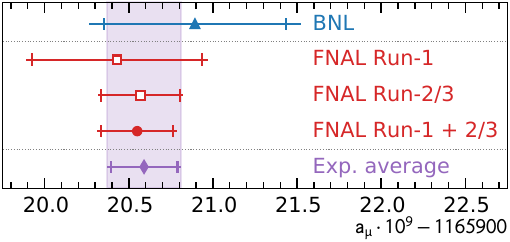}
\caption{From top to bottom:  experimental values of \amu from BNL E821, the FNAL 2021 measurement (FNAL Run-1), this measurement (FNAL Run-2/3), the FNAL combined measurement (FNAL Run-1 + 2/3), and the combined experimental average (Exp. average). The inner tick marks indicate the statistical contribution to the total uncertainties.}\label{fig:results}
\end{figure}

The measured muon magnetic anomaly
for this measurement, this measurement combined with our Run-1 result, and the combined BNL and FNAL results are
\begin{eqnarray*} 
         a_\mu^{\text{FNAL Run-2/3}} &=& 116\,592\,057(25) \times 10^{-11}~(\SI{0.21}{ppm}), \\
    a_\mu^{\text{FNAL Run-1/2/3}} &= &116\,592\,055(24) \times 10^{-11}~(\SI{0.20}{ppm}), \\
    a_\mu^{\text{Exp}} &= &  116\,592\,059(22)\times 10^{-11}~(\SI{0.19}{ppm}).
\end{eqnarray*}
These  are displayed in Fig.~\ref{fig:results}. 
Values of  $R^\prime_\mu(T_r)$ and $a_{\mu}$ with extra digits to facilitate further calculations without loss of precision due to rounding are provided in the supplement material.

\section{Comparison to Theory}
\label{sec:theory}
In recent years, all aspects of the SM theory prediction  $a_\mu^{\rm SM}$ have been scrutinized and refined with continued theoretical and computational efforts. These were summarized by the \gm Theory Initiative~\cite{Aoyama:2020ynm}, using results from Refs.~\cite{Aoyama:2012wk,Aoyama:2019ryr,Czarnecki:2002nt,Gnendiger:2013pva,Davier:2017zfy,Keshavarzi:2018mgv,Colangelo:2018mtw,Hoferichter:2019mqg,Davier:2019can,Keshavarzi:2019abf,Kurz:2014wya,Melnikov:2003xd,Masjuan:2017tvw,Colangelo:2017fiz,Hoferichter:2018kwz,Gerardin:2019vio,Bijnens:2019ghy,Colangelo:2019uex,Blum:2019ugy,Colangelo:2014qya}. While the QED and electroweak contributions are widely considered non-controversial, the SM prediction of the muon \gm is limited by our knowledge of the vacuum fluctuations involving strongly interacting particles, comprising effects called hadronic vacuum polarization (HVP) and hadronic light-by-light scattering.
The latter is currently known at a level of precision comparable to $a_\mu^\text{Exp}$, and it is the leading HVP contribution to the muon magnetic anomaly, denoted by $a_\mu^\text{HLO}$, that gives the dominant uncertainty to the SM prediction. 
These effects cannot be computed at low-energy scales due to the non-perturbative nature of QCD at large distances. It is possible to overcome this problem by means of a dispersion relation technique involving experimental data on the cross-section of electron-positron annihilation into hadrons, $e^+e^-\to \text{hadrons}$. 
In the last 20 years, the worldwide efforts of experiments working on $e^+e^-\to \text{hadrons}$ data in the energy range below a few GeV have achieved the remarkable uncertainty of
0.6\% on  $a_\mu^{\text{HLO}}$~\cite{Jegerlehner:2017gek,Aoyama:2020ynm}.
In addition, in the last few years, there has been significant progress on the first-principles calculation of $a_\mu^{\text{HLO}}$ using lattice QCD which, however, was not yet as precise as the data-driven dispersive approach compiled in~\cite{Aoyama:2020ynm}. In 2021, the BMW collaboration published the first lattice calculation of $a_\mu^{\text{HLO}}$ with sub-percent precision~\cite{Borsanyi:2020mff}. This result would move $a_\mu^{\rm SM}$ towards $a_\mu^{\rm Exp}$  and is compatible with the ``no new physics" scenario but discrepant with the dispersive approach. While the evaluation of the whole $a_\mu^{\text{HLO}}$ from the other lattice groups is in progress, excellent agreement between the different lattice groups is found for the so-called intermediate window observable~\cite{Colangelo:2022vok,Ce:2022kxy,ExtendedTwistedMass:2022jpw,FermilabLatticeHPQCD:2023jof,RBC:2023pvn}. The evaluation of this intermediate window observable shows a $4$ standard deviation discrepancy between the lattice and the data-driven computation. 
On the $e^+e^-\to \text{hadrons}$ side, in addition to the known discrepancy between KLOE~\cite{KLOE-2:2017fda,KLOE:2008fmq,KLOE:2010qei,KLOE:2012anl} and BaBar~\cite{BaBar:2012bdw,BaBar:2009wpw}, the recent CMD-3~\cite{CMD-3:2023alj,CMD-3:2023rfe} result has shown a discrepancy 
with all previous measurements used in~\cite{Aoyama:2020ynm}. The origin of this discrepancy is currently unknown and efforts are in progress to clarify the situation~\cite{Colangelo:2022jxc}.
In view of this situation, a firm comparison with the theory cannot be established at the moment.

\section{Conclusion}
We have reported a measurement of the muon magnetic anomaly to \SI{0.20}{ppm} precision, based on the first three years of data. This measurement represents the most precise determination of this quantity.
The statistical and systematic errors have been reduced by a factor of two with respect to our first measurement~\cite{Run1PRL}, due to greater than four times more data and improved running conditions, analysis procedures, dedicated measurements, and systematic studies. 
This measurement is still statistically limited and the analysis of the remaining data from three additional years of data is expected to result in an improved statistical precision by another factor of approximately two.  

\begin{acknowledgments}
We thank the Fermilab management and staff for their strong support of this experiment, as well as the tremendous support from our university and national laboratory engineers, technicians, and workshops.
Greg Bock and Joe Lykken set the blinding clock and diligently monitored its stability.

The Muon \gmtwo Experiment was performed at the Fermi National
Accelerator Laboratory, a U.S. Department of Energy, Office of
Science, HEP User Facility. Fermilab is managed by Fermi Research
Alliance, LLC (FRA), acting under Contract No. DE-AC02-07CH11359.
Additional support for the experiment was provided by the Department
of Energy offices of HEP, NP, and ASCR (USA), the National Science Foundation
(USA), the Istituto Nazionale di Fisica Nucleare (Italy), the Science
and Technology Facilities Council (UK), the Royal Society (UK),
the National Natural Science Foundation of China
(Grant No. 12211540001, 12075151), MSIP, NRF and IBS-R017-D1 (Republic of Korea),
the German Research Foundation (DFG) through the Cluster of
Excellence PRISMA+ (EXC 2118/1, Project ID 39083149),
the European Union Horizon 2020 research and innovation programme under
the Marie Sk\l{}odowska-Curie grant agreements No. 101006726,
No. 734303 and European Union STRONG 2020 project under grant agreement No. 824093 
and the Leverhulme Trust, LIP-2021-01.
\end{acknowledgments}
 
\appendix

\section{Correlations between $\mathbf{\omega_a^
{m}}$ analyses}\label{app:correlation}

Table~\ref{tab:correlations} lists the correlations coefficients between the 19 different $\omega_a$ analyses.
The largest allowed statistical differences are between the event-based analyses and the energy-based analyses. Smaller allowed statistical differences are between analyses that employ either a common
construction approach or a common histogramming method. The correlation coefficients do not account for additional allowed systematic differences between analysis methods. 

\begin{table*}[hbt!]
\begin{center}
\caption{Table of correlation coefficients based on the allowed statistical differences between the 19 different $\omega_a$ analysis approaches. They include the different reconstruction procedures and different histogramming methods. They assume a 100\% correlation of systematic uncertainties between analysis approaches.}
\label{tab:correlations}
\begin{ruledtabular}
\begin{tabular}{lrrrrrrrrrrrrrrrrrrr}
 & C\_T & E\_T & I\_T & S\_T & W\_T & B\_A & C\_A & E\_A & I\_A & S\_A & W\_A & B\_RT & E\_RT & I\_RT & B\_RA & E\_RA & K\_Q & KR\_RQ\\
\hline
B\_T & 0.967 & 0.999 & 0.967 & 0.999 & 1.000 & 0.900 & 0.871 & 0.884 & 0.867 & 0.884 & 0.884 & 0.993 & 0.995 & 0.963 & 0.895 & 0.904 & 0.765 & 0.824\\
C\_T &       & 0.967 & 1.000 & 0.965 & 0.967 & 0.891 & 0.900 & 0.875 & 0.896 & 0.874 & 0.875 & 0.961 & 0.963 & 0.996 & 0.887 & 0.895 & 0.756 & 0.815\\
E\_T &       &       & 0.967 & 0.999 & 0.999 & 0.913 & 0.885 & 0.898 & 0.880 & 0.898 & 0.897 & 0.993 & 0.996 & 0.963 & 0.909 & 0.918 & 0.753 & 0.811\\
I\_T &       &       &       & 0.965 & 0.967 & 0.897 & 0.906 & 0.881 & 0.902 & 0.880 & 0.880 & 0.961 & 0.963 & 0.996 & 0.892 & 0.901 & 0.751 & 0.809\\
S\_T &       &       &       &       & 0.999 & 0.915 & 0.886 & 0.900 & 0.882 & 0.902 & 0.899 & 0.992 & 0.995 & 0.961 & 0.911 & 0.920 & 0.751 & 0.809\\
W\_T &       &       &       &       &       & 0.915 & 0.887 & 0.899 & 0.882 & 0.899 & 0.899 & 0.993 & 0.995 & 0.963 & 0.911 & 0.919 & 0.752 & 0.810\\
B\_A &       &       &       &       &       &       & 0.994 & 1.000 & 0.994 & 0.999 & 1.000 & 0.890 & 0.886 & 0.887 & 0.991 & 0.994 & 0.688 & 0.740\\
C\_A &       &       &       &       &       &       &       & 0.994 & 1.000 & 0.993 & 0.994 & 0.862 & 0.857 & 0.896 & 0.986 & 0.988 & 0.681 & 0.732\\
E\_A &       &       &       &       &       &       &       &       & 0.994 & 0.999 & 1.000 & 0.875 & 0.871 & 0.871 & 0.991 & 0.994 & 0.676 & 0.727\\
I\_A &       &       &       &       &       &       &       &       &       & 0.993 & 0.994 & 0.858 & 0.853 & 0.892 & 0.986 & 0.988 & 0.678 & 0.729\\
S\_A &       &       &       &       &       &       &       &       &       &       & 0.999 & 0.875 & 0.871 & 0.870 & 0.990 & 0.993 & 0.677 & 0.728\\
W\_A &       &       &       &       &       &       &       &       &       &       &       & 0.875 & 0.870 & 0.871 & 0.991 & 0.994 & 0.676 & 0.727\\
B\_RT &       &       &       &       &       &       &       &       &       &       &       &       & 0.994 & 0.962 & 0.902 & 0.907 & 0.758 & 0.825\\
E\_RT &       &       &       &       &       &       &       &       &       &       &       &       &       & 0.967 & 0.895 & 0.901 & 0.767 & 0.837\\
I\_RT &       &       &       &       &       &       &       &       &       &       &       &       &       &       & 0.895 & 0.901 & 0.750 & 0.819\\
B\_RA &       &       &       &       &       &       &       &       &       &       &       &       &       &       &       & 0.994 & 0.682 & 0.743\\
E\_RA &       &       &       &       &       &       &       &       &       &       &       &       &       &       &       &       & 0.689 & 0.754\\
K\_Q &       &       &       &       &       &       &       &       &       &       &       &       &       &       &       &       &       & 0.994\\
\end{tabular}

\end{ruledtabular}
\end{center}
\end{table*}

\section{Trolley calibration constants}
\begin{figure}[h]
    \centering
    \includegraphics[width=\columnwidth]{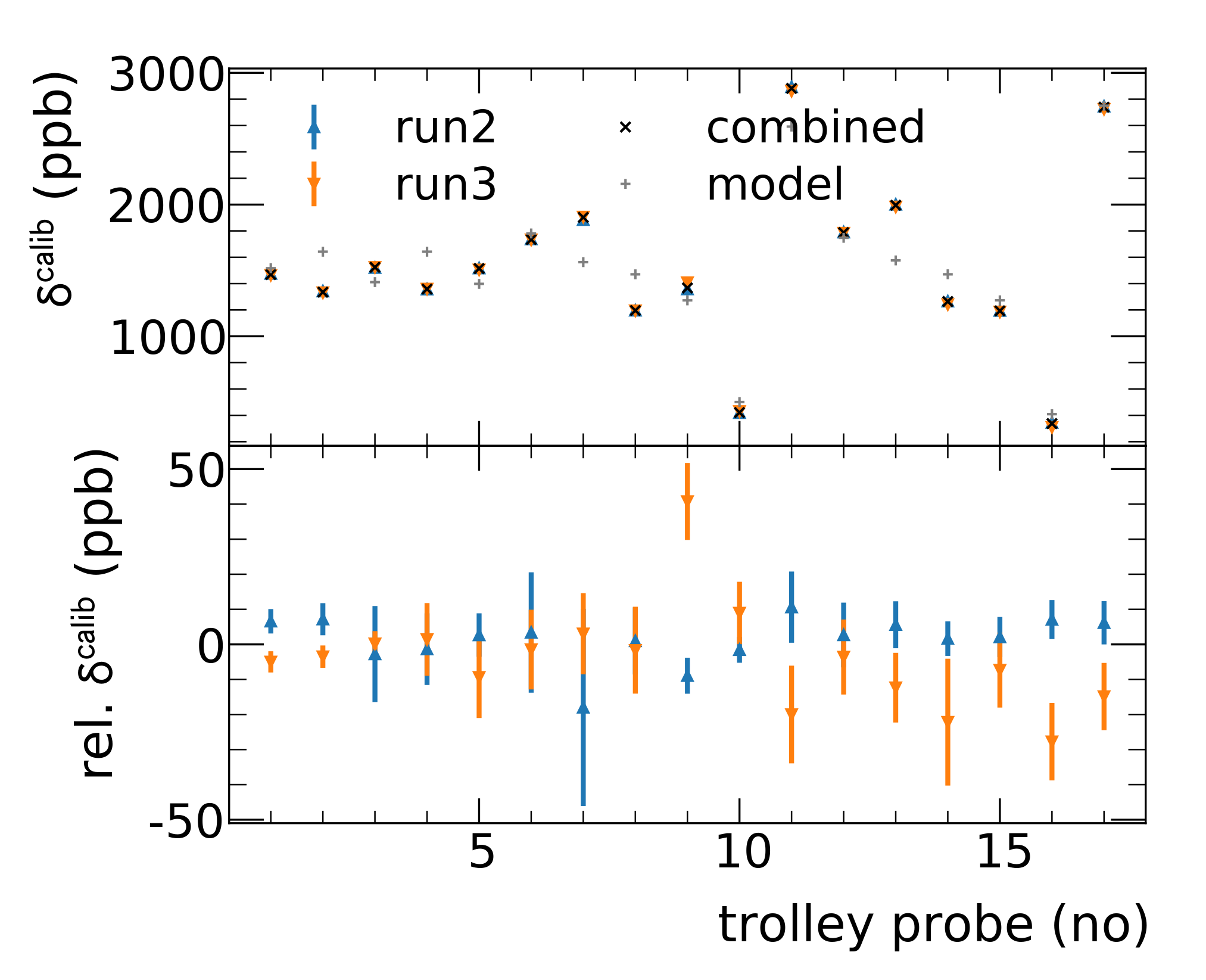}
    \caption{Top: Trolley calibration constants per trolley probe for \RunTwo (blue) and \RunThree (orange) and the combination (black). Predictions from \texttt{COMSOL} simulations (gray) with simplified geometry, which only considers the trolley shell, show qualitative consistency. Bottom: The difference of \RunTwo and \RunThree calibration constants with respect to the combined value that are used for this analysis.}
    \label{fig:calibrationConstants}
\end{figure}

The trolley calibration constants, including their contributions, are listed in Table~\ref{tb:PPCorrections}.
A graphic comparison is shown in Fig.~\ref{fig:calibrationConstants}. In addition to the \RunTwoThree average, the values and the differences from the dedicated \RunTwo and \RunThree calibration campaigns are shown, in combination with predictions from \texttt{COMSOL} simulations based on a simplified trolley geometry that only takes into account the trolley shell but not the interior details.

\begin{table*}
    \caption{Overview of trolley probe calibration constants $\delta^\text{calib}$ and individual contributions for run-2/3. All values are given in ppb.}
    \label{tb:PPCorrections}
    \begin{ruledtabular}
    \begin{tabular}{lrrrrrrrrr}
         & \multicolumn{2}{c}{$\delta^\text{fp,tr}$} & \multicolumn{2}{c}{$\delta^\text{fp,cp}$} & \multicolumn{2}{c}{$\delta^\text{av}$} & \multicolumn{2}{c}{$\delta^{s,\text{img}}$} & \multicolumn{1}{c}{$\delta_{n}^\text{calib}$} \\
        Probe & value & uncertainty & value & uncertainty & value & uncertainty & value & uncertainty & value \\ \hline
1 & \multirow{17}{*}{14.3}& \multirow{17}{*}{8} &   4.0&   4.0&   4.9&   1.9& -17.2&   8.9&  1469.0\\
  2 & &&   4.0&   4.0&  -0.2&   2.4& -17.8&   8.9&  1336.9\\
  3 & &&   3.7&   3.7&   1.8&   2.9& -17.2&   8.9&  1523.6\\
  4 & &&   4.0&   4.0&   2.8&   4.0& -17.8&   8.9&  1358.3\\
  5 & &&   4.9&   4.9&  -1.0&   3.1& -17.2&   8.9&  1514.4\\
  6 & &&   3.6&   3.6&   9.4&   4.4& -20.2&   9.1&  1734.5\\
  7 & &&   3.2&   3.2&  -9.5&   4.7& -19.4&   8.9&  1903.0\\
  8 & &&   3.2&   3.2&  -2.8&   2.9& -17.8&   8.9&  1195.8\\
  9 & &&   3.1&   3.1&   7.9&   4.0& -17.2&   8.9&  1367.2\\
 10 & &&   3.1&   3.1&   8.6&   3.4& -17.8&   8.9&  421.1\\
 11 & &&   3.1&   3.1&  19.7&   9.1& -19.4&   8.9&  2878.3\\
 12 & &&   3.7&   3.7&  40.9&   8.1& -20.2&   9.1&  1787.1\\
 13 & &&   4.4&   4.4&  -4.4&   4.4& -19.4&   8.9&  1993.8\\
 14 & &&   5.7&   5.7&   1.5&   6.1& -17.8&   8.9&  1263.9\\
 15 & &&   6.5&   6.5& -15.2&   6.5& -17.2&   8.9&  1193.0\\
 16 & &&   5.5&   5.5&  -1.0&   4.4& -17.8&   8.9&  337.2\\
 17 & &&   4.2&   4.2&   4.9&   8.3& -19.4&   8.9&  2738.5
    \end{tabular}
    \end{ruledtabular}
\end{table*} 

\bibliography{references}

\end{document}

\title{Detailed Report on the Measurement of the
  Positive Muon Anomalous Magnetic
  Moment to 0.20 ppm
  \\
  Supplemental Material
}

\maketitle

In this supplemental material, we report $\RmuprimeatTexp = \wa /
  \opprimetildeatTexp$ and \amu with extra digits, in order to avoid affecting
further elaborations with additional uncertainties due to rounding of values,
uncertainties and correlations to report them with the conventional limited
number of digits. \opprimetildeatTexp corresponds to the magnetic field $\tilde
  B$ averaged over space and time by the muons, expressed as the precession
frequency of protons in a spherical water sample at a reference temperature
$T_r=34.7\,^\circ$C.

While uncertainties on values are explicitly reported, uncertainties on
uncertainties and correlations are not to be understood to be represented by
the number of the reported digits: due to rounding of intermediate results and
estimates of systematics uncertainties, we estimate that uncertainties are
accurate to at least 1\% of their values, and correlations are accurate to at
least 1\% absolute.

\section{\RmuprimeatTexp measurements}

\subsection{FNAL (E989) measurements}

The FNAL (E989) Run 1 \RmuprimeatTexp measurement, corrected as mentioned in
the main text:
\begin{align*}
  \RmuprimeatTexp_\text{E989,\RunOne} = \dxuseValStatSyst{RmupT_E989_Run1}
\end{align*}
Above, the statistical uncertainty corresponds just to the \wa statistical uncertainty.

The FNAL (E989) \RunTwoThree measurement is:
\begin{align*}
  \RmuprimeatTexp_\text{E989,\RunTwoThree} = \dxuseValStatSyst{RmupT_E989_Run23}
\end{align*}
The statistical uncertainty includes the \wa statistical uncertainty and two relatively much smaller contribution with statistical nature:
\begin{itemize}
  \item the statistical uncertainty in the \opprimetildeatTexp measurement;
  \item the uncertainty due to time randomization of events in the \wa analyses.
\end{itemize}

In combining the FNAL (E989) \RunOne and \RunTwoThree, the systematic
uncertainties are conservatively assumed to be fully (100\%) correlated. The
result is:
\begin{align*}
  \RmuprimeatTexp_\text{E989} = \dxuseValStatSyst{RmupT_E989}.
\end{align*}

\subsection{BNL (E821) measurement}
We use the \Rmu value reported in Table I of the BNL Muon g-2 2006 final
report~\cite{BNLFinalReport}:
\begin{align*}
  {\mathcal R}_{\mu\,E821} = \dxuseValUnc{Rmu_E821}
\end{align*}
We use the reported statistical uncertainty on the magnetic anomaly $[a_{\mu\,\text{E821}} = \numprint{11659208.0}\ (5.4)\ (3.3)]$ to set the statistical contribution to ${\mathcal R}_{\mu\,\text{E821}}$, assigning the rest as systematic uncertainty.

We undo the diamagnetic shielding correction that has been applied to the BNL
(E821) \RmuprimeatTexp to obtain the published value of \Rmu:
\begin{align*}
  \RmuprimeatTexp_\text{E821} = {\mathcal R}_{\mu\,\text{E821}} / (1 - \sigmaprimeatTexp)~,
\end{align*}
where $\sigmaprimeatTexp = 2.5790\,(14)\times 10^{-5}$~\cite{Phillips:1977} is the diamagnetic shielding correction. We obtain:
\begin{align*}
  \RmuprimeatTexp_\text{E821} = \dxuseValStatSyst{RmupT_E821}.
\end{align*}

\subsection{Combined Experimental \RmuprimeatTexp measurement}

We combine the BNL and FNAL measurements assuming that there is no uncertainty
correlation and obtain:
\begin{align*}
  \RmuprimeatTexp_\text{Exp} = \dxuseValStatSyst{RmupT_exp}
\end{align*}

\section{Calculation of the muon magnetic anomaly}

The magnetic anomaly is computed as:
\begin{align*}
  \amu = \RmuprimeatTexp\ f_{\text{ext}}
\end{align*}
where
\begin{align*}
  f_{\text{ext}} = \frac{\mu'_p(T_{r})}{\mu_e(H)}\ \frac{{\mu_e(H)}}{\mu_e}\ \frac{m_{\mu}}{m_e}\ \frac{g_e}{2},
\end{align*}
and
\begin{itemize}
  \item $\mu'_p(T_{r}) / \mu_e(H)$ is the ratio of the magnetic moment of the proton in a spherical water sample at 34.7 C and the magnetic momentum of the electron in a hydrogen atom~\cite{Phillips:1977};
  \item $\mu_e(H) / \mu_e = (1-\sigma^{ep}_{\text{KI03}})$ is the ratio of the magnetic moment of the electron in a hydrogen atom and the magnetic momentum of the free electron in vacuum, obtained with a theory QED calculation \cite{Karshenboim:2003qv}, with precision limited by the number of reported digits;
  \item $m_\mu / m_e$ is the ratio of the muon and electron masses, taken from the CODATA 2018 fit, primarily driven by the LAMPF 1999 measurements of muonium hyperfine splitting~\cite{Liu:1999iz,CODATA:2018};
  \item $g_e$ is the electron gyromagnetic factor, computed from the electron anomaly $a_e = (g_e-2)/2$ world average~\cite{electronge, pdgelectronge}.
\end{itemize}

We neglect some negligible correlations on these precision constants, which
could be obtained from the CODATA fit output. The numerical values are:
\begin{align*}
\frac{\mu'_p(T_{r})}{\mu_e(H)} & =  \dxuseValUnc{muppsmuepH_34_7C_PCK77} \\
  \sigma^{ep}_{\text{KI03}}      & =  \dxuseValUncEE{sigmaep_KI03}         \\
\frac{{\mu_e(H)}}{\mu_e}       & =  \dxuseValUnc{muepHsmue_KI03}         \\
\frac{m_\mu}{m_e}              & =  \dxuseValUnc{mmusme_CODATA2018}      \\
  g_e/2 = 1 + a_{e}              & =  \dxuseValUnc{ae_PDG2023}             \\
f_{\text{ext}}                 & = \dxuseValUnc{amusRmupT}
\end{align*}
The obtained muon magnetic anomalies are:
\begin{align*}
  a_{\mu\,\text{E821}}              & =  \dxuseValStatSyst{amu_E821}       \\
  a_{\mu\,\text{E989,\RunOne}}      & =  \dxuseValStatSyst{amu_E989_Run1}  \\
  a_{\mu\,\text{E989,\RunTwoThree}} & =  \dxuseValStatSyst{amu_E989_Run23} \\
  a_{\mu\,\text{E989}}              & =  \dxuseValStatSyst{amu_E989}       \\
  a_{\mu\,\text{Exp}}               & =  \dxuseValStatSyst{amu_exp}        \end{align*}
The uncertainty of $f_{\text{ext}}$ is attributed to the systematic uncertainties.

\section{Dataset measurements}

The measurements of \oa, \opprimeatTexp and \RmuprimeatTexp for each of the Run
1, 2 and 3 datasets are provided in Tables~\ref{table:datasets-run1} and
\ref{table:datasets-run23}.
\begin{table}[tb]
  \caption{Measured  \oa, \opprimeatTexp and \RmuprimeatTexp
  for all datasets of Run 1, with total, statistical and systematic
  uncertainty, followed by the total uncertainty
  correlation matrices. There are no uncertainty correlations
  between the \oa and \opprimeatTexp measurements.}
  \label{table:datasets-run1}
  \begin{ruledtabular}
    \dxuse{table-Rmu-wa-wp-Run1-datasets-stat-syst}
  \end{ruledtabular}\vspace{1ex}

  \begin{ruledtabular}
    \dxuse{table-wa-corr-Run1-datasets}
  \end{ruledtabular}\vspace{1ex}

  \begin{ruledtabular}
    \dxuse{table-wp-corr-Run1-datasets}
  \end{ruledtabular}\vspace{1ex}

  \begin{ruledtabular}
    \dxuse{table-Rmu-corr-Run1-datasets}
  \end{ruledtabular}
\end{table}

\begin{table}[tb]
  \caption{Measured  \oa, \opprimeatTexp and \RmuprimeatTexp
  for all datasets of Run 2 and 3, with total, statistical and systematic
  uncertainty, followed by the total uncertainty
  correlation matrices. There are no uncertainty correlations
  between the \oa and \opprimeatTexp measurements.}
  \label{table:datasets-run23}
  \begin{ruledtabular}
    \dxuse{table-Rmu-wa-wp-Run23-datasets-stat-syst}
  \end{ruledtabular}\vspace{1ex}

  \begin{ruledtabular}
    \dxuse{table-wa-corr-Run23-datasets}
  \end{ruledtabular}\vspace{1ex}

  \begin{ruledtabular}
    \dxuse{table-wp-corr-Run23-datasets}
  \end{ruledtabular}\vspace{1ex}

  \begin{ruledtabular}
    \dxuse{table-Rmu-corr-Run23-datasets}
  \end{ruledtabular}
\end{table}

\bibliography{PRL-Run23}